\documentclass{aa}  

\usepackage{graphicx}
\usepackage{txfonts}
\usepackage[colorlinks = true,
            linkcolor = blue,
            urlcolor  = black,
            citecolor = cyan,
            anchorcolor = black]{hyperref}

\newcommand {\hi} {{\rm H}\,{\small\rm I}}

\newcommand {\hicap} {{\rm H}\,{\scriptsize\rm I}}

\newcommand {\kms} {\,{\rm km\,s}^{-1}}

\newcommand {\pc} {\,{\rm pc}}

\newcommand {\kpc} {\,{\rm kpc}}
\newcommand {\Mpc} {\,{\rm Mpc}}
\newcommand {\cmmq}{\,{\rm cm^{-2}}}

\newcommand {\kmskpc} {\,{\rm km\,s}^{-1}\,{\rm \kpc}^{-1}}

\newcommand {\miJyb} {\,{\rm mJy\,beam}^{-1}}

\newcommand {\de}{^{\circ}}

\newcommand {\msun}{\,{\rm M}_\odot}

\newcommand {\msunsqp}{\,{\rm M}_\odot \, {\rm pc}^{-2}}

\newcommand{\Myr}{\,{\rm Myr}}

\newcommand{\K}{\,{\rm K}}

\newcommand {\msunyr}{\,{{\rm M}_\odot\,\rm yr}^{-1}}

\newcommand{\modx}{\mathcal{M}(\mathbf{x})}
\newcommand{\avg}[1]{\left< #1 \right>} 
\newcommand{\bba}{$^{\scriptstyle 3\mathrm{D}}$B{\sc arolo}}
\def\apjl{ApJ}%
\def\apjs{ApJS}%
\def\ao{Appl.~Opt.}%
\def\aap{A\&A}%
\begin{document} 

\renewcommand{\figureautorefname}{Figure\!}
\title{HALOGAS: the properties of extraplanar HI in disc galaxies}
\titlerunning{HALOGAS: extraplanar HI in disc galaxies}
\authorrunning{A. Marasco et al.}

   \author{A. Marasco\inst{1,2},
          F. Fraternali\inst{2,3},
          G. Heald\inst{4},
          W.~J.~G. de Blok\inst{1,2,5},
          T. Oosterloo\inst{1,2},
          P. Kamphuis\inst{6},
          G.~I.~G. J\'ozsa\inst{7,8,9},
          C.~J. Vargas\inst{10},
          B. Winkel\inst{11},
          R.~A.~M. Walterbos\inst{12},
          R.~J. Dettmar\inst{6},
          \and
          E. Ju\"tte\inst{6}
          }

    \institute{ASTRON, Netherlands Institute for Radio Astronomy, Oude Hoogeveensedijk 4, 7991 PD, Dwingeloo, The Netherlands
         \and
                 Kapteyn Astronomical Institute, University of Groningen, Postbus 800, 9700 AV, Groningen, The Netherlands\\
                 \email{marasco@astro.rug.nl}
        \and
                Department of Physics and Astronomy, University of Bologna, via P. Gobetti 93/2, 40129 Bologna, Italy
        \and
        		CSIRO Astronomy and Space Science, PO Box 1130, Bentley WA 6102, Australia
        \and
        		Dept. of Astronomy, Univ. of Cape Town, Private Bag X3, Rondebosch 7701, South Africa
        \and
        		Ruhr-University Bochum, Faculty of Physics and Astronomy, Astronomical Institute, 44780 Bochum, Germany
        \and
        		Department of Physics and Electronics, Rhodes University, PO Box 94, Makhanda, 6140, South Africa
        \and
        		South African Radio Astronomy Observatory, 2 Fir Street, Black River Park, Observatory, Cape Town, 7405, South Africa
        \and
        		Argelander-Institut f\"ur Astronomie, Auf dem H\"ugel 71, D-53121 Bonn, Germany
	\and
		Department of Astronomy and Steward Observatory, University of Arizona, Tucson, AZ, U.S.A.
	\and
		Max-Planck-Institut f\"ur Radioastronomie, Auf dem H\"ugel 69, 53121 Bonn, Germany
	\and
		Department of Astronomy, New Mexico State University, Las Cruces, NM 88001, U.S.A.
             }

   \date{Received ; accepted}

 
\abstract
{
We present a systematic study of the extraplanar gas (EPG) in a sample of 15 nearby late-type galaxies at intermediate inclinations using publicly available, deep interferometric \hi\ data from the HALOGAS survey.
For each system we mask the \hi\ emission coming from the regularly rotating disc and use synthetic datacubes to model the leftover `anomalous' \hi\ flux. 
Our model consists of a smooth, axisymmetric thick component described by 3 structural and 4 kinematical parameters, which are fit to the data via a Bayesian MCMC approach.
We find that extraplanar \hi\ is nearly ubiquitous in disc galaxies, as we fail to detect it in only two of the systems with the poorest spatial resolution.
The EPG component encloses $\sim5-25\%$ of the total \hi\ mass, with a mean value of $14\%$, and has a typical thickness of a few kpc, incompatible with expectations based on hydrostatic equilibrium models.
The EPG kinematics is remarkably similar throughout the sample, and consists of a lagging rotation with typical vertical gradients of $\sim-10\kmskpc$, a velocity dispersion of $15-30\kms$ and, for most galaxies, a global inflow in both the vertical and radial directions with speeds of $20-30\kms$.
The EPG \hi\ masses are in excellent agreement with predictions from simple models of the galactic fountain powered by stellar feedback.
The combined effect of photo-ionisation and interaction of the fountain material with the circumgalactic medium can qualitatively explain the kinematics of the EPG, but dynamical models of the galactic fountain are required to fully test this framework.
}

\maketitle
%

\section{Introduction}\label{sec:intro}
Disc galaxies are characterised by the presence of a thin disc of neutral hydrogen (\hi) that can extend in radius far beyond the classical optical radius.
In the inner regions of discs, the \hi\ layers have a typical scaleheight of $100-200\pc$.
This thickness is well explained by assuming that the gas is in vertical \emph{hydrostatic equilibrium} within the galactic potential, i.e., the gas turbulent motions balance the vertical gravitational pull by the stellar disc and dark matter halo \citep{Olling95}.
In the outer regions of discs the \hi\ layer is expected to \emph{flare} due to the decrease of the vertical force, but within the optical radius the scaleheight is unlikely to vary by more than a factor two \citep{Bacchini+19}.
Thus, in the inner disc, where most of the star formation takes place, we should expect \hi\ to be confined within a few hundred parsec from the midplane.

The above expection is, in fact, not met by \hi\ observations as it was found that a number of disc galaxies keep a fraction (typically $10\%$ or more) of their \hi\ in a rather thick layer reaching up a few kiloparsecs above the midplane \citep{Swaters+97, Fraternali+01}.
In the few systems studied in detail, this layer of neutral gas turned out to show three key features: i) it has a scaleheight of typically $1-2 \kpc$, far beyond what one would expect from gas in hydrostatic equilibrium \citep[e.g.][]{Oosterloo+07}; ii) it rotates more slowly than the gas in the disc, showing a velocity gradient with height of order $-(10-20)\kmskpc$ \citep[e.g.][]{Schaap+00,Fraternali+05,Zschaechner+11}; iii) it is located mostly in the inner regions of the disc, showing a clear correspondence with the star forming disc \citep[e.g.][]{Fraternali+02,Boomsma+08}.
In this paper we generically refer to this component as \emph{extraplanar gas} (EPG), but in the literature one can find reference to it also as \emph{\hi\ halo}, as \emph{thick \hi\ disc} \citep{Kamphuis+13} or as `lagging' component \citep[because of its peculiar rotation,][]{Matthews&Wood03}.

Other than the above properties, in some galaxies, non-circular motions have also been detected. 
The EPG of NGC\,2403, UGCA\,105 and NGC\,4559 seems to have a coherent (radial) infall motion towards the centre of the galaxy \citep{Fraternali+02, Barbieri+05, Schmidt+14, Vargas+17}.
In NGC\,6946, a galaxy seen close to face-on, vertical motions of \hi\ are ubiquitously observed across the star forming disc and in clear connection to the EPG component \citep{Boomsma+08}.
The EPG of NGC\,891 shows the presence of outflow/inflow motions with an increase in velocity dispersion with height \citep{Oosterloo+07}.
On the whole, extraplanar \hi\ appears to be a somewhat separate component (both spatially and kinematically) from the thin \hi\ disc.

The presence of \hi\ at large distances from the disc has also been known in the Milky Way for decades.
\citet{Lockman84, Lockman02} found \hi\ clouds towards the inner galaxy reaching up to distances of a few hundred parsec from the midplane \citep[see also][]{DiTeodoro+18}.
Other clouds, known as high-velocity and intermediate-velocity clouds (HVCs and IVCs) \citep{Wakker&vanWoerden97}, are observed at anomalous velocities with respect to the normal rotation of the disc material.
HVCs and IVCs have distinct properties: while the former are typically located at distances of several kpc from the midplane and have sub-Solar metallicity, the latter are confined within a few kpc from the Sun and their metallicity is close to Solar \citep{Wakker01}.
This evidence, together with their different velocities may indicate different origins.
\citet[][hereafter MF11]{MF11} modelled the extraplanar layer of the Milky Way as seen in the all-sky LAB \hi\ survey \citep{Kalberla+05}.
Their model reveals that the Galactic EPG contains about $10\%$ of the total \hi, rotates with a gradient ${\rm d}v_{\phi}/{\rm d}z\simeq-15\kmskpc$ and has complex kinematics characterised by vertical and radial inflow toward the central regions of the disc.
While IVCs clearly appear as `local' features of the Galactic EPG, which is mostly built by unresolved clouds at larger distances, the HVCs seem to be a more distinct component with an origin that is still debated \citep[e.g.][]{Putman+12,Fraternali+15}.

Along with \hi, emission in H$\alpha$ and other optical lines is commonly observed around disc galaxies, probing the existence of extra-planar Diffuse Ionised Gas (DIG) layers at temperatures of $\sim10^4\K$ extending $1-2\kpc$ from the discs. 
There are several indications that these gas layers are the ionised counterpart of the \hi\ EPG: galaxies with larger SFRs show a more prominent DIG component \citep{RossaDettmar03} and the kinematics of this ionised gas is consistent with that of EPG, featuring both a lagging rotation \citep{Heald+06,Heald+07,Kamphuis+07} and non-circular motions \citep{Fraternali+04}.
Thus, the photoionisation of \hi\ EPG from bright stars in the disc is the most likely explanation for the DIG layers.

The formation of \hi\ EPG in disc galaxies has been investigated by different authors.
The mechanisms considered fall into three classes of models: equilibrium models, inflow models and galactic fountain.
The possibility that the EPG layer could be in equilibrium has been first explored by \citet{Barnabe+06} who built models of non-barotropic (baroclinic) gas layers and applied them to the observations of NGC\,891, finding that the gas temperature required to reproduce the data was $\sim10^5\K$, an order of magnitue above that of the warm \hi\ medium.
In a second step in this direction, \citet{Marinacci+10a} investigated the possibility of an equilibrium model where the random motions of the EPG are non-thermal and thus also suitable for a colder, `clumpy' component.
Within this framework, one can derive prescriptions equivalent to the hydrostatic equilibrium using the Jeans equations, with the further possibility of introducing anisotropic velocity dispersion.
The final result was that no model could fully reproduce the kinematics of NGC\,891's \hi\ EPG, but the best result could be obtained by introducing a strong anisotropy in the vertical direction, which is akin to what one finds in galactic fountain models (see below).

The second type of model proposed that the EPG layer is produced by gas accretion onto the galaxy discs from the external environment.
Galaxies are likely surrounded by large gas reservoirs as most baryons are found to be outside galaxies in the local Universe \citep[e.g.][]{Bregman07, Werk+13}.
For galaxies similar to the Milky Way, this reservoir is likely in the form of hot gas at nearly the virial temperature and contains a significant fraction of the missing baryons \citep[e.g.][]{Gatto+13,Miller&Bregman15}.
In this scenario, the extraplanar gas could be produced by the cooling of this so-called \emph{corona} in a cooling flow model.
\citet{Kaufmann+06} explored this possibility with SPH simulations and found that the kinematics of this accreting gas would in fact be similar to that observed for NGC\,891, but this idea later incurred two drawbacks.
First, it became clear that most of the cold gas in those simulations was caused by numerical effects \citep{Agertz+07, Kaufmann+09} and, in fact, unphysical.
Indeed, thermal instabilities are unlikely to develop in a corona akin to that surrounding the Milky Way \citep[][but see also \citet{Sormani+18} and references therein]{Binney+09, Joung+12}.
Second, the large mass of the extraplanar gas combined with the short dynamical time for accretion onto the disc (essentially the free fall time) would lead to accretion rates exceeding the star formation rate (SFR) of these galaxies by orders of magnitude \citep{FB08}.
Thus, it appears very unlikely that \emph{all} the extraplanar \hi\ in local galaxies could be explained by gas accretion. 
However, a fraction of EPG may well have such an origin \citep[e.g.][]{Marasco+12}.
Galaxies of lower mass are not expected to host a massive hot corona, but rather a number of `cold' ($\sim10^5\K$) gas filaments that connect the outer regions of their disc to the intergalactic space. 
These cold flows would constitute the main mode of gas accretion onto galaxies at high redshift, and should still be significant at low redshift in low mass systems. 
Again, it is unlikely that these flows, which are expected to merge with galaxy discs at large galactocentric radii, can be the origin of the extraplanar \hi, which is seen preferentially in the inner regions.

The other explanation for the presence of the EPG layer is that of a galactic fountain powered by stellar feedback \citep[e.g.][]{Shapiro&Field76, Collins+02}.
In this picture, the EPG is pushed up from the thin disc of the galaxy by the expansion of superbubbles around stellar OB associations.
Superbubbles are produced by the combined action of supernova explosions and stellar winds and can reach sizes much larger than a typical supernova bubble, thus exceeding the disc scaleheight \citep{MacLow+88}.
When a superbubble reaches the blowout, the cold material gathered in its shell is ejected into the halo region and so is its interior, which consists of rarefied hot gas.
In this scheme, the extraplanar \hi\ is made up mostly of material from the supershells \citep{Melioli+09} and partially of material from the hot bubble that is promptly cooling after the ejection \citep{Houck&Bregman90}.
In a series of papers, the kinematics of the extraplanar gas has been contrasted with the prediction of galactic fountain models with and without interaction between the fountain flow and the surrounding galactic corona \citep{FB06, FB08, Marinacci+10b}.
Both in the Milky Way and NGC\,891, the extent of the EPG layer can be well reproduced by these models, while its kinematics shows signs of interaction between the outflowing material and gas cooling from the corona.
In this scenario, the EPG would be formed mostly by fountain material with a percentage of $10-20\%$ of cooled coronal gas \citep{Fraternali+13,Fraternali17}.
The outflow velocities required by these models to reproduce the data are $\lesssim100\kms$, compatible with those measured for the high-velocity \hi\ component around the star-forming regions of nearly face-on galaxies, like NGC\,6946 \citep{Boomsma+08}.
These relatively low speeds rule out the possibility of a powerful outflow into the circumgalactic medium (CGM), suggesting a more gentle disc-halo gas circulation for the origin of the EPG.

In this paper, we use data from the HALOGAS survey \citep[][hereafter H11]{Heald+11} to investigate the presence and the properties of the EPG in a sample of 15 nearby galaxies.
This represents the first systematic investigation of the EPG properties in a sample of galaxies, and increases substantially the number of systems for which a detailed study of the EPG has been carried out.
We hereby show that the vast majority of late-type galaxies present a thick layer of EPG characterised by a slow rotation and a coherent inflow motion towards the galaxy centre.
We describe our sample and modelling methodology in Section \ref{sec:method}, and apply our method to the HALOGAS systems in Section \ref{sec:results}.
The interpretation of our findings in the context of the galactic fountain framework, together with a discussion on the limitations of our method, are presented in Section \ref{sec:discussion}.
We summarise our results in Section \ref{sec:conclusions}. 

\section{Method} \label{sec:method}
We infer the properties of the EPG by following a method analogous to that used by MF11 to study the extraplanar \hi\ of the Milky Way. 
This method is based on two main steps.
The first step consists of filtering out from the datacube the emission that comes from the regularly rotating thin \hi\ disc.
The second is to build simple parametric models of EPG, based on a limited number (7 in our case) of free parameters, that are fit to the data by producing synthetic \hi\ cubes and comparing them directly to the observations.
Our galaxy sample and the details of our method are described below. 
In Appendix \ref{app:mcmctest} we test our method on mock data.

\begin{table*}
\centering
\begin{minipage}{160mm}
\caption{Physical properties of galaxies in our sample, from H11 and \citet{Heald+12}. We also list the median kinematical inclination (INC$_{\rm BB}$) and position angle (PA$_{\rm BB}$) found with \bba\ and used in our EPG modelling.}
\label{tab:sample}
\begin{tabular}{lcccccccccc}
     \hline
     UGC & Other name & Type & dist.\ & INC$_{\rm H11}$ & INC$_{\rm BB}$ & PA$_{\rm BB}$ & D$_{\rm 25}$ & $M_B$ & $v_{\rm rot}$ & SFR\\
         &            &      & (Mpc) & ($\de$) & ($\de$) &($\de$) &($'$)& (mag) & ($\kms$) & ($\msunyr$)\\
     \hline
     1256 & NGC\,0672        &  SBcd &  	7.6 &		70 &	67.6	&64.2	&	6.4 &	-18.65 	     &  130.7 	  &  0.23\\
     1913 & NGC\,0925        &  SABd & 	 9.1 &	54 & 57.9	&284.7	&   	11.3 &  -19.66      &  102.4 &  0.77\\
     1983 & NGC\,0949        &	SAd &	11.3 &	52 &	52.5	&160.8	&	3.5 &	-17.85   & 90.9 &	0.31\\
     2137 & NGC\,1003        &	SAcd &	11.6 &	67 &	70.4	&276.3	&	6.3 &	-18.61   & 95.5 &	0.40\\
     3918 & NGC\,2403        &	SAcd &	3.2  &	62 &	62.5	&124.6	&	23.8 &-19.68  &121.9 &	0.6\\
     4284 & NGC\,2541        &	SAcd &	12.0 &	67 &	63.8	&171.8	&	7.2 &	-18.37  & 92.1 & $     0.55^a$\\
     5572 & NGC\,3198        &	SBc &	14.5 &	71 &	70.0	&214.3	&	8.8 &	-19.62  & 148.2 &	1.1\\
     7045 & NGC\,4062        &	SAc &	16.9 &	68 &	67.1	&100.1	&	4.5 &	-18.27  & 140.5 &	0.67\\
     7353 & NGC\,4258 	 &	SABbc &	7.6 &		71 &	74.0 &331.9	&	17.1 &-20.59  & 208.0 &	1.7\\
     7377 & NGC\,4274        &	SBab &	19.4 &	72 &	71.3 &279.8	&	6.5 &	-19.22  & 239.9 &	1.2\\
     7539 & NGC\,4414        &	SAc &	17.8 &	50 &	53.9	&159.7	&	4.5 &	-19.12  & 224.7 &	4.2\\
     7591b & NGC\,4448      &	SBab &	9.7 & 	71 &	73.5	&94.4	&	3.8 &	-18.43  & 221.6 &	0.056\\
     7766 & NGC\,4559        &	SABcd &	7.9 &		69 &	68.0 &323.1	&	11.3&-20.07  & 113.4 &	0.69\\
     8334 & NGC\,5055	 &	SAbc &	8.5 &		55 &	65.2 &99.4	&	13.0&-20.14  & 215.5 &	2.1\\
     9179 & NGC\,5585        &	SABd &	8.7 &		51 &	50.4	&48.2	&	5.5 &	-17.96    & 79.1 &	0.41\\
     \hline
\end{tabular}
$^{a}$ based on TIR+UV measurements from \citet{Thilker+07} and re-scaled to the distance used in this work. \citet{Heald+12} report only an upper limit based on a non-detection with IRAS $25\mu{\rm m}$.
\end{minipage}
\end{table*}

\subsection{The sample} \label{ssec:sample}
The HALOGAS sample of H11 comprises 24 late-type galaxies, partitioned into 9 nearly edge-on and 15 intermediate inclination systems.
With the exception of NGC\,2403 and NGC\,891, for which pre-existing deep \hi\ data were already available, all galaxies have been re-observed at the Westerbork Synthesis Radio Telescope (WSRT) for a total integration time per galaxy of $10\times12$ hr in order reach \hi\ column density sensitivities of a few $\times10^{19}\cmmq$.
These rank amongst the deepest interferometric \hi\ observations of galaxies in the local Universe and constitute the best dataset to study the faint EPG emission in nearby galaxies\footnote{All \hicap\ cubes and moment maps are publicly available as part of the HALOGAS Data Release 1 \citep{HALOGASDR1}.}.
Comparisons between the WSRT and GBT (single-dish) \hi-fluxes indicate that not more than $\sim2\%$ of the \hi\ masses in and around galaxies should be missed by the HALOGAS survey due to the lack of short baselines and/or sensitivity \citep{Pingel+18}.

In this work we focus on the subset of 15 galaxies seen at intermediate inclinations ($50\de<i<72\de$), that allow the separation of extraplanar gas given its peculiar kinematics.
The main physical properties of these galaxies are listed in Table \ref{tab:sample} and column density maps are shown in Figure \ref{app:maps}.
While our modelling technique can also be applied to highly inclined systems, some of the steps in the method described below would require significant modifications.
In fact, in edge-on galaxies, the EPG separation can not be done on the basis of kinematics alone (Section \ref{ssec:separation}); rotation curves must be derived with a different fitting technique and information on the gas motion in the direction perpendicular to the disc is no longer accessible (Section \ref{ssec:model}).
In order to provide a homogenous analysis throughout the entire sample, we limit our study to systems at intermediate inclination. 
Separate studies of the edge-on galaxies in HALOGAS have been undertaken \citep{Zschaechner+11, Zschaechner+12, Kamphuis+13, Zschaechner+15}.

The HALOGAS survey provides \hi\ datacubes at two different spatial resolutions, $\sim15\arcsec$ and $\sim30\arcsec$.
In this work we use exclusively the cubes at $\sim30\arcsec$.
This gives us two advantages: a higher column density sensitivity, mandatory to study the faint-level emission around galaxies, and fewer independent resolution elements to model, which alleviates the computational cost of our software.
A lower spatial resolution also washes out the small-scale fluctuations in the gas density distribution that would never be represented by our smooth, axisymmetric models, while still providing enough information to trace the global EPG parameters that we try to derive.
Dealing with fewer independent data points also implies a significant gain in computational speed.
 
Most original HALOGAS cubes are significantly oversampled, with about $\sim9$ spaxels\footnote{The term `spaxel' refers to 2D pixels in the (ra, dec) space, while `voxel' is used for 3D pixels in the (ra, dec, $v_{\rm hel}$) space, $v_{\rm hel}$ being the line-of-sight velocity with respect to the Sun.} per FWHM along the beam major axis. 
Also, the target galaxy often occupies only a small part of the total field of view.
To increase the computational speed, we have trimmed the cubes in order to discard most of the empty background and maintain preferentially the central galaxies, and re-binned them to $3$ spaxel per beam FWHM (mean on both axes), sufficient to represent the data without information loss. 
 
\subsection{Separation of extraplanar gas}\label{ssec:separation}
\begin{figure*}
\begin{center}
\includegraphics[width=0.9\textwidth]{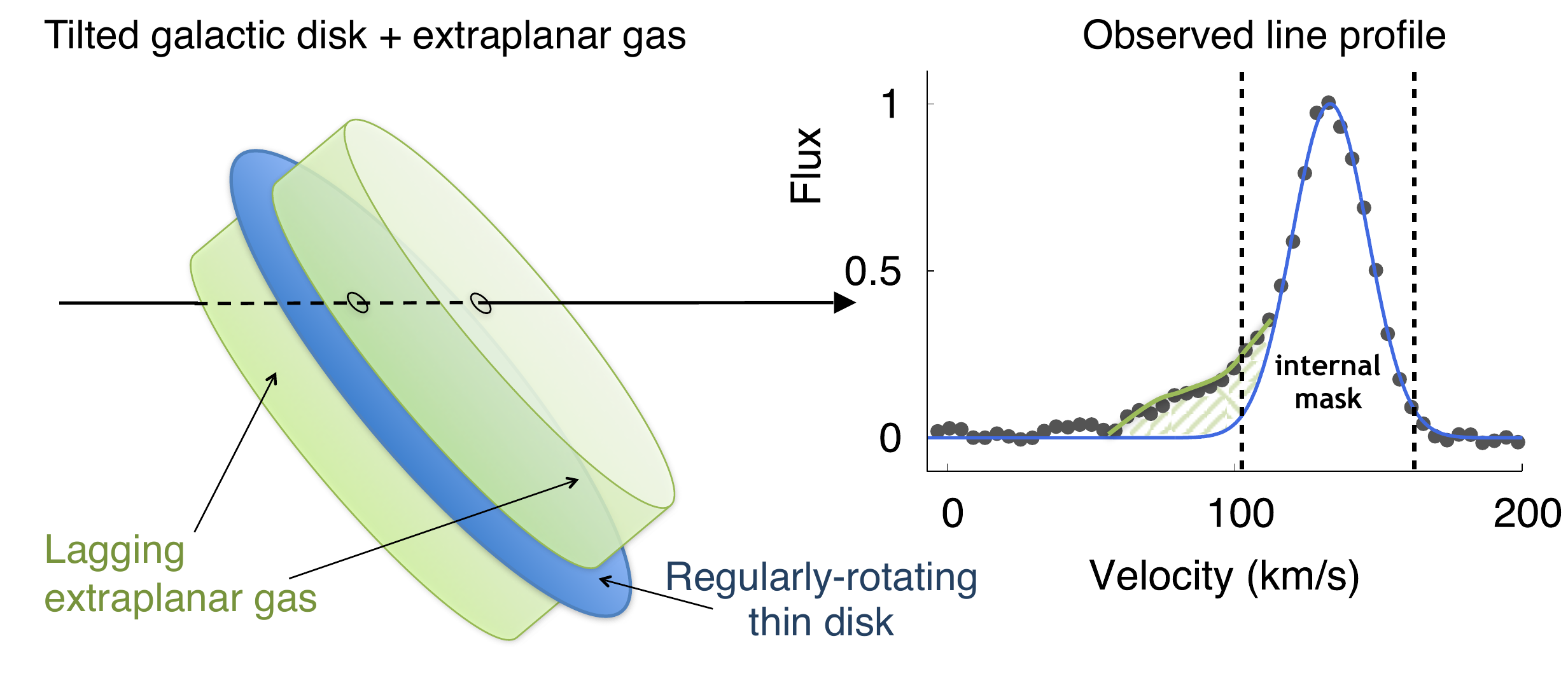}
\caption{A sketch of an observation of extraplanar gas in a galaxy seen at intermediate inclination.
The typical line profile (along the kinematic major axis of the galaxy) will be composed by a nearly Gaussian part coming from the thin disc with overlaid a tail at low rotation velocity produced by the lagging EPG layer.
The width of the disc emission is roughly symmetrical, produced by gas turbulence and well fitted by a Gaussian function (blue solid line).
The EPG separation is achieved by masking a portion of the profile with substantial contribution from this Gaussian function (`internal mask' region).
}
\label{fig:sketch}
\end{center}
\end{figure*}

The separation of the EPG emission from the disc emission is largely inspired by the technique firstly pioneered by \citet{Fraternali+02} and is based on the assumption that the kinematics of the EPG differs from that of the gas that regularly rotates within the disc.
The collisional nature of gas implies that two components with distinct kinematics along the same line-of-sight are likely to occupy distinct locations within the galaxy.

To give an example, in Fig.\ \ref{fig:sketch} we sketch how line-profiles get modified by having a layer of slowly rotating gas superimposed on the normally rotating thin disc.
If the disc is very thin, the resolution of the data is good and the inclination is not extreme (say $i< 80\de$), then the profiles are broadened only by the gas turbulence and are reasonably well described by Gaussian distributions.
The presence of a slowly rotating component shows up then as an asymmetric \emph{tail} extending towards the systemic velocity (i.e.\ at lower rotational velocities). 
This tail is clearly visible along the kinematic major axis of the galaxy and it tends to fade around the minor axis \citep[unless non-circular motions are also present, see][]{Fraternali+01}.
The line profile (dots) in Fig.\ \ref{fig:sketch} is extracted from a location along the major axis in the HALOGAS datacube of NGC\,3198 \citep[][hereafter G13]{Gentile+13}, normalized and shifted in velocity to a systemic velocity of $v_{\rm sys}=0\kms$.
The Gaussian fit (blue solid line) is obtained by excluding the prominent tail visible at velocities between $\sim50$ and $120\kms$.
The fit perfectly reproduces the high-velocity side of the profile.
A separation of the EPG can be achieved either by subtracting this Gaussian fit from the whole profile, or by masking the emission at velocities where the Gaussian fit has a high signal-to-noise ratio (e.g., the region within the vertical dashed lines in Fig.\,\ref{fig:sketch}).
The latter approach is typically more conservative and is that adopted in this study.

The above procedure represents the original disc/EPG separation strategy of \citet{Fraternali+02}.
It works well for moderately inclined galaxies seen at a high spatial resolution, so that beam-smearing effects do not influence the shape of the line profiles.
Here, we have revisited their strategy and implemented some improvements, preferentially to deal with cases where the spatial resolution is not optimal.
We firstly produce a 3D mask in order to filter out in each velocity channel the regions where \hi\ emission is absent. 
To do so, we smooth spatially the datacube by a factor of 5 (using a 2D Gaussian kernel) and set to unity (zero) all voxels with intensity above (below) a $4\sigma$ threshold, with $\sigma$ being the new rms-noise in the smoothed dataset.
We have checked by eye that, in all galaxies, this procedure outputs a mask that is generous enough to account for the entire emission coming from the galaxy while simultaneously filtering out any noise spikes occurring in the outer regions.
This \emph{external} mask is immediately applied to the original data before any additional procedure.

The next step is modelling the Gaussian part of all the line profiles, in order to describe the main \hi\ disc.
We select all profiles in the (masked) data with peaks above $4$ times the original rms-noise and fit a Gaussian function to their `upper' $40\%$ portion, i.e. we only consider flux densities above 40\% of the line peak.
Profiles with intensity peaks below the imposed $4\sigma$ threshold are excluded from the analysis and are effectively incorporated within the internal mask (see below).
Extensive experiments with visualisation tools and with mock data allowed us to establish that these values assure a proper characterisation of the Gaussian and the most efficient separation of the EPG.
We have also experimented with Gauss-Hermite polynomials and attempted to fit only the high-velocity narrower portion of the line profiles, finding the above procedure to give the best results.

We then build a cube made of all these Gaussian fits. 
This cube represents a non-parametric, approximated model for the \hi\ disc, which we use to develop an \emph{internal} mask to separate disc and EPG emission in the data.
Unfortunately, our simplistic approach to the disc modelling does not account for beam-smearing effects, which may produce deviations from Gaussianity in the shape of the line profiles even in the absence of a genuine, kinematically anomalous component.
If the spatial resolution is not optimal, these effects become a major concern especially in the inner regions of galaxies where the rotation curve has a steep gradient.
In order to mitigate these issues we further convolve the Gaussian cube with the data beam.
Tests done on synthetic data showed that this step helps in filtering out the disc emission in the innermost regions of poorly resolved systems and significantly improves the estimations of the EPG parameters (see Appendix \ref{app:mcmctest}).
Finally, we convert the Gaussian cube into a mask by setting to unity (blank) all voxels with intensity below (above) twice the data rms-noise.
Applying this internal mask to the data means effectively blanking all regions dominated by \hi\ emission coming from the regularly rotating disc: the remaining \hi\ flux is dominated by the EPG.
We illustrate this procedure on the \hi\ data of NGC\,3198 in Section \ref{ssec:3198}.

It is important to stress that the procedure above highlights only \emph{a fraction} of the emission coming from the EPG or, vice-versa, overestimates the contribution of the disc component to the total emission.
This can be easily appreciated in Fig.\,\ref{fig:sketch}, as the lagging EPG probably contributes to the total emission also at velocities larger than $120\kms$, but it is subdominant with respect to the disc contribution and virtually inseparable from the latter.
In fact, in the case of a purely lagging EPG the tail in the profiles would completely disappear along the galaxy minor axis, where no signature of rotation is visible, and one may naively conclude that EPG in galaxies is systematically more abundant along the major axis, which would certainly be a bizzarre result.
These considerations highlight the difficulties of measuring directly the properties of the EPG, like its mass, on the basis of a pure kinematical decomposition.
The approach adopted in this work attempts to bypass these limitations by building synthetic datacubes of EPG based on parametric models and fitting them to the data, having both the observed and the synthetic cubes filtered in the same manner.

\subsection{The model of the EPG layer} \label{ssec:model}
We model the EPG layer as an axisymmetric, smooth distribution characterised by three geometrical and four kinematical parameters.
The density distribution of the EPG follows the model developed by \citet{Oosterloo+07} and applied to the EPG of NGC\,891, where the surface density profile $\Sigma(R)$ is given by
\begin{equation}\label{eq:surf_halo}
\Sigma(R) = \Sigma_0 \left(1 + \frac{R}{R_{\rm g}}\right)^{\gamma} \exp \left(-\frac{R}{R_{\rm g}}\right)\,,
\end{equation}
where $\Sigma_0$ is the central surface density, $\gamma$ is an exponent regulating the density decline towards the centre and $R_{\rm g}$ is an exponential scale-length. 
For $\gamma>1$, the surface density increases with radius peaking at $R = R_{\rm g}(\gamma-1)$, and declines exponentially further out.

At a given $R$, the density distribution in the $z$ direction is given by
\begin{equation}\label{eq:vert_halo}
\rho(z) \propto \frac{\sinh(|z|/h)}{\cosh^2(|z|/h)}\,,
\end{equation}
where $h$ is the EPG scale-height.
This is an empirical formula that represents well the vertical EPG distribution in NGC\,891, and gives a density that is zero in the galactic midplane, reaching a maximum at $z=0.88h$ and declining exponentially at larger distances.
Eq.\,(\ref{eq:vert_halo}) may seem to be an unusual parametrisation for a gas density distribution, but has two main advantages.
The first is that it features a `hole' for $z\rightarrow0$, where the EPG would in fact vanish within the regularly rotating disc.
We find this preferable with respect to an exponential or a Gaussian distribution, which would leave us with the additional issue of defining a (somewhat arbitrary) separation threshold between the disc and the EPG layer in the $z$ direction.
The second advantage is purely numerical, as the inverse of the cumulative distribution function has an analytical form, which dramatically speeds up the extraction of random numbers from this distribution.
For simplicity, and to minimise the number of free parameters, we have decided not to consider a flare in the EPG distribution, but rather a radially constant scale-height $h$.
We discuss the implications of this assumption in Section \ref{ssec:limitations}.
As in MF11, the normalisation of the density distribution - which ultimately sets the mass of the EPG - is computed by re-scaling the flux of the synthetic cube to that of the data, and therefore is not a free parameter of the model.

The EPG kinematics is described by four parameters: the vertical gradient in the gas rotational speed (${\rm d}v_{\phi}/{\rm d}z$), the velocities in the radial and in the vertical directions ($v_{\rm R}$ and $v_{\rm z}$), and gas velocity dispersion $\sigma$.
Thus the EPG is allowed to rotate with a different speed with respect to the material within the disc (or to not rotate at all, for ${\rm d}v_{\phi}/{\rm d}z\ll0$).
It can globally accrete onto or escape from the galaxy, can move in/out and have a different velocity dispersion.
These simple kinematical parameters allow us to model different scenarios, from a nearly-spherical accretion with negligible angular momentum, to a galactic fountain cycle, to powerful nuclear galactic winds. 

\begin{figure}
\begin{center}
\includegraphics[width=0.47\textwidth]{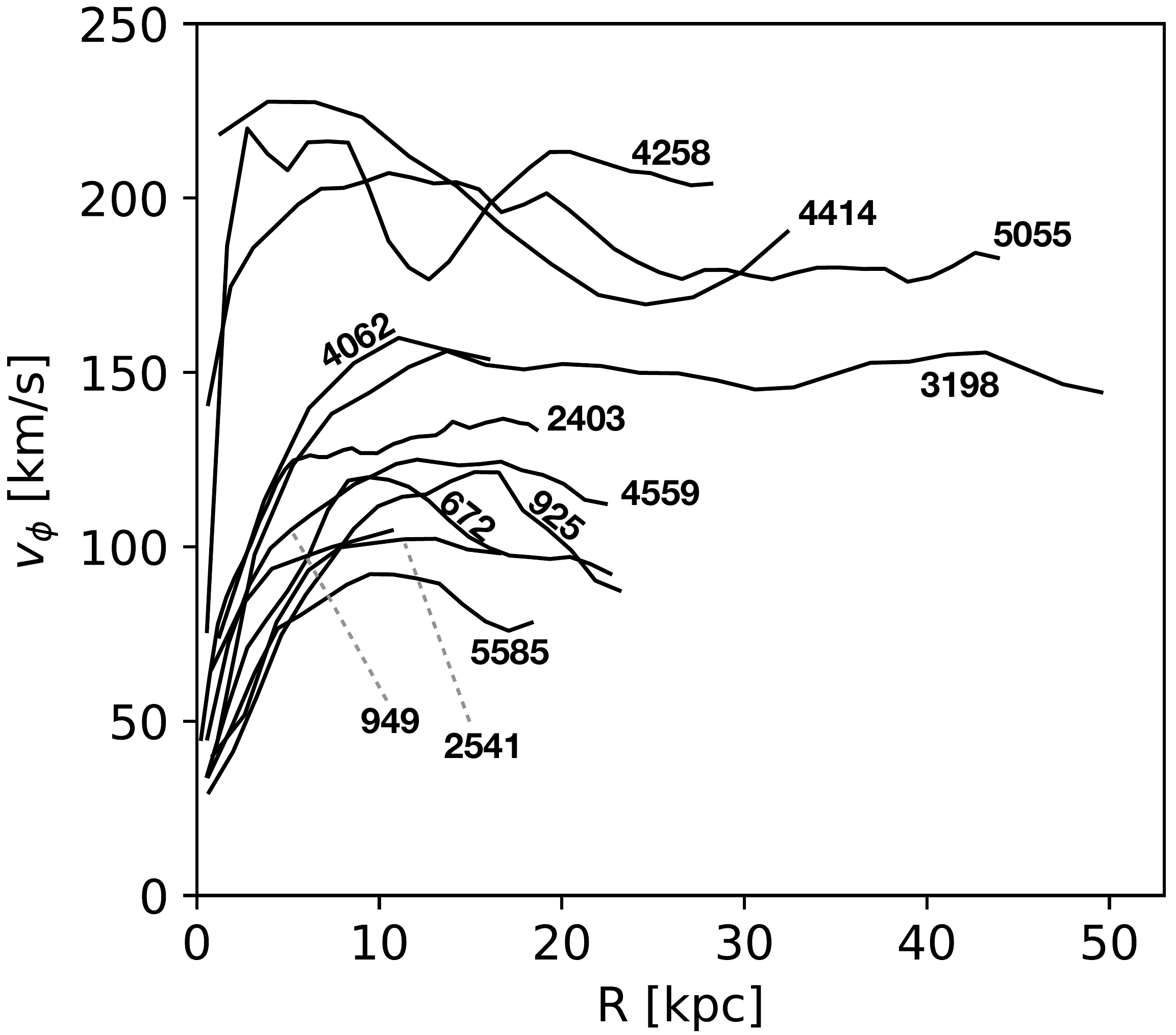}
\caption{Rotation curves for our sample of HALOGAS galaxies derived with \bba\ and used in our EPG modelling. Each system is labelled with its NGC number.} 
\label{fig:rcurve}
\end{center}
\end{figure}

Along with the rotational lag, a further ingredient is required to model the rotation of the gas: the galaxy rotation curve.
While HALOGAS galaxies are well studied nearby systems, studies of their rotation curves are rather scattered in the literature and feature different methods applied to a variety of datasets.
In the philosophy of carrying out a homogeneous analysis of the HALOGAS dataset, we have re-derived the rotation curves of all galaxies in our sample using the 3D tilted ring modelling code \bba\ \citep{Barolo}, adopting as a first estimate for their inclination the values from H11 reported in Table \ref{tab:sample} and fixing the kinematical centre to their optical centre.
In Table \ref{tab:sample} we also list the median inclination (INC) and position angle (PA) found by \bba\ for each galaxy, which will be adopted in the EPG modelling procedure (Section \ref{ssec:synthetic}).
The new inclinations are compatible with those of H11 within a few degrees, with the noticeable exception of NGC\,5055 for which our estimate ($65\de$, compared to $55\de$ of H11) is representative only for the innermost $20\kpc$ \citep[see Section \ref{ssec:epgproperties} and Fig.\,83 in][]{deBlok+08}.
In Fig.\,\ref{fig:rcurve} we show the rotation curves for the 12 systems for which we detect an EPG component: they are largely compatible with those available in the literature once re-scaled to the appropriate distance, as expected in well-resolved nearby disc galaxies.

\subsection{From the model to the synthetic cube} \label{ssec:synthetic}
For a given choice of $\mathbf{x}=(R_{\rm g}, \gamma, h, {\rm d}v_{\phi}/{\rm d}z, v_{\rm R}, v_{\rm z}, \sigma)$ we build a stochastic realisation of our model in the full 6D phase-space.
This is achieved by extracting random `particles' in a Cartesian space ($x,y,z$) - where $x$ and $y$ define the galaxy plane - following the density distribution defined by eq.\,(\ref{eq:surf_halo}) and (\ref{eq:vert_halo}).
Each particle has an associated ($v_x$,$v_y$,$v_z$) velocity vector that depends on the kinematical parameters adopted.
We stress that the particles are by no means representative of single \hi\ clouds, and are simply used as a convenient way to sample a continuous gas distribution.
In Appendix \ref{app:npart} we discuss in detail how we determine the optimal number of particles associated to each model.

The next step is to apply a rotation matrix to the particle positions and velocities in such a way that the system, to an observer located along the $z$-axis, would appear at a given INC and PA.
This is obtained by first rotating the particle set along the $x$ axis by INC degrees, and then around the $z$ axis by (PA$+90$) degrees\footnote{This assumes that the gas rotates counter-clockwise in the initial Cartesian frame.}.
The underlying assumption is that the EPG layer is \emph{not} warped, thus its sky projection can be described by single values for INC and PA, which is justified by the fact that EPG is typically observed within the inner regions of galaxies where warps are negligible (see also Section \ref{ssec:limitations}). 
Clearly, such assumption simplifies significantly the computation of the projection effects.
We account for the gas velocity dispersion by adding to the line-of-sight velocity of each particle a random (isotropic) velocity extracted from a Gaussian distribution with standard deviation given by $\sigma$.
We note that here $\sigma$ represent a \emph{total} velocity dispersion along the line-of-sight, and incorporates all microscopic (e.g. temperature) and macroscopic (e.g. cloud-to-cloud random motions) effects that can affect the broadening of the \hi\ profiles.

Using the galaxy distance $D$, this particle set is then transferred into a sky (ra,dec,$v_{\rm hel}$) 3D grid with voxel size equal to that in the observed datacube.
At this stage, the intensity in a given voxel of the synthetic cube has still no physical units, and simply represents the number of particles in that voxel.
For a given \hi\ mass $M_{\rm HI}$, we convert all intensities into Jy/beam units by multiplying them by $(M_{\rm HI}/\msun)\,I_{\rm tot}^{-1}\,\Lambda^{-1}$, where $I_{\rm tot}$ is the sum of all intensities in the cube and $\Lambda$ is a conversion factor given by
\begin{equation}\label{eq:lambda}
\Lambda = 2.067\times10^5 \left(\frac{\Delta_{\rm v}}{\kms}\right) \left(\frac{D}{\Mpc}\right)^2 \left(\frac{\Delta_{\rm ra}\Delta_{\rm dec}}{{\rm arcsec}^2}\right) \left(\frac{{\rm arcsec}^2}{B_{\rm maj} B_{\rm min} }\right)\,
\end{equation}
where $\Delta_{\rm ra}$ and $\Delta_{\rm dec}$ are the spaxel size, $\Delta_{\rm v}$ is the channel separation, and $B_{\rm maj}$ and $B_{\rm min}$ are the FWHM along the beam major and minor axes \citep[see eq.\,(3) and (5) in][]{Iorio+17}. 
The resulting cube is then spatially smoothed to the resolution of the observations, using a 2D Gaussian kernel with axis ratio and orientation given by the $B_{\rm maj}$, $B_{\rm min}$ and $B_{\rm pa}$ keywords in the data header.
To better match the velocity resolution of our syntehtic cube to that of the HALOGAS data we further perform Hanning smoothing, convolving consecutive velocity channels with a 0.25-0.50-0.25 triangular kernel.
This has a negligible impact on our results, given the typical velocity dispersion of the EPG ($15-30\kms$) compared to the typical velocity resolution in HALOGAS (FWHM of $\sim8\kms$, corresponding to a standard deviation of $3.4\kms$).

\begin{table}
 \centering
   \caption{EPG parameters for the toy-models shown in Fig.\,\ref{fig:toy_pv}. A description of the various parameters can be found in Table\,\ref{tab:parameters}.}
   \label{tab:toy_param}
   \setlength\tabcolsep{3.0pt}
   \begin{tabular}{lccccccc}
     \hline
     name & $R_{\rm g}$ & $\gamma$ & $h$ & ${\rm d}v_{\phi}/{\rm d}z$ & $v_{\rm z}$ & $v_{\rm R}$ & $\sigma$\\
     	      &\scriptsize{[$\kms$]}& &\scriptsize{[$\kpc$]} & \scriptsize{[$\kmskpc$]}& \scriptsize{[$\kms$]} & \scriptsize{[$\kms$]} & \scriptsize{[$\kms$]}\\
     \hline
     \texttt{fiducial} & $2$ & $3$ & $1.0$ & $-15$ & $0$ & $0$ & $20$\\
     \texttt{slowrot} & $2$ & $3$ & $1.0$ & $-30$ & $0$ & $0$ & $20$\\
     \texttt{fastrot} & $2$ & $3$ & $1.0$ & $-8$ & $0$ & $0$ & $20$\\
     \texttt{vert\_in} & $2$ & $3$ & $1.0$ & $-15$ & $-20$ & $0$ & $20$\\
     \texttt{vert\_out} & $2$ & $3$ & $1.0$ & $-15$ & $+20$ & $0$ & $20$\\
     \texttt{rad\_in} & $2$ & $3$ & $1.0$ & $-15$ & $0$ & $-20$ & $20$\\
     \texttt{rad\_out} & $2$ & $3$ & $1.0$ & $-15$ & $0$ & $+20$ & $20$\\
     \texttt{highdisp} & $2$ & $3$ & $1.0$ & $-15$ & $0$ & $0$ & $40$\\
     \texttt{lowdisp} & $2$ & $3$ & $1.0$ & $-15$ & $0$ & $0$ & $10$\\
     \texttt{Rg\_H} & $3$ & $3$ & $1.0$ & $-15$ & $0$ & $0$ & $20$\\
     \texttt{Rg\_L} & $1$ & $3$ & $1.0$ & $-15$ & $0$ & $0$ & $20$\\
     \texttt{gamma\_H} & $2$ & $4$ & $1.0$ & $-15$ & $0$ & $0$ & $20$\\
     \texttt{gamma\_L} & $2$ & $2$ & $1.0$ & $-15$ & $0$ & $0$ & $20$\\
     \texttt{thicker} & $2$ & $3$ & $2.0$ & $-15$ & $0$ & $0$ & $20$\\
     \texttt{thinner} & $2$ & $3$ & $0.5$ & $-15$ & $0$ & $0$ & $20$\\   
     \hline
   \end{tabular}
\end{table}

\begin{figure*}
\begin{center}
\includegraphics[width=0.96\textwidth]{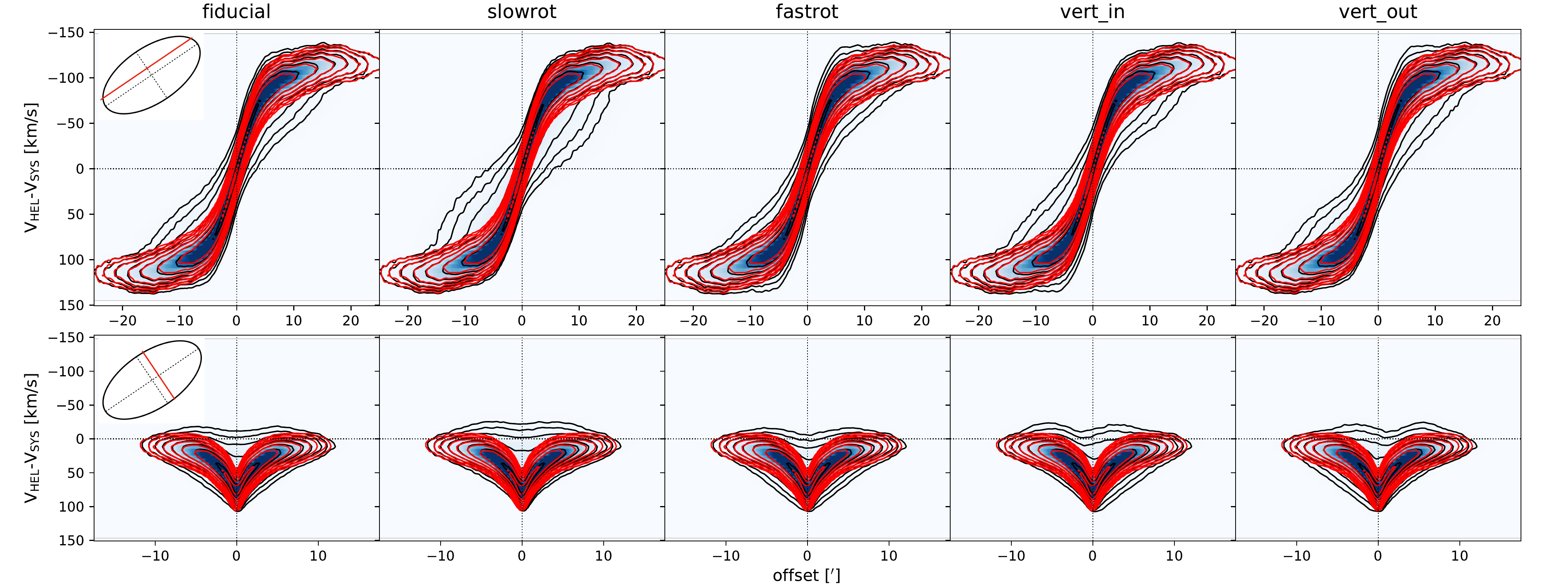}
\includegraphics[width=0.96\textwidth]{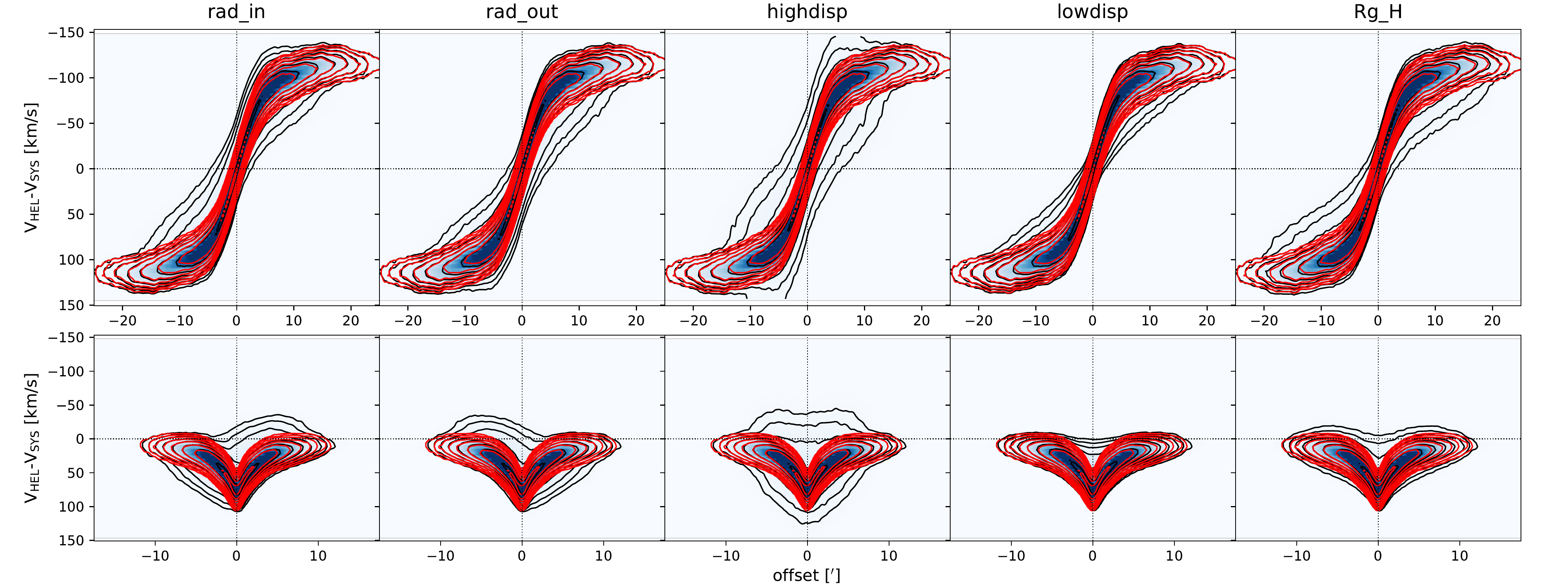}
\includegraphics[width=0.96\textwidth]{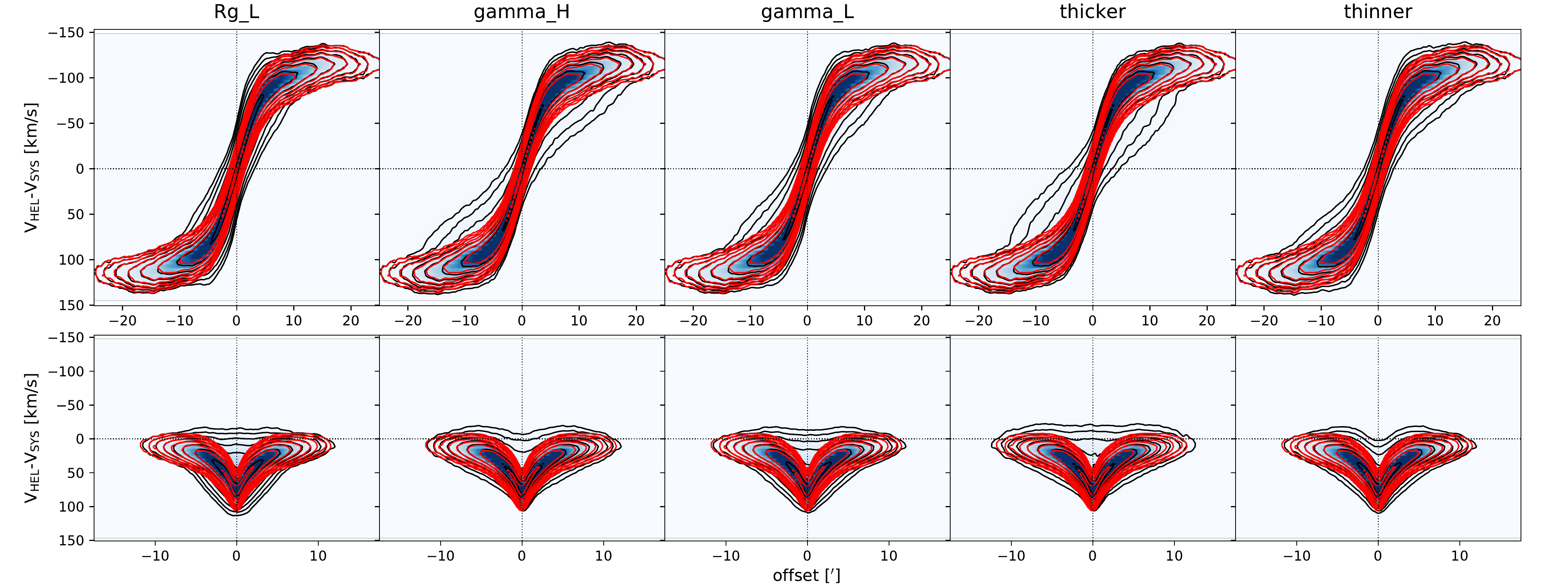}
\caption{Position-velocity (pv) slices for the disc+EPG toy models discussed in Section \ref{ssec:synthetic} and based on the parameters listed in Table\,\ref{tab:toy_param}. For each model we show a pv slice parallel to the major (on the top) and to the minor (on the bottom) axes. The insets in the top-left panels sketch the slice directions. Axis-offset is $2'$ in both cases. The slice thickness is 3 spaxels, corresponding to a resolution element. Black contours are spaced by powers of 2, the outermost being at an intensity level of $0.4\miJyb$. Red contours show the emission from the disc alone, without the contribution of the EPG.} 
\label{fig:toy_pv}
\end{center}
\end{figure*}

We now illustrate the effect of the various model parameters on the synthetic cubes.
For this purpose we build a series of toy-models made of two components, a thin disc and a thick EPG layer, the latter enclosing $15\%$ of the total \hi\ mass which we take as $3\times10^9\msun$ using NGC\,2403 as a reference.
The disc is kept fixed in all models and follows the rotation curve of NGC\,2403; the radial surface density and the vertical density profiles follow Gaussian distributions \citep[e.g.][]{Martinsson+13} centred at $(R,z)\!=\!(0,0)$ with standard deviations of $7.3\kpc$ and $0.1\kpc$ respectively,
The properties of the EPG layer varies model by model and are listed in Table\,\ref{tab:toy_param}.
All systems are seen at an inclination of $60\de$. 
We have adopted the same resolution and grid size of the VLA cube of NGC\,2403 to build the synthetic data ($30\arcsec\simeq0.5\kpc$), in order to show the signatures of the various parameters in a well-resolved, clean case.

Fig.\,\ref{fig:toy_pv} shows, for each model, a position-velocity (pv) diagram parallel to the major axis (on top) and another parallel to the minor axis (bottom), both with an axis-offset of $+2'$.
The red contours highlight the \hi\ emission from the disc alone, without the EPG contribution, and are the same in all panels.
In all models the EPG emerges as a faint, kinematically distinct \hi\ feature which is extended preferentially towards the systemic velocity, producing the so-called `beard' \citep{Schaap+00} in the major axis pv slices.
Clearly, different parameters leave different imprints on the synthetic cube.
As expected, the effects of varying the EPG rotation can be better appreciated along the major axis (see models \texttt{slowrot} and \texttt{fastrot}), while radial motions are readily visible in slices parallel to the minor axis (models \texttt{rad\_in} and \texttt{rad\_out}).
We note that in all pv-slices the approaching and the receding sides are perfectly symmetric by construction, except for the cases where radial or vertical motions are present.
Position-velocity plots exactly along the major axis (i.e. no offset, not shown here) are symmetric in all cases.

\begin{figure}
\begin{center}
\includegraphics[width=0.45\textwidth]{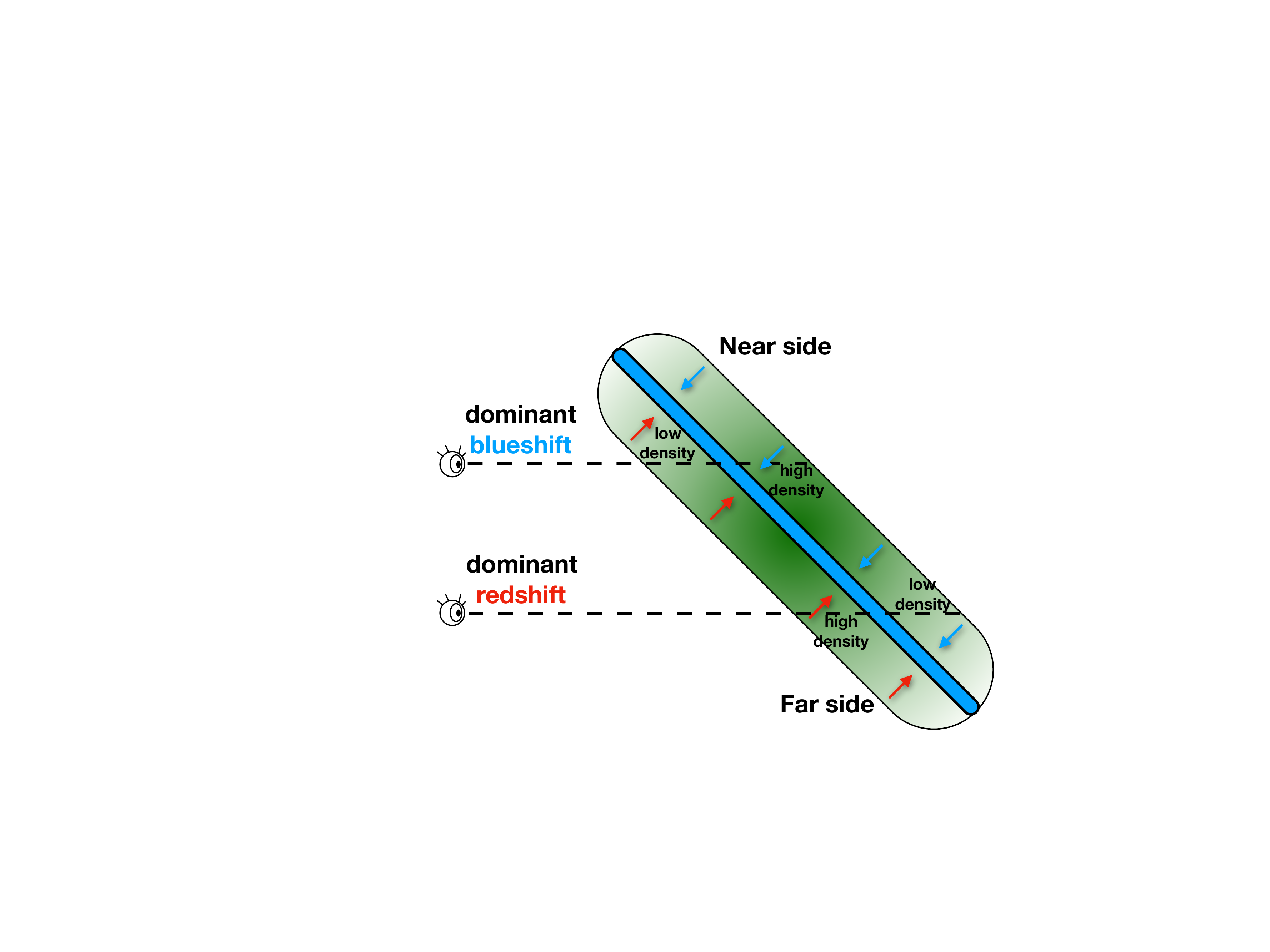}
\caption{Sketch showing how vertical motions produce a distinct signature on the inferred \hicap\ kinematics depending on the EPG thickness, inclination, orientation and surface density profile. In this case, the presence of a vertical inflow (red and blue arrows) and a radially declining surface density profile (shades of green) produce a redder or bluer-shifted signal depending on which galaxy side is considered.} 
\label{fig:sketch_vertical}
\end{center}
\end{figure}

Other effects, like those produced by varying $v_{\rm z}$, are more subtle.
Given the symmetry of our models, in an optically-thin regime one would expect that gas that moves perpendicularly to the midplane, either outflowing from it ($v_{\rm z}>0$) or inflowing onto it ($v_{\rm z}<0$), leaves the same signature on the cube.
An intuitive example is that of a face-on galaxy: gas in the foreground moving towards the observer (i.e., moving away from the disc) would be indistinguishable from background gas accreting onto the galaxy.
Quite remarkably, Fig.\,\ref{fig:toy_pv} demonstrates instead that even the sign of vertical motions can be distinguished, as one can appreciate by comparing models \texttt{vert\_in} and \texttt{vert\_out}.
We show why this is possible in Fig.\,\ref{fig:sketch_vertical}, where we sketch the case of a galaxy seen at an intermediate inclination where the EPG is  inflowing vertically towards the disc and has a radially declining \hi\ surface density. 
For a given inclination, thickness and - most noticeably - radial density profile of the EPG, a sight-line piercing through the system will encounter different densities above and below the midplane, which would cause a preferentially blue-shifted or red-shifted \hi\ signal.
Since the EPG kinematics, thickness and density profile are all fit to the data simultaneously, we are confident that our approach can exploit the subtle signatures in the data and correctly recover the gas vertical motions.
The fact that we derive coherent kinematical properties in most HALOGAS galaxies (Section \ref{ssec:epgproperties}) indicates that this is the case (see also Appendix \ref{app:mcmctest}).

Fig.\,\ref{fig:toy_pv} also highlights possible degeneracies in the parameter space.
The effect of varying the vertical rotational gradient is somewhat degenerate with varying the halo thickness, meaning that thicker, rapidly rotating halos appear similar to thinner, slowly rotating ones (see also MF11).
Another possible degeneracy is that between vertical and radial motions, as both produce similar asymmetries in the pv-slices.
In general, the signature of these parameters on the data is not unique but depends on the surface density profile and on the thickness of the EPG, for reasons analogous to those discussed above for the vertical motions.
Therefore, in order to infer their values from the data, it is of fundamental importance to fit both the EPG geometry and the EPG kinematics at the same time by employing a method that can automatically account for the degeneracy in the parameter space.
We describe our fitting strategy in the Section below.

Finally, we stress that the sign of both radial and vertical motions is ultimately determined from the knowledge of the near/far sides of the system.
For instance, in the example sketched in Fig.\,\ref{fig:sketch_vertical}, the signature of vertical motions would be the same if we considered vertical outflows rather than inflows and switched the two galaxy sides.
In other words, if the `true' 3D orientation of the galaxy is unknown, the values of $v_{\rm R}$ and $v_{\rm z}$ can be determined up to a sign.
We determine the near/far side of each HALOGAS galaxy using the commonly adopted assumption of \emph{trailing} spiral arms, which are readily recognisable in all systems from optical/UV images.
After fitting the model to the data as described in the Section below, we change the sign of the inferred ($v_{\rm R}$, $v_{\rm z}$) couple, depending on whether the 3D orientation of our model matches that of the galaxy under consideration. 

\subsection{Fitting the model to the data} \label{ssec:fit}
We use a Bayesian approach to determine the optimal parameters for the EPG of each galaxy in our sample.
The probability $\mathcal{P}$ of our parameters $\mathbf{x}$ given some data $\mathcal{D}$ is given by
\begin{equation}\label{eq:bayes}
\mathcal{P}(\mathbf{x}|\mathcal{D}) \propto
        \mathcal{P}(\mathcal{D}|\mathbf{x})\,\mathcal{P}(\mathbf{x})
\end{equation}
(Bayes' theorem) where $\mathcal{P}(\mathcal{D}|\mathbf{x})$ is the likelihood function and $\mathcal{P}(\mathbf{x})$ is the prior.

As in \citet{Marasco+17}, we write the likelihood as 
\begin{multline}\label{eq:likelihood}
 \mathcal{P}(\mathcal{D}|\mathbf{x}) \propto \prod^{\rm n.voxels} \exp \left(-\frac{|\modx-\mathcal{D}|}{\epsilon} \right) = \\ 
 = \exp \left(-\sum^{\rm n.voxels}\frac{|\modx-\mathcal{D}|}{\epsilon} \right) = \exp\left(-\mathcal{R}(\mathbf x)/\epsilon\right),
\end{multline}
where $\modx$ indicates the model cube obtained with a given choice of parameters, $\mathcal{R}(\mathbf x)$ is the sum of the \emph{absolute} residuals between the model and the data, $\epsilon$ is the uncertainty in the data (assumed to be constant over the whole dataset) and the sum is extended to all voxels outside the internal mask (see Section \ref{ssec:separation}).
The external mask, which sets to zero the intensities in the regions outside the galaxy, is not applied to the synthetic cubes in order to penalise models where the EPG emission extends too far out either in radius or in velocity.

Eq.\,(\ref{eq:likelihood}) does not represent a standard $\chi^2$ likelihood, which is instead based on the sum of the \emph{squared} residuals.
Our choice is driven by the necessity of giving a larger weight to the faint-level emission around galaxies, with respect to what a $\chi^2$ likelihood would provide.
We have experimented with other residuals - $(\mathcal{D}-\mathcal{M})^2$ or $|\mathcal{D}-\mathcal{M}|/\rm{max}(\mathcal{D},\mathcal{M})$ - finding the absolute residuals to perform better on the basis of visual inspection of position-velocity diagrams.
Note that the use of absolute residuals has featured in previous works \citep[e.g. MF11,][]{Marasco+12}, and is the `default' fitting method in \bba.
 
The value of $\epsilon$ in eq.\,(\ref{eq:likelihood}) is particularly relevant as it affects the width of the posterior distributions, which give the uncertainty on the fit parameters (see also Section \ref{sec:uncertainties}).
In principle $\epsilon$ should be the rms-noise of the datacube, corrected for the fact that adjacent voxels are correlated.
However, given that our model is a simple axisymmetric approximation of the EPG, the requirement that it fits the data to the noise level is not realistic and would lead to a severe underestimation of the uncertainty associated to our parameters. 
To account for this, we use instead an \emph{effective} uncertainty that takes into account the deviation from axisymmetry in the EPG component, computed as follows.
Consider $I(i,j,k)$ to be the intensity in a given voxel with ra, dec and velocity coordinates given by $i$, $j$ and $k$, where $I(0,0,0)$ is the intensity at the galaxy centre and at the systemic velocity.
In a pure axisymmetric system we expect that $\delta I \equiv I(i,j,k)-I(-i,-j,-k)\simeq0$ or, more precisely, that the standard deviation of the distribution of $\delta I$, $\sigma_{\delta I}$, is equal to the rms-noise in the data.
We can then use $\sigma_{\delta I}$ as an estimator for the deviation from pure axisymmetry.
While other methods to quantify \hi\ asymmetries in galaxies have been proposed in the past \citep[e.g.][]{Holwerda+11, vanEymeren+11, Giese+16}, our approach is better suited for the current study as it uses simultaneously the information on the spatial and kinematic distribution of the gas.
Using the centre and systemic velocity of each galaxy, we compute $\sigma_{\delta I}$ using all voxels that have not been filtered out by either the external or the internal masks, obtaining values ranging typically from $1$ to $8$ times the rms-noise in the data.
We then set $\epsilon\!=\!\sigma_{\delta I}\times n_{\rm vpr}$, $n_{\rm vpr}$ being the number of voxel per resolution element, approximated by
\begin{equation}\label{eq:npix}
n_{\rm vpr} \simeq 2 \times \frac{\pi}{4\ln(2)} \frac{B_{\rm min} B_{\rm maj}}{\Delta_{\rm ra} \Delta_{\rm dec}}
\end{equation}
where the factor $2$ at the beginning accounts for Hanning smoothing.
As we have $n_{\rm vpr}\sim20$, the values of $\epsilon$ ranges from $20$ to $160$ times the data rms noise.

\begin{table}
 \centering
   \caption{Summary of the geometrical and kinematical parameters for our EPG model and of the priors adopted. All priors are uniform within the range indicated.}
   \label{tab:parameters}
   \setlength\tabcolsep{3.0pt}   
   \begin{tabular}{lccc}
     \hline
     Parameter & description & prior range & units \\  
     \hline
     $R_{\rm g}$  &  eq.\,(\ref{eq:surf_halo}) & $(0,50]$ & $\kpc$\\
     $\gamma$  & eq.\,(\ref{eq:surf_halo}) & $[-50,50]$ & $-$\\	
     $h$  & eq.\,(\ref{eq:vert_halo}) & $(0,50]$ & $\kpc$\\
     ${\rm d}v_{\phi}/{\rm d}z$ & rotational gradient & $[-100,100]$ & $\kmskpc$\\
     $v_{\rm R}$ & radial velocity & $[-300,300]$ & $\kms$\\	
     $v_{\rm z}$ & vertical velocity & $[-300,300]$ & $\kms$\\	
     $\sigma$ & velocity dispersion & $(0,100]$ & $\kms$\\	
     \hline
   \end{tabular}
\end{table}

Table \ref{tab:parameters} summarises the model parameters and our choice of priors, all uniform within reasonable ranges.
We sample the posteriors with an affine-invariant Markov Chain Monte Carlo (MCMC) method, using the \texttt{python} implementation by \citet{emcee}.
We use 100 walkers and a number of chain steps varying from 1000 to 2000, depending on the galaxy in exam.
The chains are initialised by distributing the walkers in a small region around a minimum in the parameter space, which is determined via a Downhill Simplex minimisation routine \citep{NelderMead65}\footnote{This typically returns a \emph{local} minimum. The MCMC chains quickly depart from this initial position.}.
As discussed in Section \ref{ssec:model}, the EPG mass of each model is set by normalising the flux in the synthetic cube to that of the data, and is stored at each step of the chains. 
At the end of the process, the chains are inspected by eye in order to determine the initial `burn-in' chunk that must be discarded\footnote{This can vary from 200 to $\sim1000$ steps.}, while the remaining portion is used to sample the posterior probability.

In Appendix \ref{app:mcmctest} we test our fitting strategy on artificial data and show that the various degeneracies discussed in Section \ref{ssec:synthetic} are not complete: our method can recover the various input parameters with (very) good accuracy, thanks to the simultaneous fit to the EPG morphology and kinematics.

\section{Results}\label{sec:results}
\subsection{The EPG of NGC\,3198}\label{ssec:3198}
To illustrate our method, we first treat in detail a single, reference system: NGC\,3198.
This galaxy was already studied by G13 using the same HALOGAS data as in the current study.
G13 experimented with different models and concluded that the \hi\ emission from this galaxy could not be reproduced by a single disc component, but it required the contribution of a thick ($3\kpc$ scale-height) layer of EPG featuring a vertical rotational lag of $-15\kmskpc$ in the approaching side and $-7\kmskpc$ in the receding side.
The EPG, which was assumed to have the same surface density profile of the whole gas distribution, would account for $\sim15\%$ of the total \hi\ content.

\begin{figure}
\begin{center}
\includegraphics[width=0.5\textwidth]{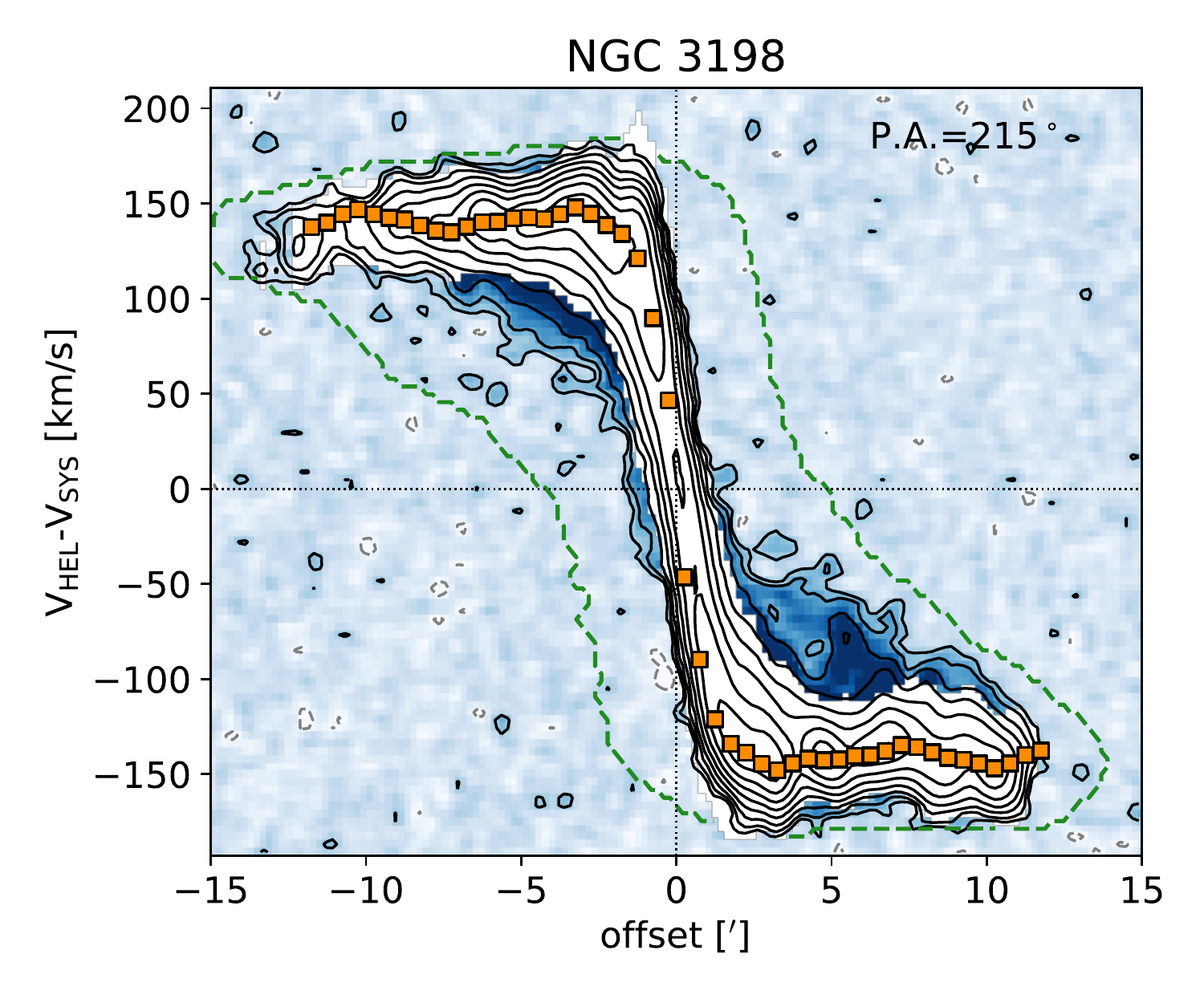}
\caption{Position-velocity slice along the major axis of NGC\,3198 ($1'\simeq4.2\kpc$ at the distance adopted). The white area shows the emission from the thin \hicap\ disc (internal mask), derived as discussed in Section \ref{ssec:separation}. The region outside the green dashed contour (external mask) is not associated to the galaxy and is set to zero in the data. The orange squares show the rotation curve derived with \bba. Black contours are spaced by powers of 2, the outermost being at an intensity level of $0.28\miJyb$ ($2\sigma_{\rm noise}$). A negative contour ($-2\sigma_{\rm noise}$) is shown as a dashed grey contour.} 
\label{fig:n3198_mask}
\end{center}
\end{figure}

We follow the method described in Section \ref{ssec:separation} to filter out the emission from the thin disc and from regions not associated to the galaxy.
To illustrate the effect of this procedure, in Figure \ref{fig:n3198_mask} we show a pv-slice along the major axis of the galaxy where we have highlighted the regions associated to the internal (white area) and to the external (green dashed contour) masks.
Clearly, the masking of the disc leaves out a low-velocity tail of \hi\ emission (the so-called `beard') preferentially from within the inner regions of the galaxy.
As discussed above, the internal mask is applied to both the data and the models, while the external mask is used to filter out from the data potential \hi\ sources not associated to the galaxy but it is not applied to the models.

Figure \ref{fig:n3198_mask} also shows the rotation curve derived with \bba\ (orange squares).
\bba\ uses a tilted ring approach to model the whole \hi\ emission in the 3D data.
As the emission is vastly dominated by the thin disc component, we expect that the presence of the EPG has a negligible impact on the inferred rotation curve.
Still, in order to give a higher weight to the bright emission from the disc, we have configured \bba\ so that it uses $\chi^2$-like residuals to infer the quality of the fit.
Note that, since we use axisymmetric templates, unlike G13 we do not attempt to model the approaching and the receding sides of the galaxy separately but fit the whole system simultaneously.
Aside from this, our rotation curve is perfectly compatible with that determined by previous studies \citep[][G13]{Begeman89,deBlok+08}.

\begin{figure*}
\begin{center}
\includegraphics[width=0.8\textwidth]{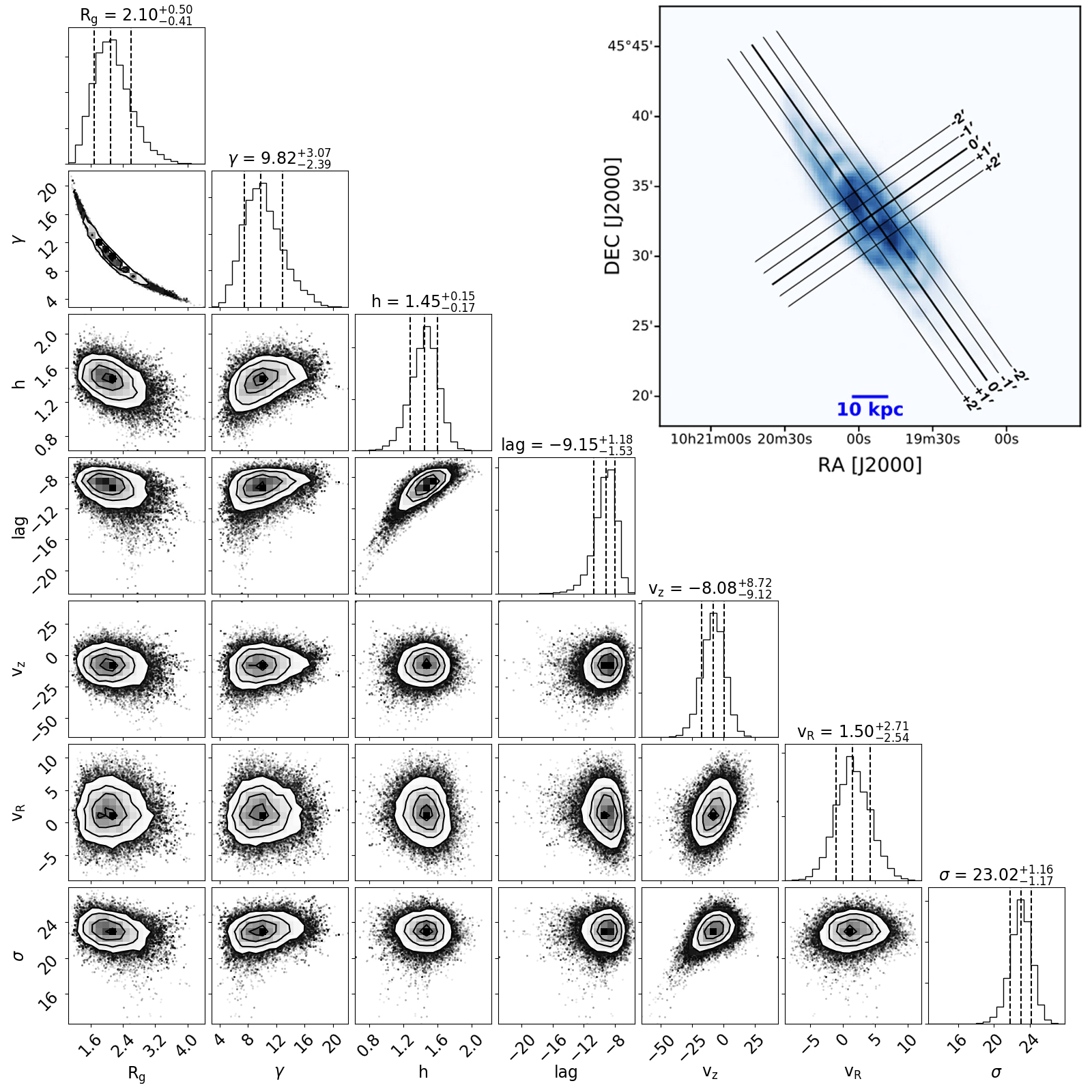}
\includegraphics[width=1.0\textwidth]{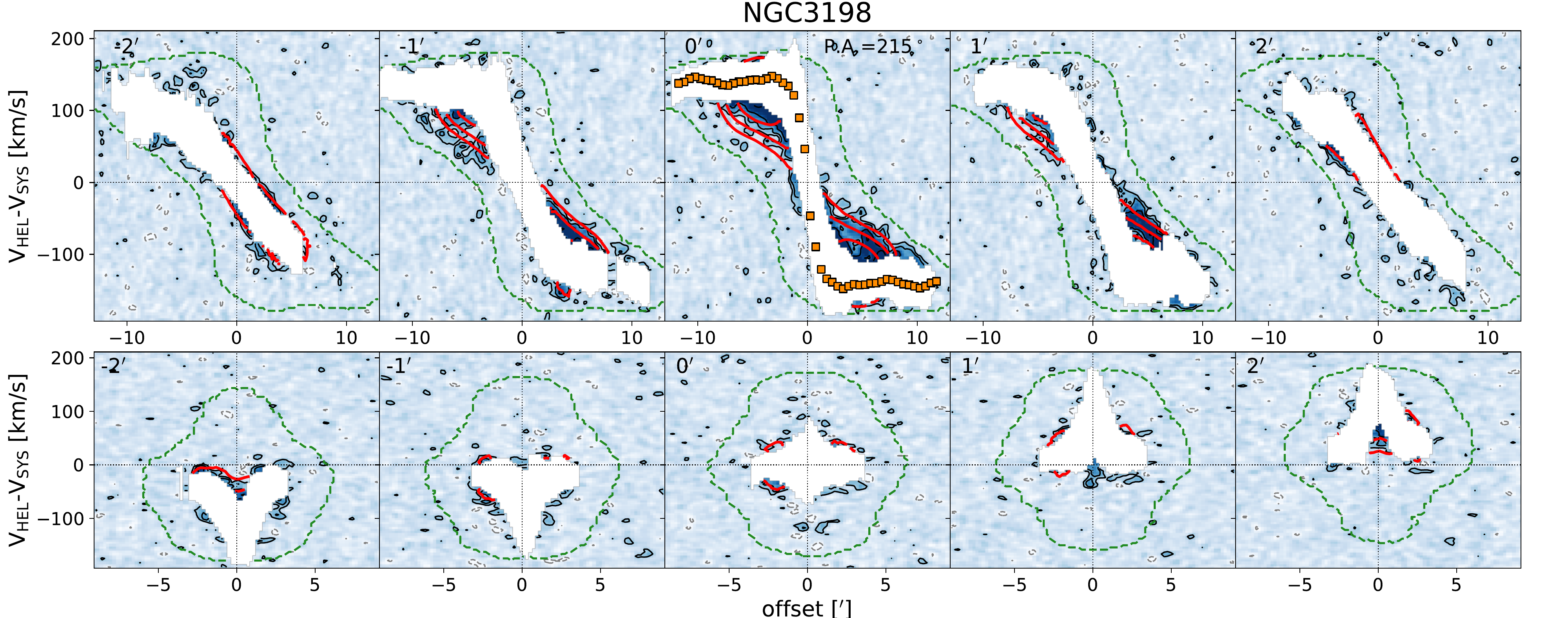}
\caption{\emph{Top:} corner-plots showing the correlation between the various parameters (shaded regions, with contours at arbitrary iso-density levels) along with their marginalised probability distribution (histograms on top) for the EPG of NGC\,3198. \emph{Top-right inset}: total \hicap\ map of NGC\,3198 showing the cuts parallel to the major and minor axes used in pv-slices below. \emph{Bottom:} pv-slices for the data (black contours) and for our best-fit model (red contours). Slice off-sets are indicated on the top-left of each panel. The white area and the green dashed contour indicate the internal and external mask, respectively. Orange squares show the rotation velocities determined with \bba. Contours are spaced by powers of 2, the outermost being at an intensity level of $0.28\miJyb$ ($2\sigma_{\rm noise}$). A negative contour ($-2\sigma_{\rm noise}$) is shown for the data as a dashed grey line.} 
\label{fig:corner_n3198}
\end{center}
\end{figure*}

Once the data have been masked and the rotation curve has been derived, we can proceed to model the leftover \hi\ emission with our MCMC routine.
As our EPG modelling technique requires a single estimate for the inclination and position angle of the entire galaxy, for consistency we adopt the median values determined by \bba\ (INC$=70\de$, PA$=214\de$, see Table \ref{tab:sample}), which are compatible with those reported by \citet{deBlok+08}.
In the top panel of Fig.\,\ref{fig:corner_n3198} we show the corner-plots for the EPG parameters of NGC\,3198, along with their marginalised posterior distribution.
All posteriors are unimodal, indicating that there is a well-defined choice of parameters that best reproduce the properties of the EPG in this galaxy.
Hereafter we quote the median of the posterior distributions as best-fit values, and take the 16th and 84th percentiles as a nominal uncertainty.
The corner-plots also show some partial degeneracy, especially between $R_{\rm g}$ and $\gamma$ and between the rotational lag and the scale-height (which we have already anticipated with our toy models in Section \ref{ssec:synthetic}), which increases the uncertainty associated to these parameters.
We find a value of $-9.2\pm1.4\kmskpc$ for the rotational lag, a precise estimate that falls well within the range quoted by G13 for the two sides of the galaxy, while the magnitude of the vertical and radial velocities are compatible with zero within their uncertainty.
As for the mass fraction of the EPG ($f_{\rm EPG}$), derived as the ratio between the \hi\ flux in the unmasked EPG model and the total flux in the unmasked data cube, we find a value of $8.6\%$, in slight tension with respect to the $10-20\%$ quoted in G13. 
Also, we infer a scale-height of $\simeq1.4\kpc$ that is about half their estimated value. 
Arguably, there are substantial differences between our model/method and those of G13.
We discuss this further in Section \ref{ssec:previous}.

The bottom panel of Fig.\,\ref{fig:corner_n3198} compares the data (black contours) and our best-fit model (red contours) via a series of pv-slices parallel to the major (top panels) and minor (bottom panels) axes.
The orientations and separations of the various slices are shown in the top-right inset of Fig.\,\ref{fig:corner_n3198}, overlaid on top of the total \hi\ map (see Appendix \ref{app:maps}).
In the data, the EPG seems to be systematically more abundant in the approaching side of the galaxy (see major-axis slices at $\pm1'$), a feature that our axisymmetric model clearly cannot reproduce.
Apart from this difference, there is excellent agreement between the model and the data. 
The former reproduces very well the emission from the low-velocity gas, visible in the major-axis slices. 
The absence of prominent asymmetric features in the minor-axis slices, evident in both the data and the model, testify the lack of global inflow/outflow motions. 

\subsection{EPG properties across the HALOGAS sample}\label{ssec:epgproperties}
We now present the results for the rest of the HALOGAS sample.
Unless specified differently, the procedure adopted is the same as that described for NGC\,3198 in Section \ref{ssec:3198}.

\begin{figure*}[tbh!]
\centering
\includegraphics[width=1.0\textwidth]{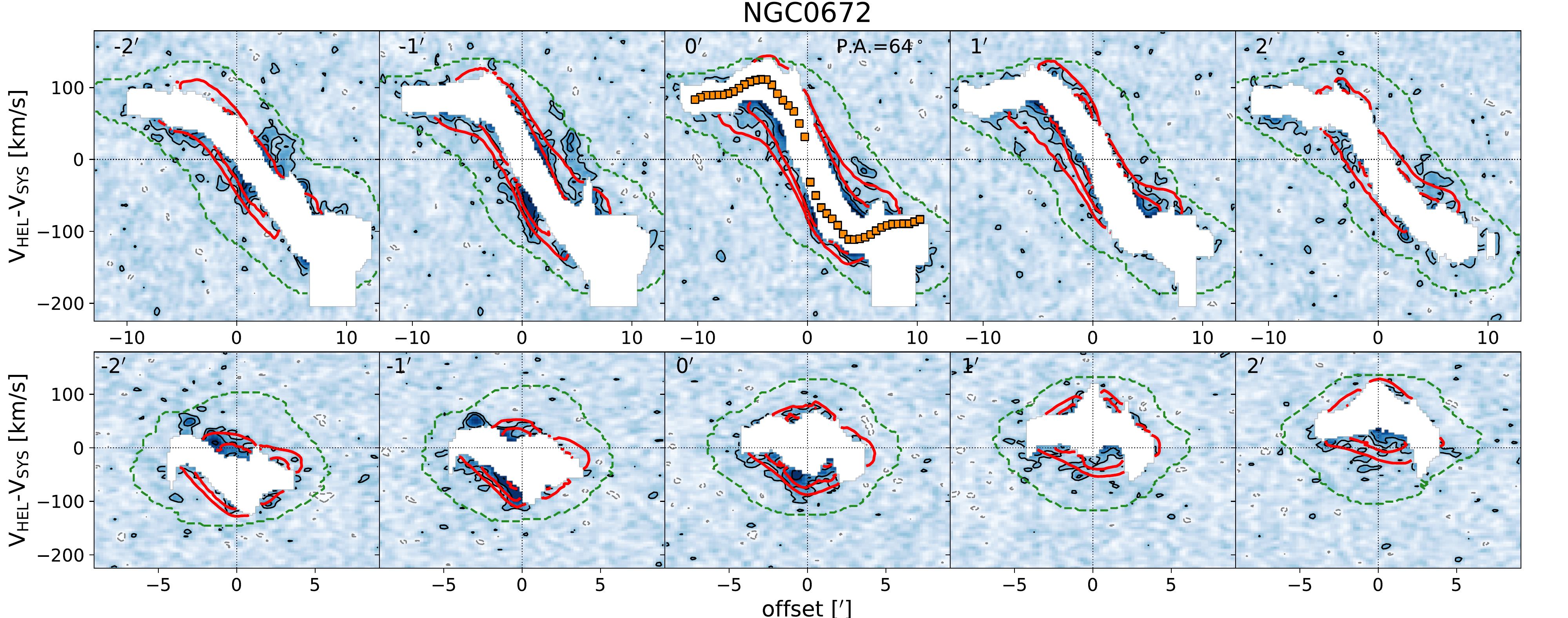}
\includegraphics[width=1.0\textwidth]{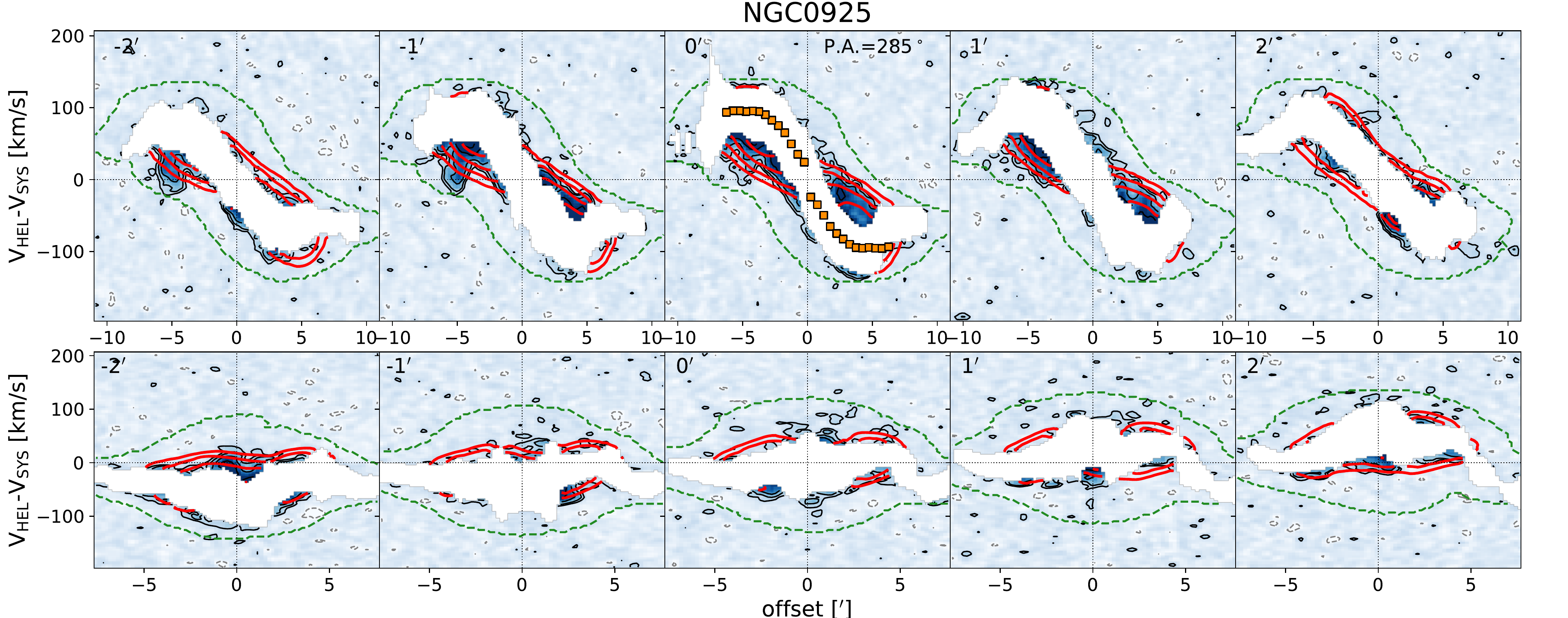}
\includegraphics[width=1.0\textwidth]{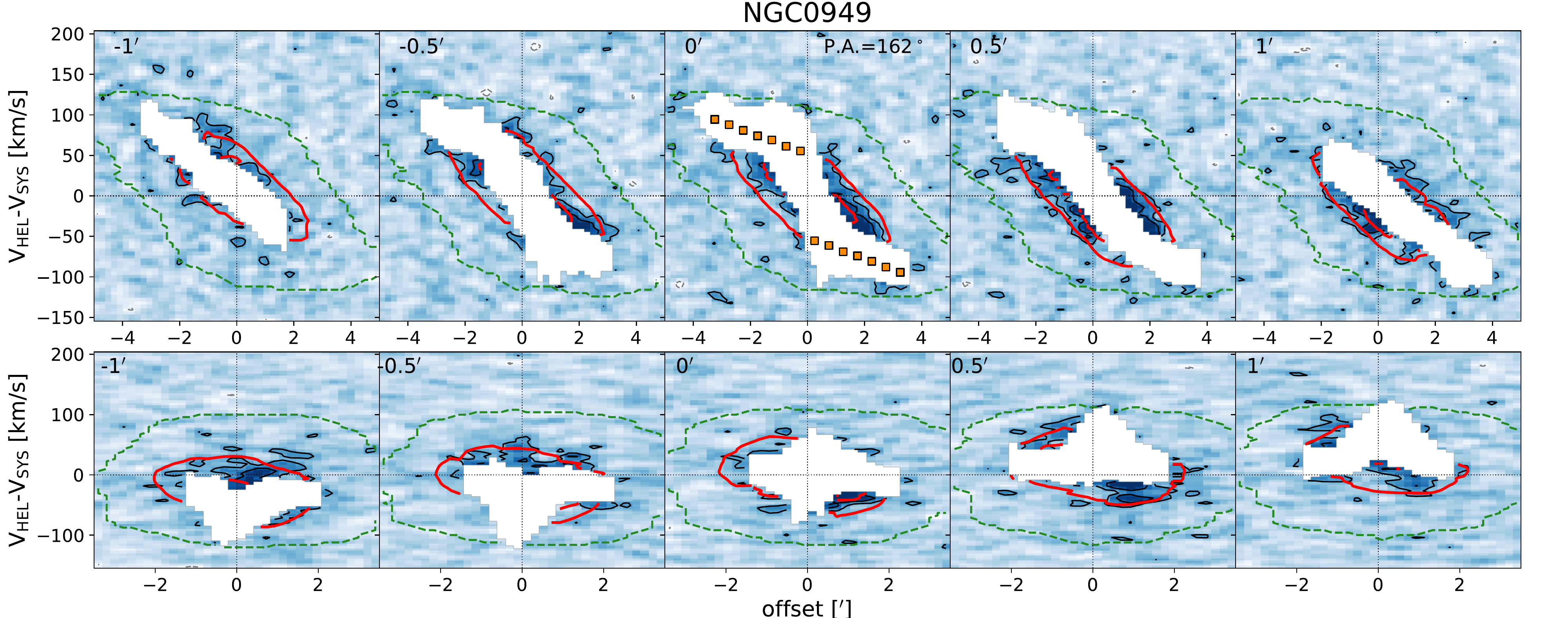}
\caption{Position-velocity slices for the galaxies in our sample (see description for the bottom panel of Fig.\,\ref{fig:corner_n3198}). Individual cases are discussed in the text.}
\label{fig:allpv}
\end{figure*}
\addtocounter{figure}{-1}

\begin{figure*}[tbh!]
\centering
\includegraphics[width=1.0\textwidth]{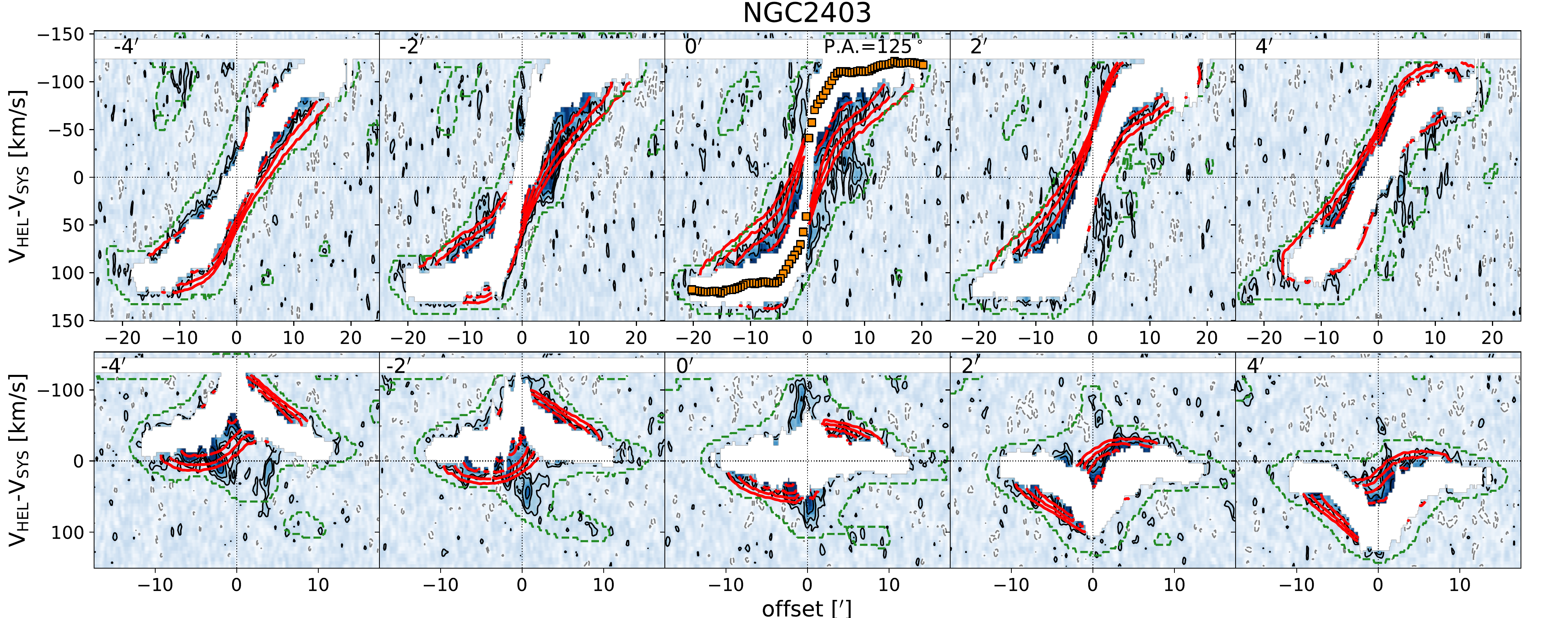}
\includegraphics[width=1.0\textwidth]{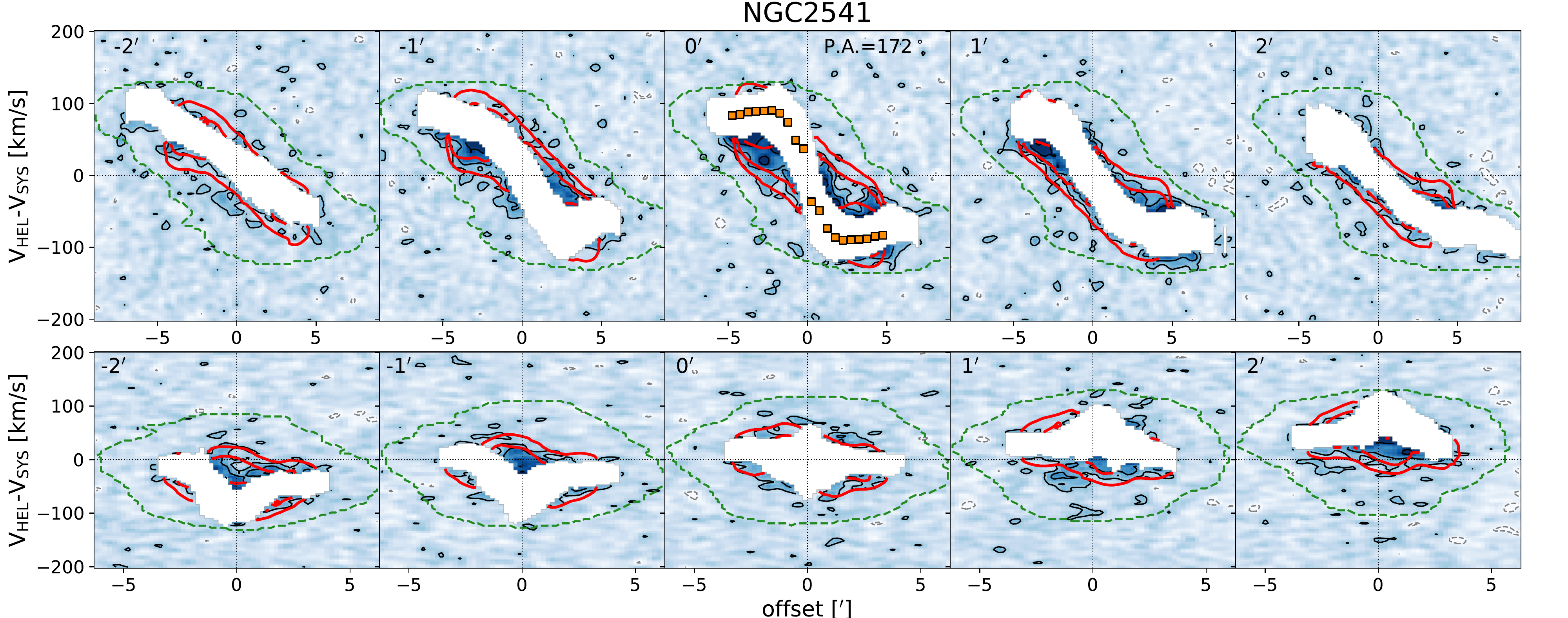}
\includegraphics[width=1.0\textwidth]{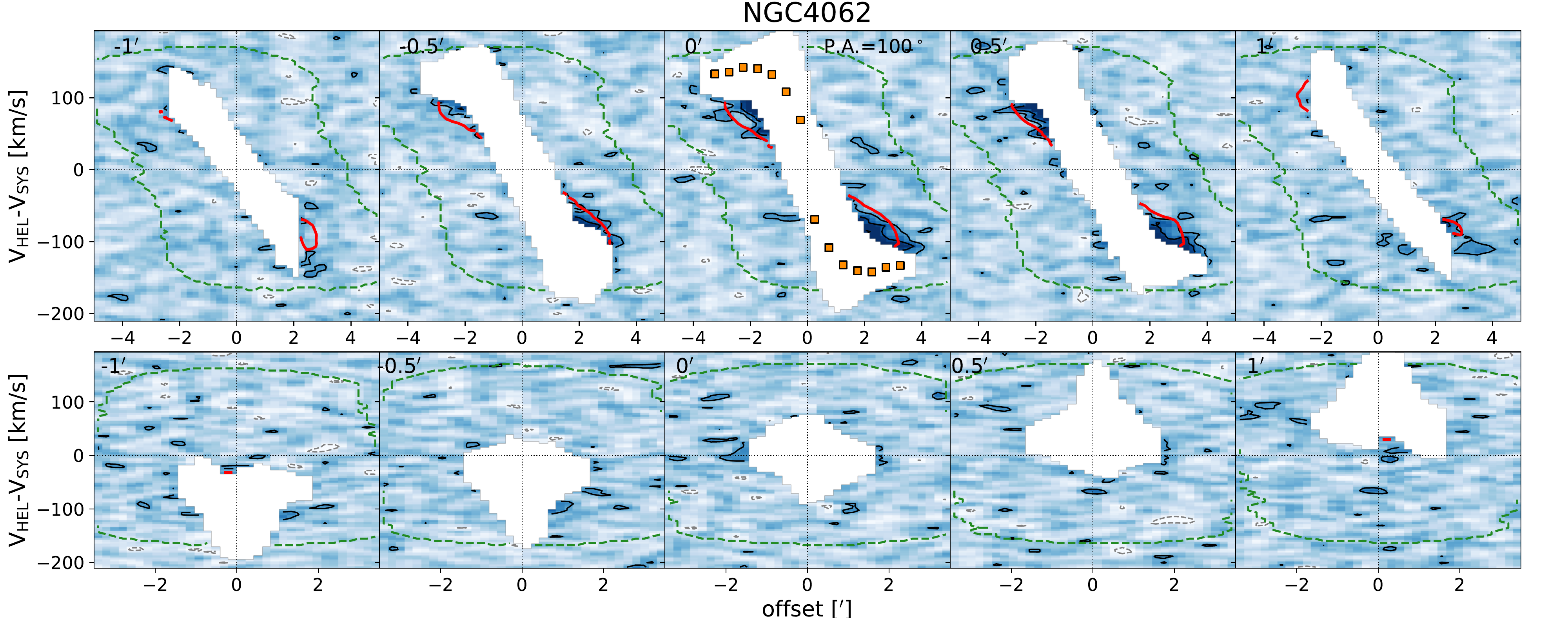}
\caption{Continued}
\label{fig:allpv}
\end{figure*}
\addtocounter{figure}{-1}

\begin{figure*}[tbh!]
\centering
\includegraphics[width=1.0\textwidth]{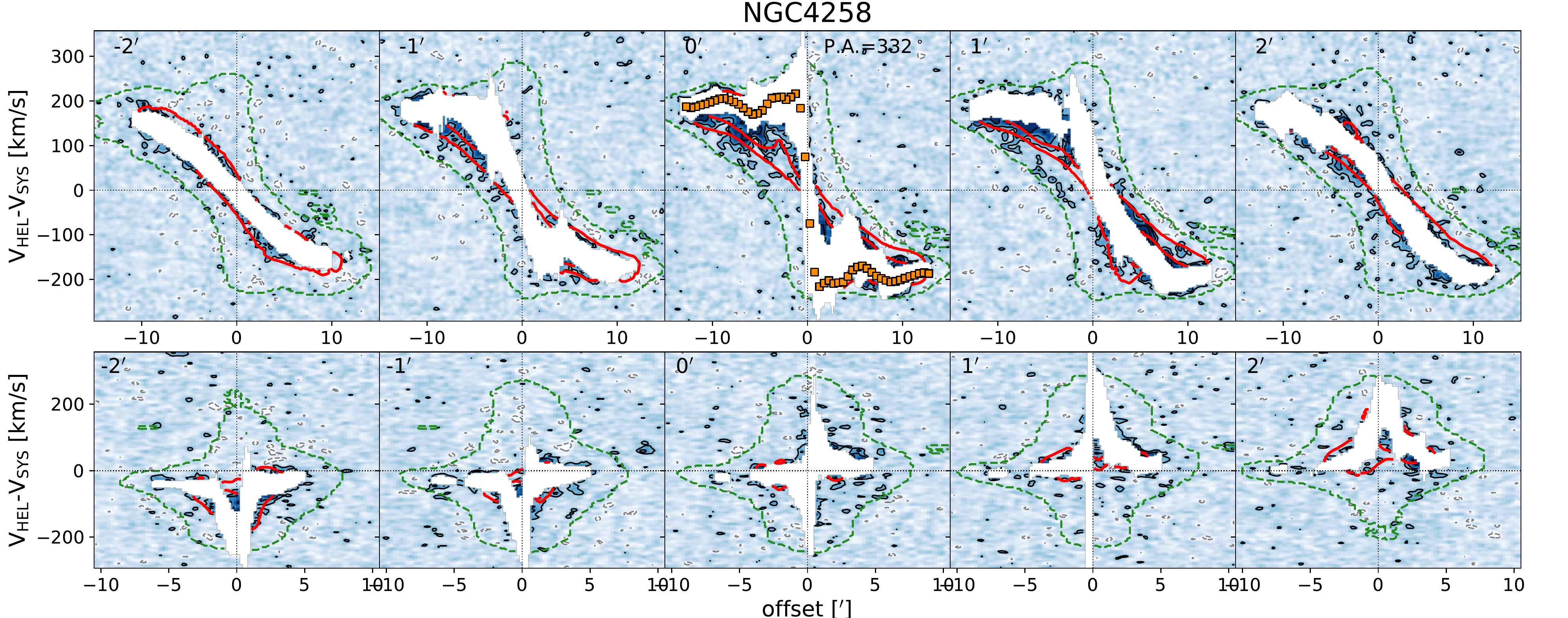}
\includegraphics[width=1.0\textwidth]{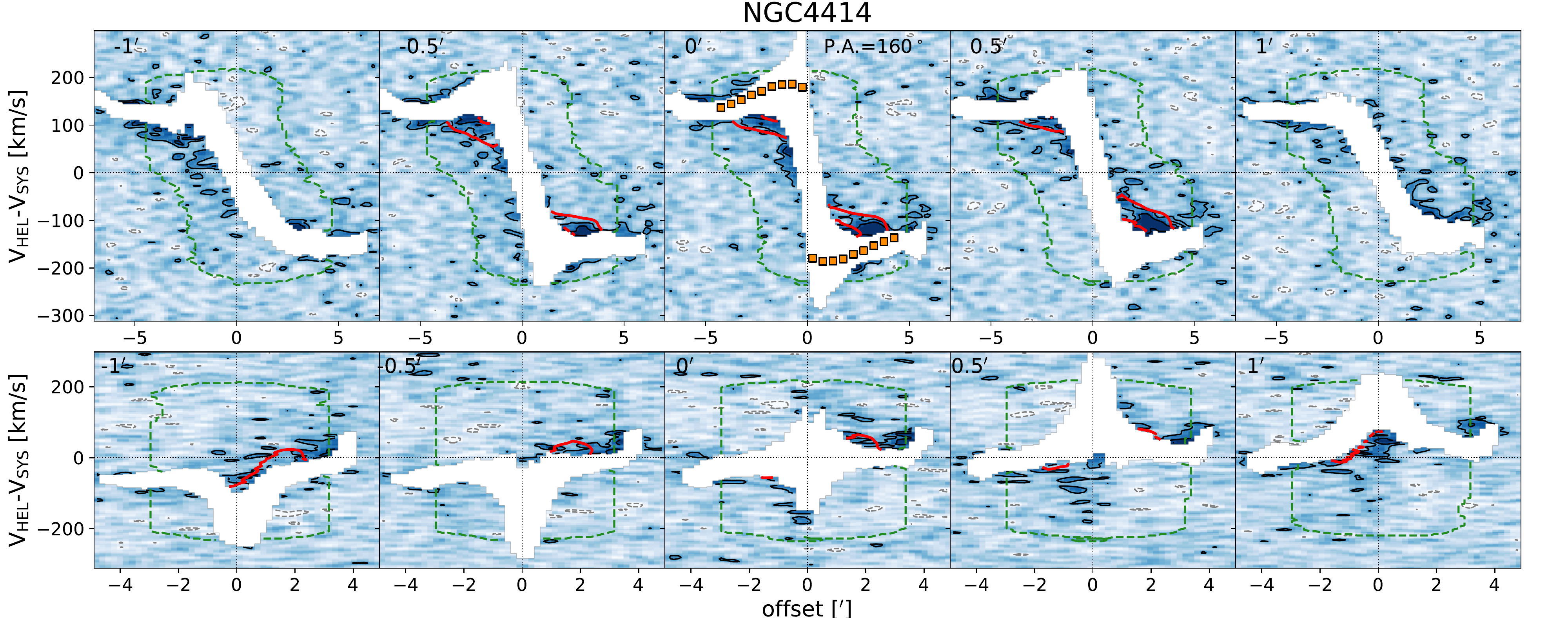}
\includegraphics[width=1.0\textwidth]{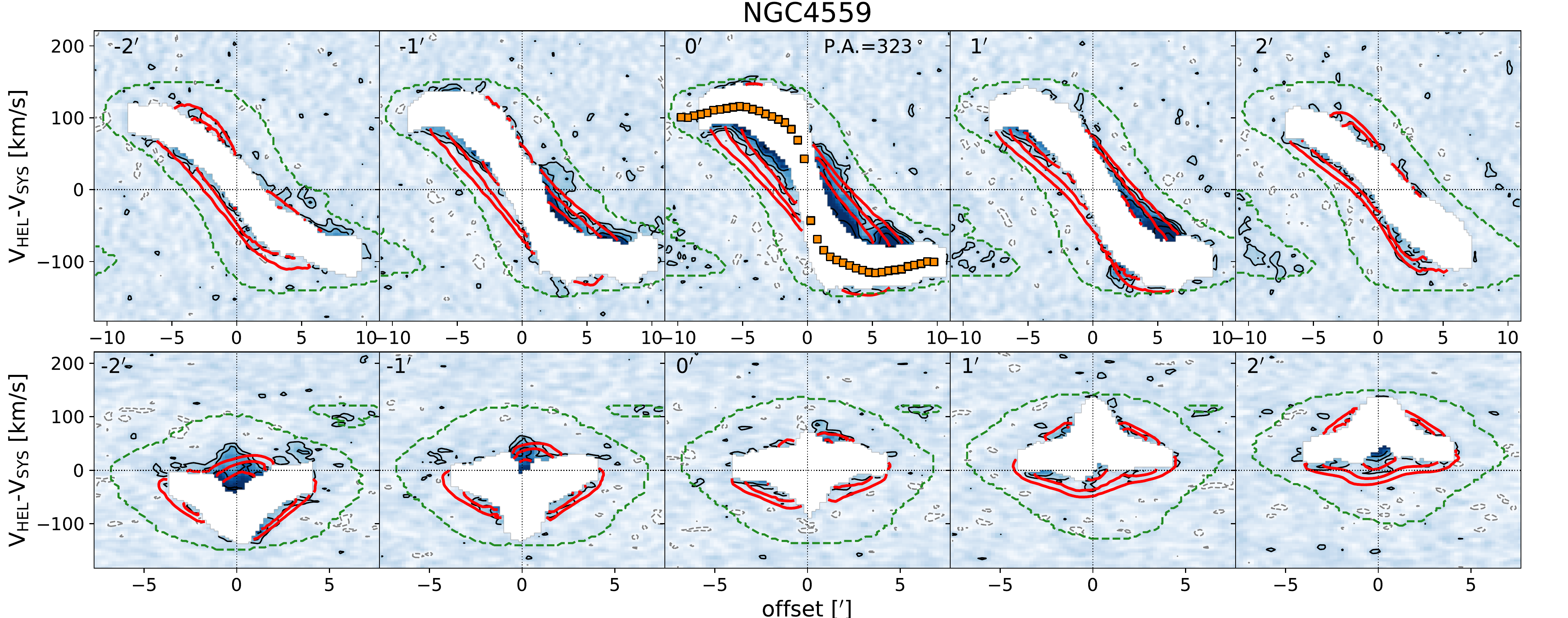}
\caption{Continued}
\label{fig:allpv}
\end{figure*}
\addtocounter{figure}{-1}

\begin{figure*}[tbh!]
\centering
\includegraphics[width=1.0\textwidth]{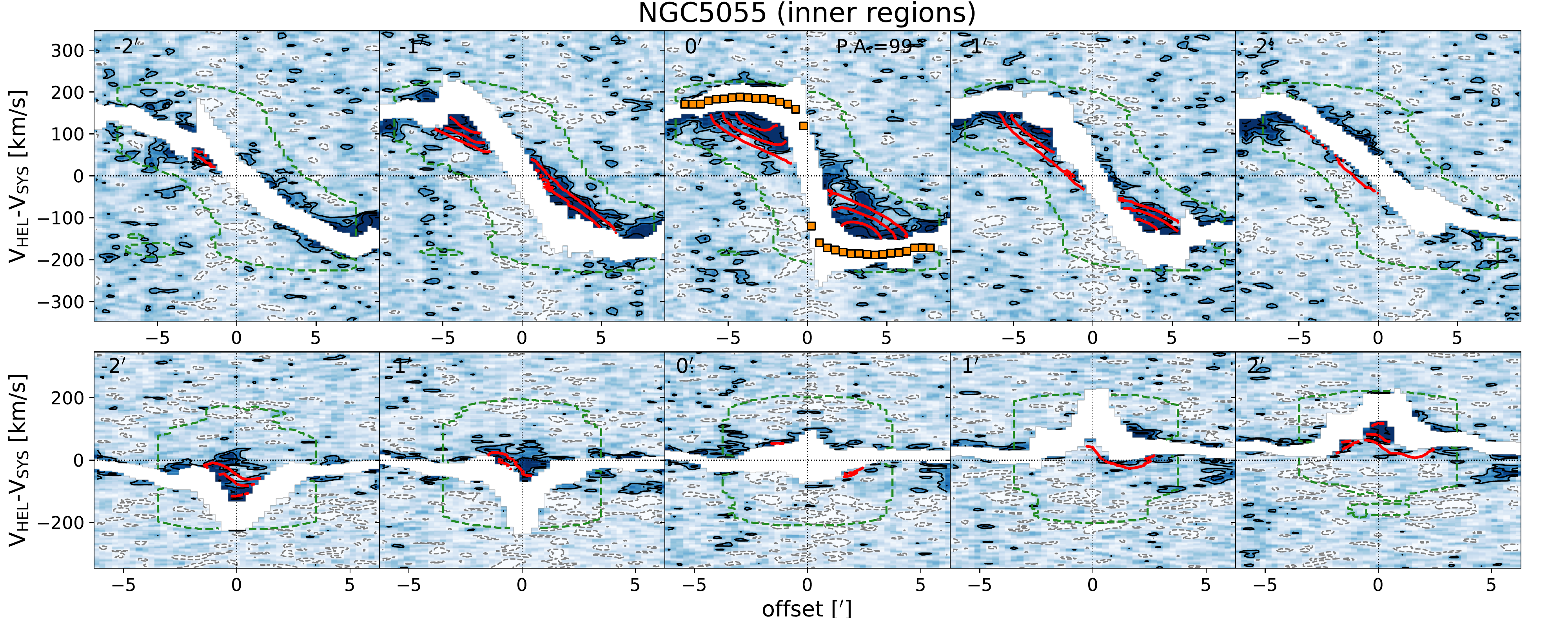}
\includegraphics[width=1.0\textwidth]{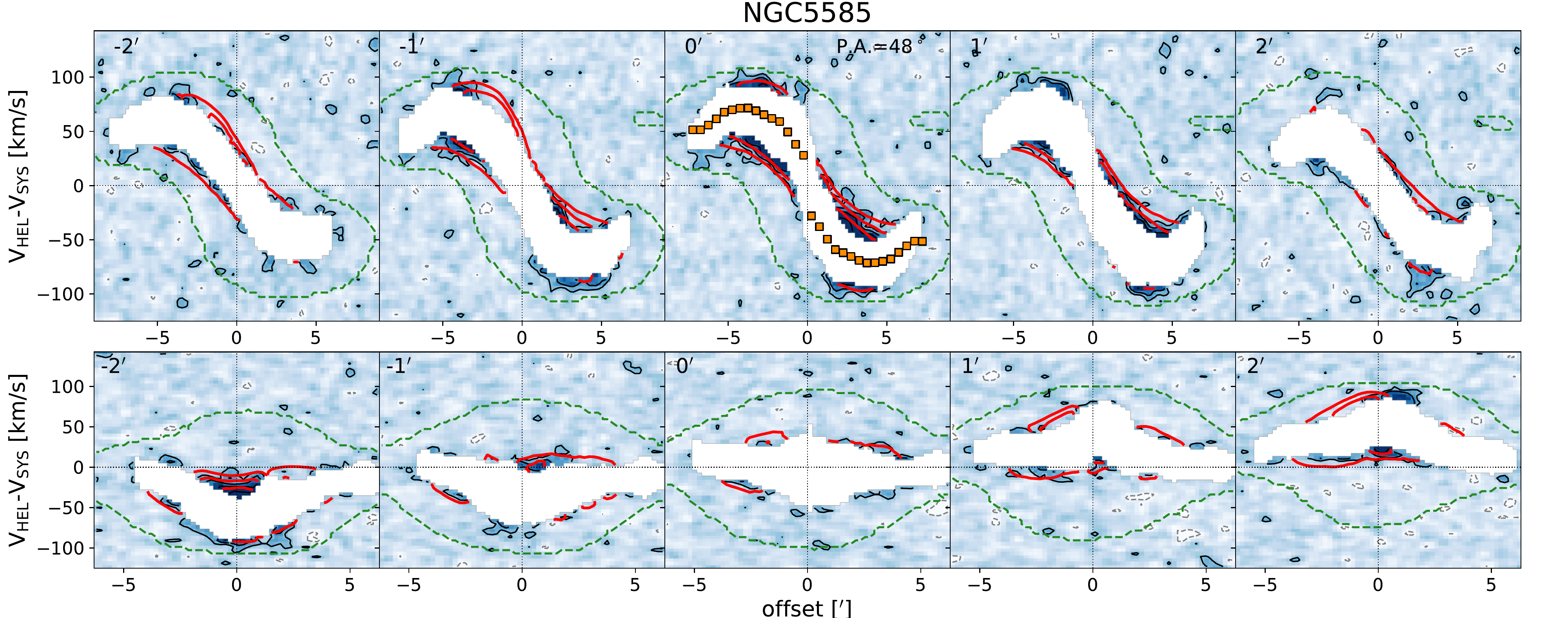}
\caption{Continued}
\label{fig:allpv}
\end{figure*}

Three galaxies out of 15 were left out from the analysis: NGC\,1003, NGC\,4274 and NGC\,4448.
A visual inspection of the \hi\ datacube of NGC\,1003 revealed the presence of a strong line-of-sight warp in the eastern part of the galaxy, where the orientation becomes nearly edge-on. 
This effect seems to be less prominent on the western side.
This line-of-sight warp, already noticed by \citet{Heald+11b}, also explains the appearance of the total \hi\ map (Fig.\,\ref{fig:maps}), which suggests a much higher inclination than that inferred from the optical image ($70\de\!-\!75\de$).
As our EPG separation method fails when consecutive rings overlap along the line of sight, we decided to discard this system from our study.

In NGC\,4274 and NGC\,4448, instead, the masking of the \hi\ disc leaves virtually no EPG emission to work with.
These are two of the systems with the fewest resolution elements; their datacubes show very asymmetric \hi\ profiles which - according to our mask - are largely caused by beam-smearing rather than by the presence of kinematically anomalous gas.
Our method relies on the modelling of a kinematically distinct component and can not be used here, but we can not exclude that some EPG may actually be present in these systems.
More advanced modelling techniques, based on fitting simultaneously both the disc and the EPG component to the data, may be preferred in these cases, but are beyond the scope of this paper.
 
\begin{table*}
\caption{Best-fit parameters and associated uncertainty for the EPG of the galaxies studied in this work.}
\label{tab:propertiesEPG}
\centering
\def\arraystretch{1.1}
\resizebox{\textwidth}{!}{%
\begin{tabular}{lccccccccccc}
\hline\hline\noalign{\vspace{5pt}}
     Galaxy & Beam & Independent & $M_{\hi,\, {\rm EPG}}$  &  $f_{\rm EPG}$ & $R_{\rm g}$ & $\gamma$ & $h$ & ${\rm d}v_{\phi}/{\rm d}z$ & $v_R$ & $v_z$ & $\sigma$\\
                & (arcsec) & voxels & ($10^8\msun$)             & 			    & ($\kpc$)       &             & ($\kpc$) & ($\kmskpc$) &($\kms$)&($\kms$)&($\kms$)\\
         (1)   &    (2)     &    (3)     &       (4)			    &      (5)             &   (6)             &     (7)    &    (8)      &        (9)         &   (10)     &  (11)     &  (12)\\   
\noalign{\smallskip}
\hline\noalign{\vspace{5pt}}
     NGC\,0672$^a$ & 42x32 & 936 & 4.9 & 0.14& 6.3$\pm$0.4 & 0.1$\pm$0.1 & 1.5$\pm$0.1& -8.4$\pm$0.5  & -25.8$\pm$1.4& -35.4$\pm$2.4&  25.3$\pm$0.9 \\
     NGC\,0925$^a$ & 38x33 & 870 & 7.6 & 0.14& 0.9$\pm$0.1 &13.3$\pm$2.1& 2.0$\pm$0.1& -9.7$\pm$0.7  &  3.3$\pm$3.5  & -17.8$\pm$5.6&  21.6$\pm$1.3 \\
     NGC\,0949 & 39x35 & 190 & 1.6 & 0.27& 3.9$\pm$0.8 & 0.5$\pm$0.5 & 1.6$\pm$0.1& -18.0$\pm$1.4& -29.6$\pm$3.5& -19.1$\pm$3.5&  27.7$\pm$1.6 \\
     NGC\,2403 & 30x29 & 4797 & 5.9 & 0.22& 2.3$\pm$0.1 & 2.6$\pm$0.3 & 0.9$\pm$0.1& -11.7$\pm$0.5& -21$\pm$0.9   & -9.4$\pm$1.2  &  15.2$\pm$0.4 \\
     NGC\,2541 & 34x34 & 634 & 7.4 & 0.15& -                   & 4.9$\pm$1.0 & 2.8$\pm$0.1& -6.5$\pm$0.3  & -24.3$\pm$1.9& -15.9$\pm$2.3&  27.4$\pm$1.0 \\
     NGC\,3198 & 35x33 & 637 & 9.8 & 0.09& 2.1$\pm$0.5 &9.8$\pm$2.8&1.4$\pm$0.2& -9.2$\pm$1.4   & -1.5$\pm$2.6  & 8.1$\pm$8.9   &  23.0$\pm$1.2 \\
     NGC\,4062 & 39x34 & 148 & 2.1 & 0.11&  -		       & 5.0$\pm$2.2& 2.5$\pm$0.1& -9.4$\pm$1.0   &-25.8$\pm$5.8& -23$\pm$7.5    &  34.6$\pm$2.0 \\
     NGC\,4258 & 33x33 & 1422 & 6.6 & 0.11& 6.6$\pm$1.1 & 2.3$\pm$0.4& 1.4$\pm$0.1&-10.0$\pm$0.5   &-17.3$\pm$1.3& -27.2$\pm$3.8&  24.8$\pm$0.7 \\
     NGC\,4414$^b$ & 39x33 & 194 & 5.1 & 0.12&  - 		  &-3.7$\pm$1.9& 0.5$\pm$0.3&-58$^{+24}_{-56}$&-39.5$\pm$4.8& -13.1$\pm$14.2&  26.3$\pm$2.9\\
     NGC\,4559 & 41x32 & 926 & 6.0 & 0.13& 1.9$\pm$0.3 & 4.0$\pm$0.9& 1.7$\pm$0.1& -8.2$\pm$0.6    &-19.8$\pm$2.5& -33.9$\pm$4.1   &  21.3$\pm$1.2 \\
     NGC\,5055$^b$ & 36x33 & 491 & 2.6& 0.04& 1.1$\pm$0.2 & 8.3$\pm$1.5& 0.6$\pm$0.1& -40$^{+8}_{-13}$ &-13.7$\pm$3.9& -39.5$\pm$8.6&  20.2$\pm$2.4 \\
     NGC\,5585 & 34x33 & 368 & 3.0 & 0.12& 2.5$^{+1.9}_{-0.9}$ & 2.6$\pm$1.8& 1.4$\pm$0.3& -2.3$\pm$0.8    & 0.4$\pm$2.4 & 5.4$\pm$4.0   &  18.5$\pm$1.0 \\
\noalign{\vspace{2pt}}\hline
\noalign{\vspace{5pt}}
\multicolumn{12}{p{1.2\textwidth}}{\textbf{Notes.} (1) NGC name; (2) FWHM of the synthesised beam; (3) Number of independent EPG voxels modelled, computed as the number of voxels in the masked dataset with intensity above twice the rms-noise divided by the number of voxels per resolution element (eq.\,\ref{eq:npix}); (4) \hicap\ mass of the EPG component, with an uncertainty of $\sim20\%$; (5) ratio between $M_{\rm EPG}$ and the total \hi\ mass; (6-7) surface density parameters (eq.\,\ref{eq:surf_halo}), a `-' indicates that the parameter is unconstrained; (8) scale height (eq.\,\ref{eq:vert_halo}); (9) vertical rotational gradient; (10) velocity in the direction perpendicular to the rotational axis (positive values mean outflow); (11) velocity in the direction perpendicular to the disc (positive values mean outflow); (12) velocity dispersion. \newline$^a$ Interacting systems.\newline$^b$ Poorly-fit systems, which are also warped. The fit is limited to (and $f_{\rm EPG}$ is computed within) the innermost $20\kpc$.}\\
\end{tabular}}
\end{table*}     

Figure \ref{fig:allpv} shows the pv slices for the remaining HALOGAS galaxies and for their best-fit model.
Some individual cases require further discussion.
\begin{itemize}
\item \textbf{NGC\,0672.} 
This system is interacting with a companion, IC\,1727, an irregular dwarf which strongly contaminates the \hi\ flux at negative (approaching) velocities (see also maps in Appendix \ref{app:maps}).
The rotation curve of NGC\,0672 has been derived using only the receding side of the disc, and a significant portion of the approaching side has been blanked and added to the internal mask (as it appears from the white square regions in the slices along the major axis).
An \hi\ filament is visible in the major axis slices at $-2'$ and $-1'$ at about the systemic velocity, possibly caused by the interaction between the two galaxies. The filament is not reproduced by our model, as expected.
Note that in this galaxy we have used $\gamma>0$ as an additional prior.
The best-fit model would naturally prefer a negative $\gamma$ (i.e., a larger central concentration for the EPG), which slightly improves the pv-slice along the major axis but at the cost of producing no emission in the slices at $\pm2'$ offsets, where anomalous \hi\ is clearly visible.

\item \textbf{NGC\,0925.}
This galaxy shows an extended \hi\ tail towards the south, possibly caused by a recent interaction with a very faint system \citep[][H11]{Sancisi+08}.
This tail seems to be efficiently masked by our procedure, as can be seen from the extent of the white regions in minor axis pv-slices, and the model appears to fit reasonably well the leftover anomalous emission.

\item \textbf{NGC\,0949.}
The tilted ring fit with \bba\ indicates clearly the presence of a warp ($>10\de$) in both inclination and position angle.
Given the limited number of resolution elements (7 per side), we have decided to fix the INC and PA to median values for both the rotation curve and the EPG modelling.
We have verified that the EPG mass and kinematics do not change significantly if we use a rotation curve that accounts for the warp.


\item \textbf{NGC\,2403.} 
The anomalous gas of this galaxy has been studied in detail first by \citet{Schaap+00} and then by \citet{Fraternali+02}, and represents the prototypical case for a slow rotating \emph{and} inflowing EPG.
We compare our findings with those of \citet{Fraternali+02} in Section \ref{ssec:previous}.
The overall \hi\ emission is contaminated by the Galactic \hi\ foreground in a few channels at negative velocities, which we have blanked (white horizontal stripes).
Also here a prominent filament is visible in minor axis pv-slices at offsets $\le0$ and along the major axis \citep[see also][]{deBlok+14}. As expected, the filament is not reproduced by our model.

\item \textbf{NGC\,2541.}
In this galaxy, too, \bba\ finds a warp in both INC and PA, with magnitudes of $\sim5\de-10\de$.
Our model seems to reproduce well the emission from the anomalous component, which is fairly axisymmetric and smooth.

\item \textbf{NGC\,4062.}
The anomalous gas is barely visible in pv slices parallel to the major axis, and is virtually absent along the minor axis.
Given the limited resolution and number of voxels to fit, the application of our model to this system must be taken with some skepticism.
However, tests on mock data have shown that even in this condition our model can recover the correct parameter within a $2\sigma$ uncertainty, as we discuss in Appendix \ref{app:mcmctest}.

\item \textbf{NGC\,4258.}
This galaxy features a complex kinematical pattern, with streaming motions in the innermost few kpc (as visible from the markedly asymmetric pv slice along the minor axis) and a $\sim15\de$ warp in both PA and INC distributed along its entire disc.
It also contains an active nucleus and it is classified as a Seyfert 2 object.
Despite this complexity, the properties of its EPG appear to be analogous to those of the rest of the sample.

\item \textbf{NGC\,4414.}
As already noticed by \citet{deBlok+14b}, this galaxy has substantial ($\sim10\de$) warp in position angle, and its rotation curve indicates the presence of a prominent stellar bulge.
As for NGC\,5055 (another warped galaxy, see below), modelling the whole system resulted in a very poor fit and we decided to focus on the innermost $20\kpc$ region, where variations in PA are less severe.
Our model seems to miss a significant fraction of the anomalous \hi\ flux, especially along the major axis (at about the systemic velocity) and in parallel slices at $\pm1'$ offsets.

\item \textbf{NGC\,4559.}
The EPG of NGC\,4559 was already studied by \citet{Barbieri+05} and \citet{Vargas+17}, with which we compare our findings in Section \ref{ssec:previous}.
The pv-slices parallel to the major axis show that the EPG is systematically more abundant on the approaching side, a feature that our axisymmetric model can not reproduce. As discussed, our model gives pv-slices along the major axis (offset of $0'$) that are symmetric by construction.

\item \textbf{NGC\,5055.}
This system is well known to host a prominent warp in its outskirt \citep[e.g.][]{Battaglia+06}, which can be traced accurately with \bba\ \citep[see Fig.\,4 in][]{Barolo}.
Modelling the EPG of the whole dataset resulted in a very poor fit, and we decided to focus on the innermost $20\kpc$ where the INC and PA are relatively stable.
In fact, anomalous \hi\ flux may spuriously arise at $R>20\kpc$ due to the overlap of consecutive annuli along the line of sight, given the rapid variation of the INC and PA in these regions. 
Despite this treatment, a significant fraction of the anomalous \hi\ emission, including a filament visible along the major axis, does not seem to be reproduced by our model: it is likely that the EPG mass of this galaxy is underestimated.
We also note that the background noise in the datacube is not constant and is difficult to treat properly.
We discuss this `difficult' system in more detail in Section \ref{ssec:limitations}.
\end{itemize}

The best-fit parameters for the EPG of these 12 galaxies are listed in Table \ref{tab:propertiesEPG}, along with their associated $1\sigma$ uncertainty, computed as half the difference between the 16th and 84th percentiles of their probability distribution.
The nominal uncertainty on the EPG mass that we derive from the marginalised posterior is small, typically around $5\%$, and is not quoted in Table \ref{tab:propertiesEPG}.
More realistic errors come from the uncertainty on the distances, for which we assume a fiducial value of $10\%$ (corresponding to a $20\%$ error on the mass).
The EPG fraction, $f_{\rm EPG}$, varies from $9\%$ (NGC\,3198) to $27\%$ (NGC\,0949) with a mean value of $14\%$, and is fairly consistent with the values typically quoted in the literature \citep[e.g.][]{Fraternali10}.
This confirms that \emph{the EPG is a common feature of late-type galaxies}.

For three galaxies (NGC\,2541, NGC\,4062 and NGC\,4414) we find a flat posterior in the $R_{\rm g}$ parameter, meaning that the EPG scale radius for these systems is unconstrained.
Interestingly, this has little impact on the EPG mass, mostly because of the flux normalisation and because the model is always truncated at the outermost radius measured in the rotation curve (we do not extrapolate the EPG mass beyond that radius).
With the exception of NGC\,5055 and NGC\,4414, which are both poorly fit by our model, the EPG scale-height is always confined within $1\!-\!3\kpc$.
The EPG is more turbulent than the gas in the disc, with velocity dispersions varying from $15$ to $35\kms$, a factor of $1.5-3$ higher than the \hi\ in the disc.
Even accounting for this larger velocity dispersion, if the EPG were in hydrostatic equilibrium we would expect scale-heights of $\lesssim1\kpc$ \citep[e.g.][]{Bacchini+19}. 
Hence the EPG is not in hydrostatic equilibrium, and this has fundamental implications for its origin (see Section \ref{ssec:origin} for further discussions).

All galaxies show a \emph{negative gradient} in rotational velocity, i.e., the EPG systematically rotates at a lower speed than the disc.
The rotational lag varies from a few to $\sim-20\kmskpc$, with a median value of $-10\kmskpc$, meaning that rotation is still the main motion of this anomalous component.
The only two outliers in this scheme are again NGC\,5055 and NGC\,4414, with a lag of $\approx-50\kmskpc$ (but notice the large errors), which is compensated by the small scale-height given the (partial) degeneracy between these two parameters.
Remarkably, once we have determined the near/far sides of these systems and subsequently adjusted the sign of $v_{\rm R}$ and $v_{\rm z}$, we find that a)
all galaxies are compatible with having $v_{\rm R}\!\leq\!0$ within the uncertainties; b) 11 galaxies out of 12 are compatible with having $v_{\rm z}\!\leq\!0$ within the uncertainties.
That is, the EPG appears to be \emph{globally inflowing towards the galaxy centre} with typical speeds of $20\!-\!30\kms$.

\begin{figure*}
\begin{center}
\includegraphics[width=0.65\textwidth]{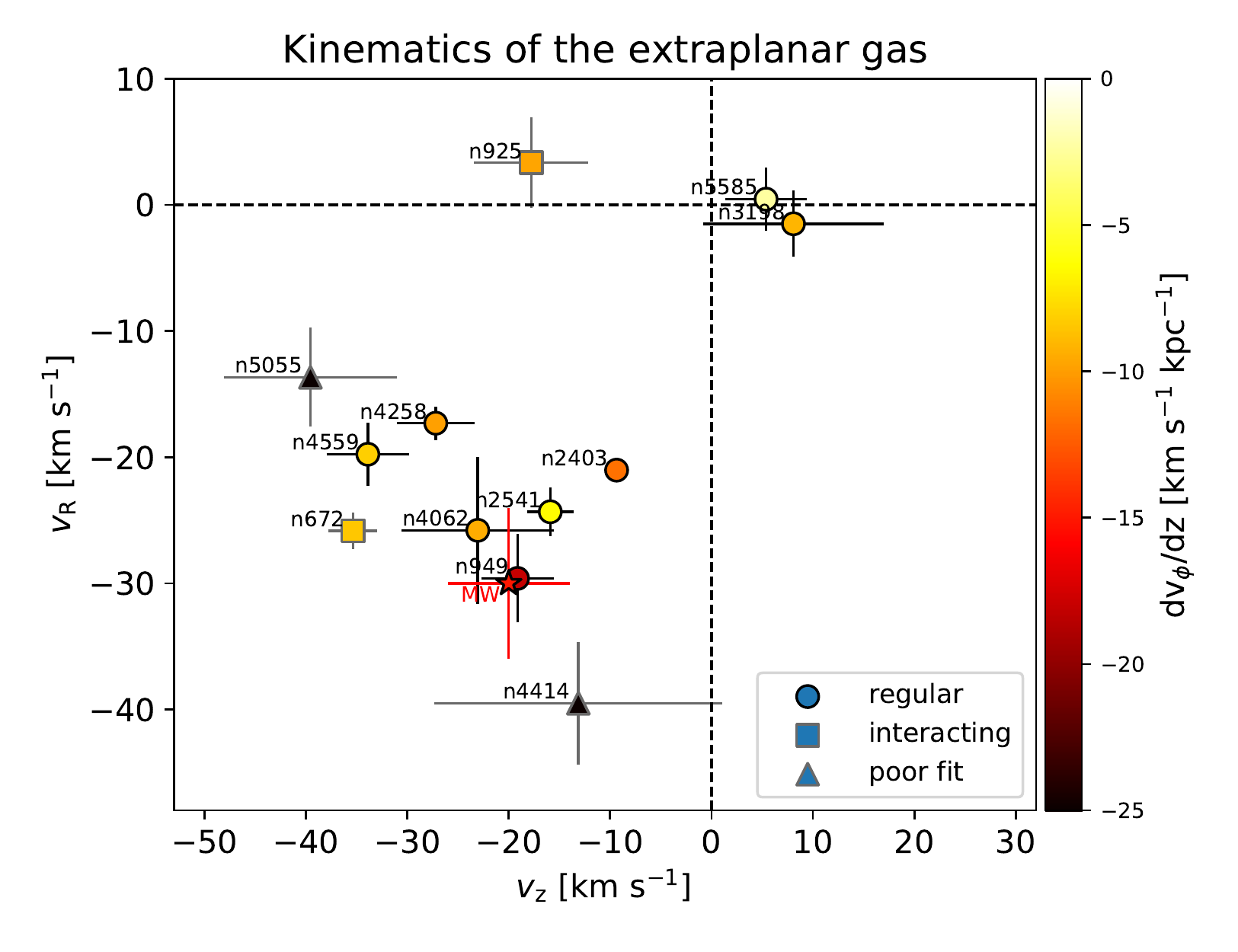}
\caption{Kinematics of the EPG in the HALOGAS galaxies as derived from our models. EPG tends to be globally inflowing with a speed of $20-30\kms$. Interacting and poorly fitted galaxies are shown with separate symbols. Values for the Milky Way (MW) are from MF11.}
\label{fig:kinematics}
\end{center}
\end{figure*}

Figure \ref{fig:kinematics} summarises the kinematical properties of the systems analysed.
Galaxies that cluster in the bottom-left region of the plot ($v_{\rm R}<0$, $v_{\rm z}<0$) have also more precise kinematical measurements, as it appears from the error-bars.
The fact that several late-type galaxies show similar kinematical properties for their EPG strongly supports the idea of a common physical origin for this \hi\ component.
We discuss this further in Section \ref{ssec:origin}.

\section{Discussion} \label{sec:discussion}
\subsection{Limitations of our method}\label{ssec:limitations}
Our method represents an important step forward in the analysis of kinematically anomalous gas in galaxies.
While we tailored our procedure around systems at intermediate inclinations, it is in principle possible to extend our fitting strategy to edge-on galaxies once a method for filtering the emission from the disc is decided.
However, our approach comes with a number of limitations that we discuss below.

Two key assumptions are that the EPG is smooth and axisymmetric.
These are clearly not valid in the cases where the kinematically anomalous gas is built by a few, isolated \hi\ complexes scattered around the disc.
This does not seem to be a typical scenario, but there are cases where isolated \hi\ complexes/filaments are visible and can have an impact on the fitting procedure.
An example is NGC\,2403, where an anomalous, bright \hi\ filament is visible in most of the pv-slices parallel to the minor axis.
This filament is kinematically detached from the rest of the smooth EPG component and is likely to have a separate origin \citep[e.g.][]{deBlok+14}.
Even if it accounts only for a few per cent of the total EPG \hi\ mass, Its presence may impact the modelling by artificially increasing the EPG thickness, rotational lag or velocity dispersion.
Note in fact how, in the pv-slice along the major axis, the model appears to slightly `overshoot' the data.

Much more common are the cases where the EPG appears to be more abundant on one side of the galaxy.
A striking example is that of NGC\,4559, where most of the anomalous \hi\ is located on the approaching side of the disc as it appears from the major axis pv-slice.
As we discussed, in an axisymmetric model the pv along the major axis is always symmetric.
In this particular case, our model attempts to find a compromise between the two galaxy sides, overestimating (underestimating) the emission from the receding (approaching) region.

In a scenario where the EPG is produced by a galactic fountain cycle, we may expect that the anomalous gas originates preferentially from the regions of intense star formation, like spiral arms. 
While this may introduce an additional asymmetry in the EPG distribution, in principle the relative speed between the disc material, the spiral arm pattern and the halo gas should wash out any imprint of the original ejection region.
Total maps of the kinematically anomalous gas, in fact, do not show hints of a spiral structure \citep[e.g.][]{Fraternali+02,Boomsma+05}, although asymmetries may be present \citep[also in the ionised component, see ][]{Kamphuis+07}.
Another feature that a galactic fountain is expected to produce is a flare in the EPG distribution \citep[e.g.][]{Marasco+12, Marasco+15}. 
In order to model such feature we should include an additional free parameter to describe the radial dependency of $h$.
In the philosophy of keeping the model as simple and generic as possible, in this work we have not explored the possibility of a flare, but it is a feature that we plan to introduce in future studies.

Another limitation of our model is the assumption of a single choice of INC and PA for the whole galaxy, i.e., we assume that the EPG component is not warped.
In our sample, at least six galaxies show a substantial ($\sim10\de$ or more) warp in their \hi\ distributions: NGC\,0949, NGC\,1003, NGC\,2541, NGC\,4258, NGC\,4414 and - the most extreme case - NGC\,5055.
In principle, a warped disc should not produce asymmetries in the line profiles unless consecutive rings overlap significantly along the line of sight, and it is expected to be automatically filtered out by our masking procedure.
In addition, if the EPG component originates from a galactic fountain, we expect it to be confined within the inner (star-forming) disc, where variations in INC and PA are weak.
In practice, applying our procedure to the whole cubes of NGC\,5055 and NGC\,4414 resulted in very poor fits to the anomalous \hi\ emission of these galaxies (while this does not seem to be the case for the other warped systems).
The fit improves when we limit the model to the innermost $\sim20\kpc$, where the warp is less pronounced.
However, even when we fit only the inner parts, we find that a) the model does not reproduce a significant fraction of the anomalous \hi\ flux; b) the best-fit values for the rotational gradient and for the scale-height are distinct from those of the rest of the sample.
It is possible that a fraction of the EPG in these two galaxies has a different origin with respect to the others, perhaps related to the same physical process that have built the warp.
Our model may not be appropriate to describe all the anomalous \hi\ in these systems.
In general, uncertainties in the adopted value of inclination have little impact on our results and seem to affect exclusively the inferred rotational gradient.
We have experimented by varying the fiducial INC of NGC\,3198 and NGC\,4062 by $\pm5\de$, adjusting their rotation curve accordingly, finding a variation of about $\pm1.5\kmskpc$ in the best-fit $dv_{\phi}/dz$ while all the other parameters (including the EPG masses) remained unaffected.

An important question is how robust - or model dependent - our results are, with particular focus on the EPG mass.
We have already discussed how modelling the anomalous \hi\ emission directly in the 3D data domain is fundamental to derive the EPG mass, given that the total \hi\ flux measured after the disc subtraction largely depends on the kinematics of the EPG itself and on the galaxy inclination.
That said, our model is parametric and it is clear that a different parametrisation can lead to different results.
For instance, in our parametrisation of the vertical density distribution (eq.\ref{eq:vert_halo}), $\rho(z)\propto\exp(-|z|/h)$ for $|z|\!\gg\!h$ and the density profile deviates from an exponential law only close to the midplane, where a depression occurs to accomodate for the thin disc.
As discussed, a pure exponential layer would make the boundaries between the disc and the EPG ambiguous and is not used in this work.
At any rate, the surface density that one obtains by neglecting the inner depression, i.e. by assuming an exponential profile all the way down to the midplane, would be a factor of $2$ larger than what we currently infer.
As a test, we have tried to fit NGC\,3198 using an EPG model with exponential surface and vertical density profiles, rather than the profiles described by eq.\,(\ref{eq:surf_halo}) and (\ref{eq:vert_halo}), finding a factor 2 difference in the EPG mass.
The best-fit scale height and the kinematical parameters, though, were compatible with those derived in our original parametrisation within 2$\sigma$ on both measures.
Interestingly, in this new parametrisation we find a much stronger degeneracy between scale-height and rotational lag (already noticed by G13), and wider and more asymmetric posterior probability distributions.
We have repeated this test on NGC\,4062, for which our results are supposedly less robust given the limited resolution and extent of the EPG emission in this galaxy.
The results show a $50\%$ larger EPG mass but, quite remarkably, the same kinematical parameters of our previous determination.
This strengthen our findings for this particular galaxy and indicates that, in general, the inferred EPG kinematics do not depend strongly on the model parametrisation.

Additional uncertainties come from the choice of the flux-density threshold at which we `clip' our disc model to produce the mask.
We have tested whether our results for NGC\,3198 vary if we change our fiducial threshold ($2\times$ the data rms-noise) to a smaller ($1\times$ the rms-noise) or to a larger  ($3\times$ the rms-noise) value, finding that in all cases the kinematical parameters and the scale-height remain compatible within the error-bars.
$R_{\rm g}$ and $\gamma$ are instead less stable, leading to some fluctuations in $f_{\rm EPG}$ ($0.07$, $0.09$ and $0.10$ for thresholds of $1\times$, $2\times$ and $3\times$ the rms-noise).
Uncertainties on the distances lead to similar or larger error-bars on the \hi\ masses.
A more in-depth discussion on the error-bars associated to our measurements is presented in Section \ref{sec:uncertainties}.

Finally, we would like to draw attention to our disc/EPG separation method, which is based on building a Gaussian model for the thin disc emission that is then used to filter out the disc contribution in the data.
As for warps, we expect that our masking procedure is robust against large-scale non-circular motions (like a global radial flow) within the thin \hi\ disc, given that their effect would be to shift the velocity centroids of the \hi\ lines without affecting their shape.
Local non-circular motions within the disc on scales comparable to or smaller than the beam size would instead affect the shape of the profiles and produce `anomalous \hi' that would be incorporated in our EPG modelling.
While it is difficult to quantify the incidence of these effects, we do not expect them to have a systematic impact on our results.
Systematics are instead caused by beam-smearing: even in a razor-thin disc, line-profiles can deviate from a Gaussian if the spatial resolution is poor.
Galaxies like NGC\,4275 and NGC\,4448 show remarkably asymmetric line profiles, but such asymmetries are consistent with being produced by beam-smearing rather than by a distinct, anomalous \hi\ component.
Experiments made with mock \hi\ cubes indicate that even a small contamination from the disc emission can strongly affect the recovery of the EPG parameters, typically leading to lower values of ${\rm d}v_{\phi}/{\rm d}z$, $h$ and $\sigma$.
This is why we have adopted a conservative mask, achieved by further smoothing the Gaussian model with the instrumental beam: in many cases this leaves little \hi\ flux to work with (or no flux at all for NGC\,4275 and NGC\,4448) but it is a `necessary evil'.
In general, a more complete modelling approach would consist of fitting simultaneously both the disc and the EPG components to the data.
Such method may provide a better alternative to infer the EPG properties in poorly resolved galaxies, at the cost of a higher number of free parameters with respect to the current method.
In the near future, MeerKAT will provide deep \hi\ data at a higher angular resolution for several systems, significantly improving our understanding of the EPG properties in galaxies.

\subsection{Reliability of the uncertainties in the EPG parameters} \label{sec:uncertainties}
\begin{figure*}
\begin{center}
\includegraphics[width=1.0\textwidth]{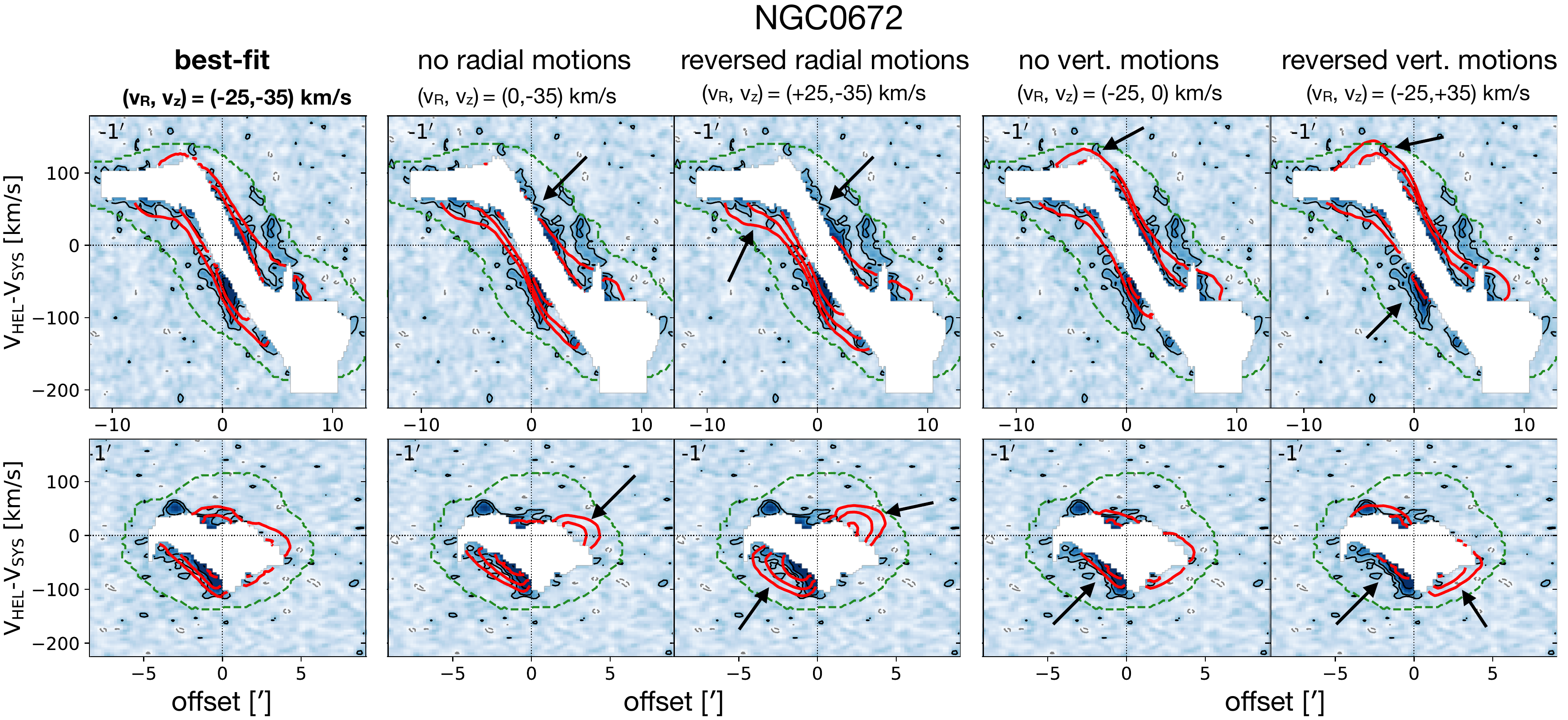}
\caption{Comparison between the best-fit model for the EPG of NGC\,0672 (first column) and other models that use the same parameters but different values for $v_{\rm z}$ and $v_{\rm R}$, as labelled on top of each panel. For simplicity we show only the pv-slices at $1'$-offset. Contours are the same as in Fig.\,\ref{fig:corner_n3198}. Arrows show the locations where the various models are inferior to the best-fit one.}
\label{fig:n672_vzvr}
\end{center}
\end{figure*}

\begin{figure}
\begin{center}
\includegraphics[width=0.48\textwidth]{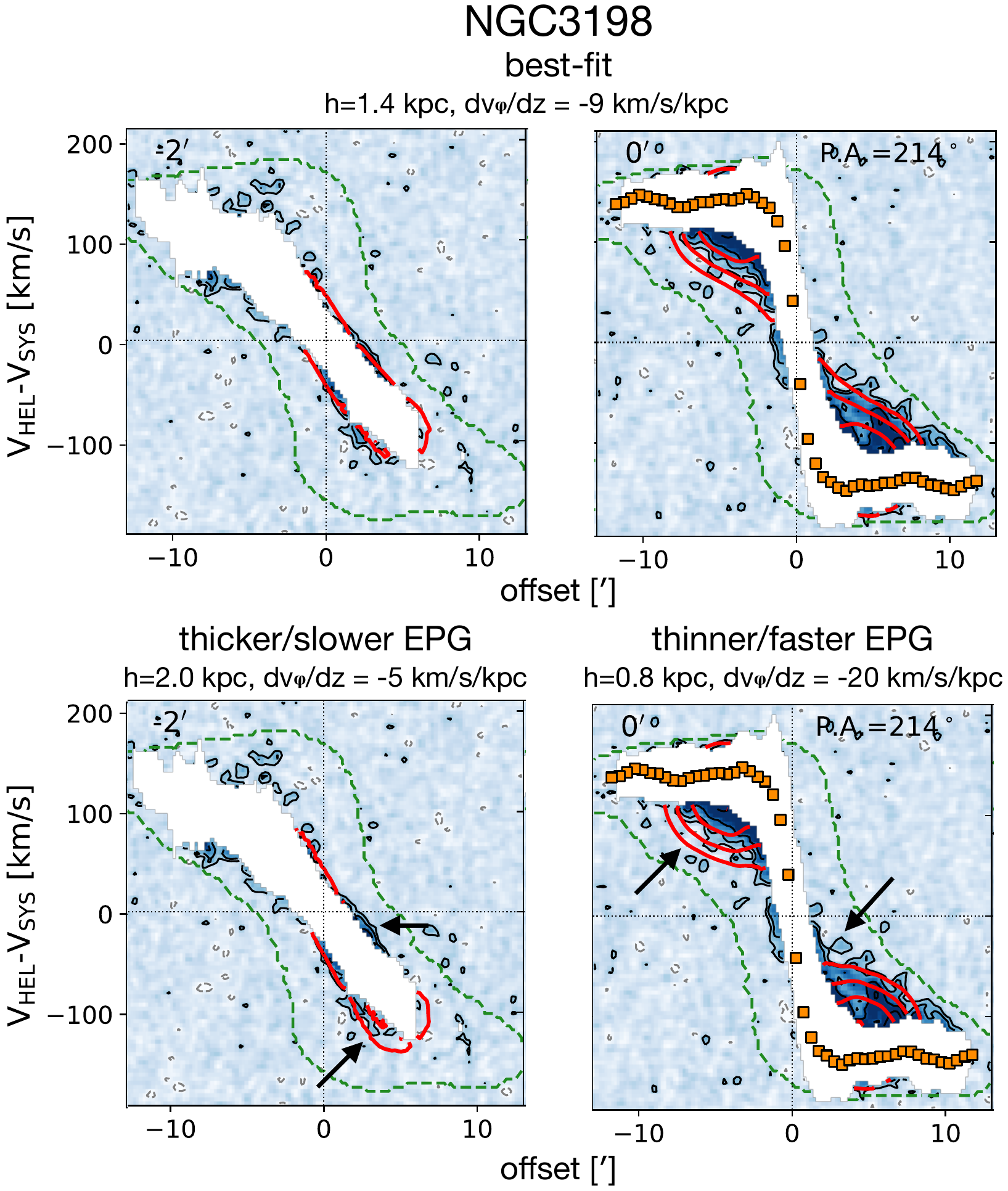}
\caption{Comparison between the best-fit model for the EPG of NGC\,3198 (top panels) and two models that use the same parameters but different values for $h$ and $dv_{\phi}/dz$ (bottom panels). Contours are the same as in Fig.\,\ref{fig:corner_n3198}. Arrows show the locations where the models are inferior to the best-fit one. We have focussed on pv-slices where the differences are more evident.}
\label{fig:n3198_hlag}
\end{center}
\end{figure}

Our fitting approach is extremely sensitive to tiny differences between models and data.
Despite our careful treatment of the likelihood (Section \ref{ssec:fit}), the a-posteriori probability of our models drops very quickly as the parameters depart from their best-fit values.
As a consequence, models that are labeled to differ by several $\sigma$ from the best-fit one may still appear as `good' at a visual inspection.
To illustrate this, in Fig.\,\ref{fig:n672_vzvr} we compare the best-fit model for the EPG of NGC\,672, for which we have found a clear evidence of vertical and radial inflow with $(v_{\rm z}, v_{\rm R})=(-35,-25)\kms$ and error-bars of a few $\kms$, with models that have the same best-fit parameters but where $v_{\rm z}$ or $v_{\rm R}$ are either zeroed or reverted.
While the best-fit model (first column in Fig.\,\ref{fig:n672_vzvr}) clearly outperforms those where the $v_{\rm R}$ or $v_{\rm z}$ have `wrong' sign (third and fifth column), it is only marginally better than those with no radial or vertical motions (second and fourth column). 
The latter in particular appears almost identical to the best-fit models, except for a few locations indicated by the arrows. 
The situation would not differ if we had chosen a different pv-slice.
However, given the nominal error-bars associated to our measurements, a model with $v_{\rm z}\!=\!0$ is at $\sim15\sigma$ from the best-fit values! 

The example discussed represents a somewhat extreme case, but similar considerations apply to other parameters as well. 
In Fig.\,\ref{fig:n3198_hlag} we compare the best-fit EPG for NGC\,3198 (top panels) with a model featuring a thicker, more slowly rotating EPG (bottom-left panel) and with a model where the EPG is thinner and the lag is more pronounced (bottom-right panel).
These models deviates by $\sim4-5\sigma$ from the best-fit one and are selected in the direction of the thickness-lag degeneracy. 
Also in this case the differences with respect to the best-fit model are marginal.

In general, it is hard to distinguish by eye models that differ by less than $\sim3\sigma$ from their quoted best-fit values.
In this work we used a statistical approach to the modelling and preferred not to adopt any arbitrary by-eye criterion to define the uncertainty associated to the estimated parameters.
The examples illustrated above may help the reader to have a better perspectives on the error-bars quoted in this study.

\subsection{Comparison with previous work}\label{ssec:previous}
The presence of anomalous \hi\ features around the disc of late-type galaxies has been known for several decades, and we refer the reader to the review of \citet{Sancisi+08} for a more complete discussion on the topic.
However, it was only in the last $\sim20$ years that more focussed studies on the anomalous \hi\ component were carried out for a limited number of systems, where deep \hi\ data were available.

NGC\,891 is arguably the best-studied case of EPG in galaxies \citep{SancisiAllen79,Swaters+97,Oosterloo+07}.
Edge-on systems like this have the considerable advantage that the disc/EPG separation can be done directly on the total \hi\ map rather than resorting to anomalies in the gas kinematics. 
Consequently, once the boundaries between disc and EPG have been set, the EPG mass can be readily computed and does not suffer from sensitivity drops along the minor axis as in galaxies at intermediate inclinations.
It has been suggested that the large EPG fraction of NGC\,891 \citep[$28\%$ of the total,][]{Marinacci+10a} indicates that at least part of its anomalous gas may originate from an interaction with the companion UGC\,1807.
While the typical $f_{\rm EPG}$ of our HALOGAS sample is $\sim14\%$, for NGC\,2403 and NGC\,0949 we measure $22\%$ and $27\%$ respectively. 
This suggests that the case of NGC\,891 is not unique.

Another relevant study was that of \citet{Fraternali+02}, who studied the EPG of NGC\,2403 using VLA \hi\ observations.
They used a kinematical disc/halo decomposition - on which the procedure adopted in this work is largely based - and inferred the presence of an EPG component with $f_{\rm EPG}\sim10\%$, characterised by a lagging rotation and a radial inflow with speeds of $10-20\kms$.
These properties were determined by applying a tilted ring model to the velocity field of the anomalous \hi\ emission.
We used the same data as in \citet{Fraternali+02} and reached very similar conclusions.
Our 3D modelling, though, allows us to determine the EPG kinematics with a greater accuracy: we find lagging rotation with ${\rm d}v_{\phi}/{\rm d}z$ of $\sim-12\kmskpc$ and a global inflow in both the radial ($-20\kms$) and vertical ($-10\kms$) directions. 
We also predict an EPG fraction of $22\%$, $\sim$ twice the value inferred by Fraternali et al. 
We believe that this difference is fully due to our modelling scheme and that our results are more reliable.

\citet{Barbieri+05} used WSRT \hi\ observations to study the anomalous gas in NGC\,4559, finding evidence for a thick ($2\!<\!h\!<\!3\kpc$), slowly rotating component with $f_{\rm EPG}$ of $\sim10\%$, and with possible hints of a radial inflow at the level of $\sim15\kms$.
Similar results were obtained by \citet{Vargas+17} who used the same HALOGAS data as in the current study and performed an analysis similar to that of G13 for NGC\,3198 (see below).
Our results are in agreement with these findings, and we can even confirm the presence of radial (and vertical) inflow motions.
We stress, though, that this system is significantly lopsided in both its density distribution and kinematics, which makes it difficult to model it via axisymmetric templates.
As shown by \citet{Vargas+17}, most of the anomalous \hi\ flux in this galaxy is spatially coincident with the regions of star formation as traced by H$\alpha$ and UV images, suggesting a galactic fountain origin for the EPG.

As already discussed (Section \ref{ssec:3198}), G13 determined an \hi\ mass and a scale-height for the EPG of NGC\,3198 that are about twice the values determined in this work.
Since we used the same data of G13, the difference must be in the method.
Firstly, G13 used a two-component model comprising a disc and an EPG layer to model the whole \hi\ dataset.
While in principle this approach is more powerful than that adopted here, in practice - given the large number of parameters involved - it relies on modifying by hand the model parameters one by one until a satisfactory reproduction of the data is achieved.
This complicates the assessment of the uncertainty associated to the various parameters and, most importantly in this particular case, of their degeneracy.
Secondly, G13 used an exponential vertical profile and a surface density that is a re-scaled version of that of the total \hi\ component.
As discussed in Section \ref{ssec:limitations}, fitting the EPG of NGC\,3198 with a double-exponential layer results in twice the EPG mass and a strong correlation between ${\rm d}v_{\phi}/{\rm d}z$ and $h$: a model with ${\rm d}v_{\phi}/{\rm d}z\!=\!-10\kmskpc$ and $h\!=\!3\kpc$ (as in G13) is indistinguishable from one with ${\rm d}v_{\phi}/{\rm d}z\!=\!-60\kmskpc$ and $h\!=\!0.4\kpc$, as we could verify directly in the pv slices.
Interestingly, the best-fit velocity dispersion for the EPG in this parametrisation is $\sigma\sim12\kms$ (with a large error-bar), equal to the fiducial value adopted by G13 but half of our current estimate.
These considerations highlight the importance of properly taking into account the correlations in the parameter space.

The HALOGAS \hi\ data of NGC\,4414 were already studied by \citet{deBlok+14b}, who noticed the presence of a prominent U-shaped warp in the system's outskirt.
They used different techniques to estimate the fraction of EPG in the innermost $\sim20\kpc$, finding values ranging from $3\%$ up to $12\%$ depending on the method adopted.
Our estimate ($12\%$) agrees with their quoted upper limit, but we reiterate that the quality of our fit is poor and our model may be not optimal to describe the EPG in this galaxy.
 
MF11 have used 3D models to infer the global properties of the EPG of the Milky Way.
They built a thin disc model to filter out the \hi\ emission from the regularly rotating disc and fit the leftover emission with the same model that we adopt here, although they had fixed $R_{\rm g}$ and $\gamma$ to some fiducial values in order to minimise the number of free parameters.
They were the first to highlight that the Galactic \hi\ halo features a lagging rotation ($-15\kmskpc$) and a global inflow towards the disc with speeds of $20-30\kms$, properties that - according the our new results - are quite common in late-type galaxies.

It is interesting to compare our findings for systems at intermediate inclinations with those reported in the literature in edge-on galaxies.
Edge-on systems constitute an ideal laboratory to study the vertical distribution and rotational lag of the EPG, provided that line-of-sight warps are properly taken into account \citep[e.g.][]{Kamphuis+13}.
A compilation of EPG studies in edge-on systems is offered in \citet{Zschaechner+15}, who highlighted how the rotational lag is maximised close to the galaxy centre, reaching values up to $-40\kmskpc$, and gradually weakens outwards, flattening beyond $\sim\!R_{25}$ at values $\gtrsim-10\kmskpc$.
As we do not model the radial variation of ${\rm d}v_{\phi}/{\rm d}z$, the values that we infer are more sensitive to the outer parts of the EPG layer, where most of the \hi\ flux is enclosed.
In this respect, our measurements for the rotational lag are compatible with those of \citet{Zschaechner+15}.
Instead, the EPG scale-heights listed in \citet{Zschaechner+15} are smaller than those determined in this work by a factor $\sim2-3$.
A possible explanation for this discrepancy is again the different vertical density profile adopted (exponential vs our Eq.\,\ref{eq:vert_halo}), as discussed in Section \ref{ssec:limitations}.

\subsection{The origin of the EPG}\label{ssec:origin}
\begin{figure*}
\begin{center}
\includegraphics[width=0.49\textwidth]{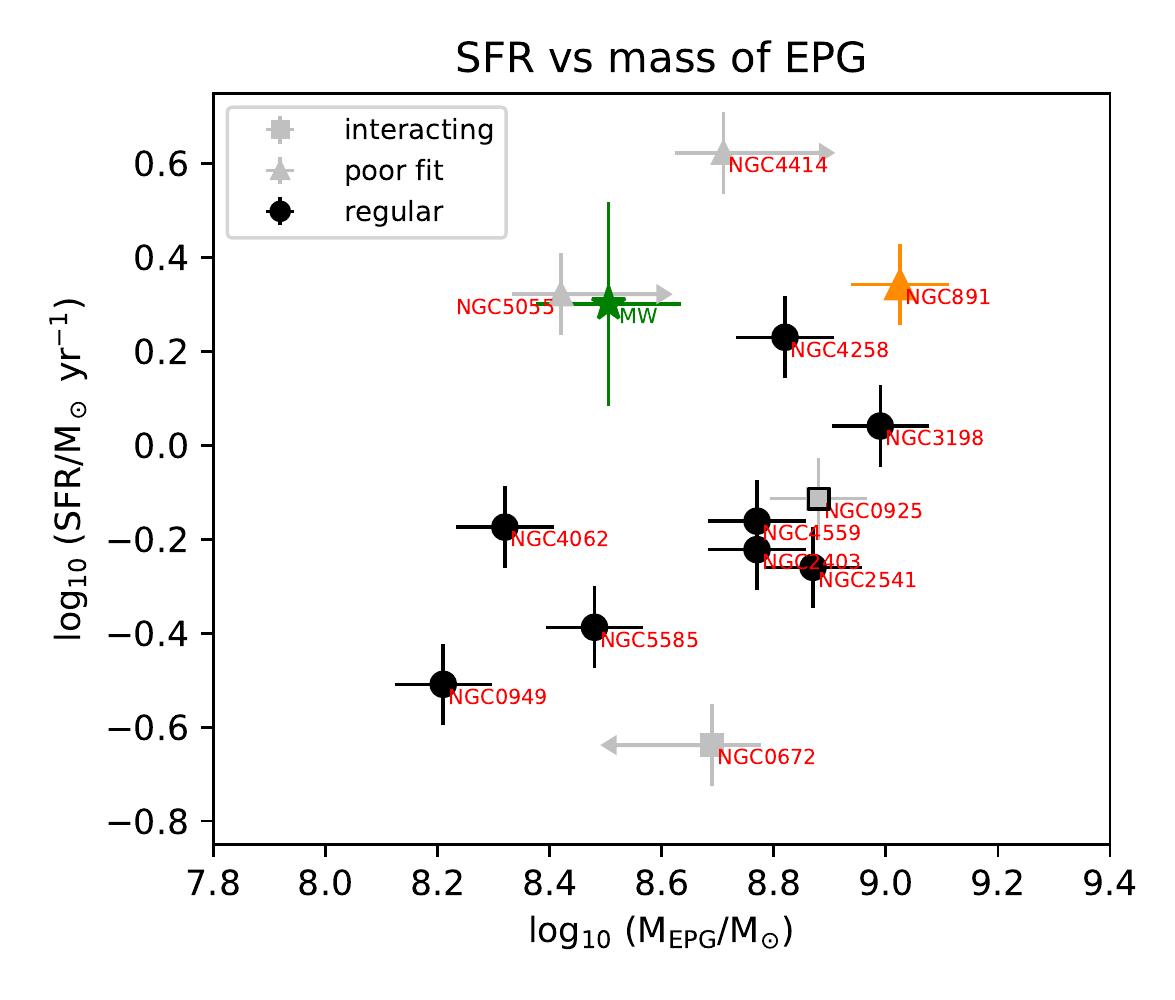}
\includegraphics[width=0.49\textwidth]{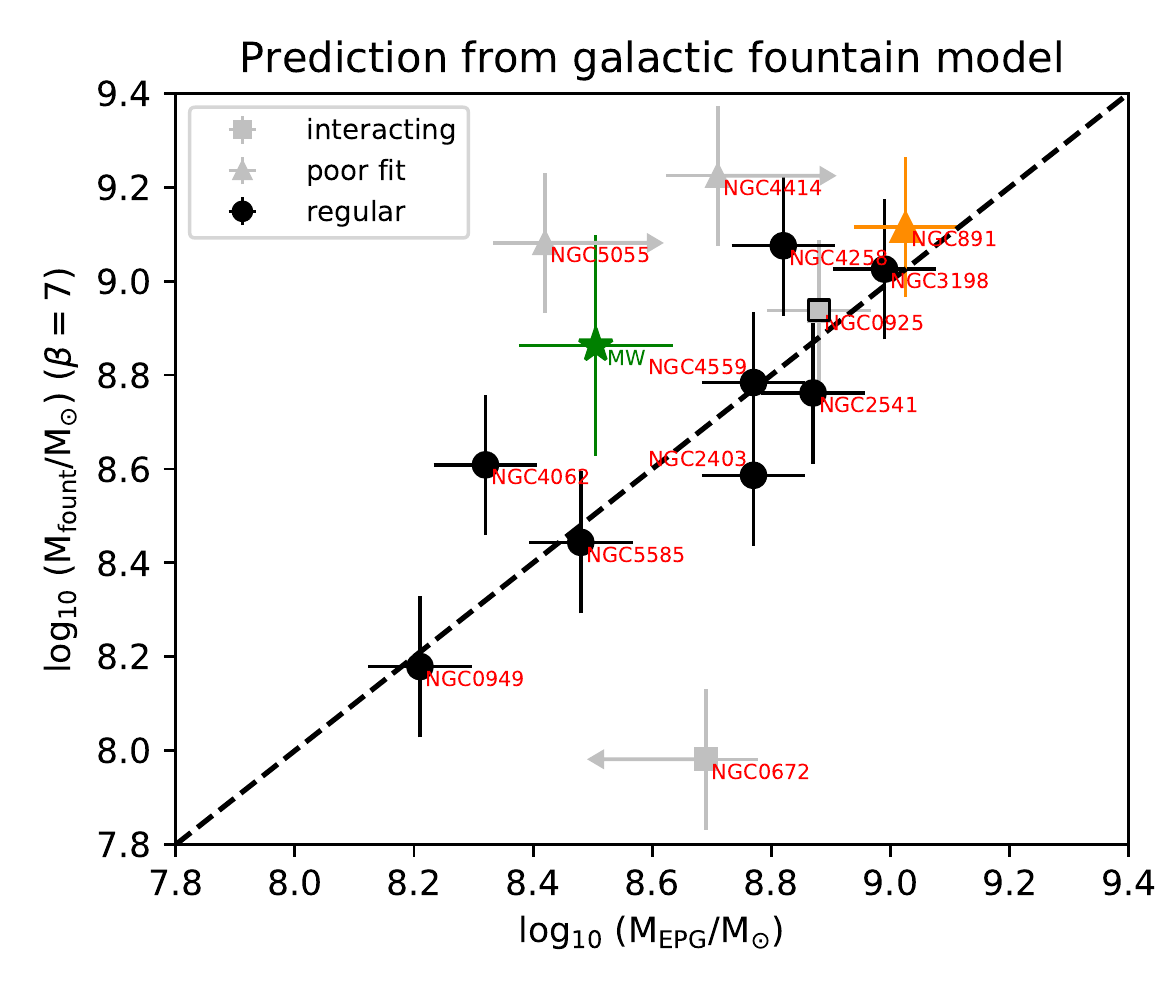}
\caption{\emph{Left panel}: SFR vs EPG mass for our HALOGAS sample. \emph{Right panel}: theoretical EPG mass predicted by a galactic fountain model (eq.\,\ref{eq:mfount_approx} with $\beta=7$) vs that inferred from the data. The two panels span the same range ($\sim1.5$ dex) on both axes. Error-bars are determined by assuming a $10\%$ uncertainty on the distances and a $20\%$ uncertainty on the total SFR, in addition to the uncertainty on the $R_{\rm SFR}$ discussed in the text. Interacting galaxies and systems with a poorly fit EPG are shown with separate symbols. We have also included NGC\,891\citep[M$_{\rm EPG}$ from][]{Marinacci+10a} and the Milky Way, for which we used $M_{\rm EPG}\!=\!3.2\pm1.0\times10^8\msun$ (MF11), and assumed fiducial values for SFR, R$_{\rm SFR}$ and $v_{\rm flat}$ of $2\pm1\msunyr$, $2.5\pm0.5\kpc$ and $240\pm10\kms$ respectively.}
\label{fig:fountain_prediction}
\end{center}
\end{figure*}

The best-fit parameters found for the EPG give us fundamental clues on the origin of this component.
Our analysis reveals that in all HALOGAS systems the EPG `knows' about the presence of the disc, as it rotates along with it at a (slighty) lower speed, with a ${\rm d}v_{\phi}/{\rm d}z$ that varies from system to system but has a typical value of $-10\kmskpc$.
In addition, in most galaxies the EPG appears to be globally inflowing in both the radial and vertical directions, with velocities around $20-30\kms$.
This peculiar kinematics, along with a significant thickness ($1\!<\!h\!<\!3\kpc$), imply that the EPG properties are not compatible with predictions based on hydrostatic equilibrium models \citep[e.g.][]{Barnabe+06,Marinacci+10a,Bacchini+19}.
Non-stationary models are therefore required to explain the properties of this component.

A possibility is that the whole EPG originates from the spontaneous condensation of the innermost layers of the CGM, leading to a continuous, smooth accretion of pristine gas onto the disc.
In galaxies where we find $v_{\rm z}<0$, we can compute the infall rate of the EPG, $\dot{M}$, as $1.4\,M_{\rm HI, EPG}\,v_{\rm z}\,\avg{z}^{-1}$, where the factor $1.4$ accounts for the helium and $\avg{z}\simeq1.32h$ is the median EPG displacement from the midplane.
We find inflow rates ranging from 3 to about 50 times the SFR of each system, with a median $\dot{M}$/SFR of $9$.
This implies that, if the EPG were produced solely by external gas accretion, galaxies would double their cold gas reservoir in a mere few hundred Myrs, indefinitely growing in mass and size at a very fast pace.
From the theoretical point of view, it is very unlikely that such a disparity between gas accretion and star formation rate can lead to a realistic galaxy population. 
These considerations, along with theoretical arguments against the spontaneous condensation of the CGM \citep[e.g.][]{Binney+09,Nipoti10}, strongly disfavour this scenario.
We do not exclude, though, that a \emph{small} fraction of the EPG may come from external accretion (see below).

Following the work firstly pioneered by \citet{FB06} and then followed up by \citet{FB08} and \citet{Marasco+12}, we can readily interpret all the discussed features in the framework of the galactic fountain cycle.
In this context the EPG is built by material ejected from the disc by stellar feedback that travels through the halo region, eventually returning back to the disc on timescales of several tens of Myrs.
The fountain material is ejected from regions of active star formation and thus, at the beginning of its journey through the halo, can be photo or shock-ionised and therefore invisible in the \hi\ phase.
This produces an observational bias in \hi\ against the fountain material that leaves the disc ($v_{\rm z}>0$) in favour of that returning back to it ($v_{\rm z}<0$), which justifies the vertical velocities inferred for most galaxies.
A higher velocity dispersion for the fountain material is expected due to fountain cloud-to-cloud relative motions and, for single clouds, to ram-pressure stripping from the hot CGM.

The fountain clouds inevitably develop some rotational lag as a consequence of the angular momentum conservation \citep[for details see][]{Fraternali17}, but this is typically insufficient to explain the measured values of ${\rm d}v_{\phi}/{\rm d}z$.
In addition, simple ballistic orbits for clouds ejected perpendicularly to the disc show that the fountain material moves preferentially outwards ($v_{\rm R}>0)$, in disagreement with our measurements.
Both these issues can be solved by assuming that the fountain clouds interact with a CGM having locally a \emph{lower specific angular momentum}: as the clouds exchange momentum with the CGM, they lose rotational speed and flow backwards, acquiring a net $v_{\rm R}<0$ in the latest part of their trajectory \citep[see Fig.\,6 in][]{FB08}.
In this more refined model, the fountain orbits are characterised by positive values of ($v_{\rm z}$,$v_{\rm R}$) in the ascending part and by negative ($v_{\rm z}$,$v_{\rm R}$) in the descending part, while mixed signs for the radial and vertical velocities are less common.
This pattern is consistent with that shown in Fig.\,\ref{fig:kinematics}, where the top-left and bottom-right regions are practically empty (the exception being NGC\,925, which is still compatible with $v_{\rm z}\simeq0$ and is an interacting system).
Photo-Ionisation in the early (ascending) phase can explain the bias towards the descending portion of the fountain, hence the position of most galaxies in Fig.\,\ref{fig:kinematics}, while systems like NGC\,5585 and NGC\,3198 may be characterised by a different distribution in their ionisation flux that drives them towards the opposite side of the plot.
The effect of photo-ionisation could be investigated further in H$\alpha$ spectroscopic cubes \citep[e.g. from Fabry-P\'erot observations, or with WEAVE,][]{weave}, using the same modelling approach adopted in this work, by looking at signatures of opposite ($v_{\rm z}$,$v_{\rm R}$) as compared to the HALOGAS results, although part of the outflowing gas can be very hot and thus not visible in H$\alpha$.

It is not straightforward to predict a priori a timescale for the recombination of the fountain gas, as this depends on the initial launching conditions and on the combined effect of different ionising mechanisms.
A lower limit can be set by the typical recombination timescales for the warm interstellar medium in absence of any source of radiation, which range between $0.1-1\Myr$ depending on the electron density \citep{Spitzer78}.
A rough upper limit can be set by the lifetime of OB associations (of the order of $10\Myr$) under the assumption that the gas is fully photo-ionised, although extragalactic UV background may be important above the disc by somewhat increasing this value.
Early hydrodynamical calculations for thermally-powered galactic fountains \citep{Houck&Bregman90} have also suggested a few tens of $\Myr$ as the time required by the hot and rarefied material ejected into the halo to condense and form discrete dense clouds.
Finally, $17\Myr$ is the best-fit recombination timescale in the fountain model of \citet{Marasco+12} required to reproduce the Galactic EPG \hi\ data.
All these timescales are typically smaller than the fountain orbital times ($30-70\Myr$, see eq.\,\ref{eq:torb} below).

\citet{Marasco+12} successfully modelled the EPG in the Milky Way using a dynamical model of the galactic fountain interacting with the CGM.
As the kinematics of the Galactic EPG is similar to that of most HALOGAS systems (Figure \ref{fig:kinematics}), we may expect that the same model can be applied to the galaxies studied in this work.
This will be the subject of a future study.

To further test the galactic fountain framework, we compare the EPG masses reported in Table \ref{tab:propertiesEPG} to the mass expected to be produced by a galactic fountain cycle, which we write in a general form as
\begin{equation}\label{eq:mfount}
M_{\rm fount} = 2 \pi \int_0^{+\infty} \Sigma_{\rm fount}(R) R {\rm d}R
\end{equation}
where $\Sigma_{\rm fount}$ is the gas surface density of the fountain component.
Given that the fountain cycle is powered by stellar feedback, we expect that $\Sigma_{\rm fount}$ depends on both the star formation rate density $\Sigma_{\rm SFR}$ and on the vertical orbital time $t_{\rm orb}$:
\begin{equation}\label{eq:sigma_fount}
\Sigma_{\rm fount}(R) = \beta\,\Sigma_{\rm SFR}(R) \times t_{\rm orb}(R)
\end{equation}
where $\beta$ is the \emph{mass-loading factor} and represents the amount of gas that joins the fountain cycle per unit star formation rate.

For a flat rotation curve in an isothermal potential, the orbital time is nearly a linear function of radius and can be roughly approximated as
\begin{equation}\label{eq:torb}
\frac{t_{\rm orb}(R)}{\Myr} \simeq 25\left(\frac{R}{\kpc}\right) \left(\frac{v_{\rm flat}}{100\kms}\right)^{-1}\,.
\end{equation}
Eq.\,(\ref{eq:sigma_fount}) can be further simplified by assuming that $\Sigma_{\rm SFR}$ is an exponential function of radius:
\begin{equation}\label{eq:sigma_SFR}
\Sigma_{\rm SFR}(R) = \frac{\rm SFR}{2\pi R^2_{\rm SFR}} \exp(-R/R_{\rm SFR})
\end{equation}
where the value of the total SFR for each galaxy is taken from Table \ref{tab:sample}.
Combining eq.\,(\ref{eq:mfount}), (\ref{eq:torb}) and (\ref{eq:sigma_SFR}) leads to a simple approximation for the fountain mass:
\begin{equation}\label{eq:mfount_approx}
\frac{M_{\rm fount}}{\msun} \simeq 5\times10^9 \left(\frac{\beta \times \rm SFR}{\msunyr}\right) \left(\frac{R_{\rm SFR}}{\kpc}\right) \left(\frac{v_{\rm flat}}{\kms}\right)^{-1}\, .
\end{equation}
As for $R_{\rm SFR}$, we use the results of \citet{Leroy+08} who studied the FUV+NIR star formation density profiles for a sample of 23 nearby galaxies and found $R_{\rm SFR}\simeq(0.22\pm0.06)R_{25}$.

The left panel of Fig.\,\ref{fig:fountain_prediction} compares SFRs and EPG masses for our galaxy sample.
While these two quantities are arguably related in a galactic fountain framework, only a mild correlation appears to be present (Pearson coefficient of $0.23$, or $0.66$ using only the `regular' systems).
The correlation improves significantly (Pearson coefficient of $0.44$, or $0.84$ for the regular systems) when we compare the EPG masses with the theoretical prediction from eq.\,(\ref{eq:mfount_approx}), where we used a constant ad-hoc mass-loading factor $\beta\!=\!7$ (right panel of Fig.\,\ref{fig:fountain_prediction}) which we discuss below.
Now most HALOGAS systems align well on the one-to-one line, notable exceptions being NGC\,5055, NGC\,4414 and NGC\,672.
As discussed already, the EPG masses of the first two systems are likely underestimated by our model, while that of NGC\,672 may be augmented by the interaction/merging with its (massive) companion.
The trend is remarkably tight if one focuses solely on the non-interacting, well-fitted HALOGAS galaxies, corroborating the validity of the galactic fountain framework.
Interestingly, NGC\,925 follows very well the trend of the non-interacting systems despite the presence of a low-mass nearby perturber.
We have used distinct symbols in Fig.\,\ref{fig:fountain_prediction} to emphasise that the magnitude of perturbation expected for NGC\,925 is small compared to that of NGC\,672.

The value of the mass loading factor adopted in eq.\,(\ref{eq:mfount_approx}) deserves further discussion.
In general, $\beta$ is very difficult to constrain observationally and we have fixed its value to $7$ in order to maximise the agreement between the observed and the theoretical EPG masses.
Remarkably, this empirically estimated value is perfectly compatible with those determined by \citet{FB06} by applying their dynamical fountain models to NGC\,891 and 2403: they found ratios between the fountain outflow rate and the star formation rate ranging from $4$ to $9$, depending on the details of the gravitational potential.
It is interesting to combine our estimate for $\beta$ with the typical EPG inflow rate determined at the beginning of this Section. 
If stellar feedback ejects material at a typical rate of $7$ times the SFR, and the inflow rate of the EPG is $\sim9$ times the SFR, there is a \emph{net gas accretion} onto the disc at (approximately) the same rate as the star formation.
While these numbers represent only a crude estimate for the actual outflow/inflow rates, the scenario described is compatible with expectations from a fountain-driven gas accretion framework \citep{FB08,Marasco+12,Fraternali17}, where fountain clouds trigger the condensation and stimulate the subsequent accretion of a small fraction of the CGM.
This accreting gas plays a key role in the galaxy evolution by replenishing the material used in the process of star formation \citep[e.g.][]{Sancisi+08}.

As discussed, NGC\,4274 and NGC\,4448 have been excluded from our analysis as our masking procedure leaves little \hi\ emission to work with.
The theoretical EPG masses predicted by our fountain model for these two systems are $\log(M_{\rm EPG}/\msun)=8.85$ and $7.02$ respectively, with an uncertainty of $0.15$ dex.
The small mass predicted for NGC\,4448 comes from its exceptionally small star formation rate ($0.056\msunyr$) and scale length ($1.2\kpc$).
One may ask whether EPG layers with such masses would be actually detectable in the data or, equivalently, what is the upper limit on the EPG mass for these two systems.
Answering these questions is hard, given that the detectability of EPG at a given mass depends on its kinematics and density distribution, which are unknown.
As a test, we have produced synthetic cubes for a series of EPG layers of different masses, using as representative scale-height and kinematical parameters the median values of the sample ($h=1.6\kpc$, ${\rm d}v_{\phi}/{\rm d}z=-10\kmskpc$, $v_{\rm z}=-18\kms$, $v_{\rm R}=-20\kms$, $\sigma=25\kms$) and experimenting with different values for $R_{\rm g}$ and $\gamma$. 
We verified visually the detectability of these models in the pv-slices of NGC\,4274 and NGC\,4448.
The results of these experiments show that, for $\log(M_{\rm EPG}/\msun)\gtrsim8.5$ ($7.4$), anomalous \hi\ always appears in the pv-slices of NGC\,4274 (NGC\,4448) regardless the details of the surface density distribution.
The quoted detection limits correspond to about $40\%$ and $60\%$ of the total \hi\ content of NGC\,4274 and NGC\,4448, well above the typical EPG fractions measured for the rest of the sample.
With respect to our predictions, NGC\,4274 is somehow poor of anomalous \hi\ while NGC\,4448 seems consistent with the fountain model, i.e., its fountain mass from Eq.\,(\ref{eq:mfount_approx}) is below the detection limit.

\section{Conclusions}\label{sec:conclusions}
In this work we have analysed a sample of 15 late-type galaxies at intermediate inclination using publicly available \hi\ data from the HALOGAS survey \citep{Heald+11}. 
The unprecedented depth of these data allowed us to study the properties of the extraplanar gas (EPG), the faint, kinematically anomalous \hi\ component that lies at the interface between the disc and the large-scale circumgalactic medium (CGM).
This is the first time that a systematic and homogeneous study of the EPG has been carried out on a sample of galaxies.

In order to separate the \hi\ emission from the EPG, we have built Gaussian models for the \hi\ line profiles and used them to filter out the disc contribution in the datacubes. 
The leftover, kinematically anomalous \hi\ emission has been modelled by building synthetic \hi\ cubes of idealised EPG and fitting them to the data via a Bayesian MCMC approach.
Our models use 3 parameters to describe the EPG density distribution (eq.\,\ref{eq:surf_halo} and \ref{eq:vert_halo}) and 4 parameters for its global kinematics (vertical rotational lag ${\rm d}v_{\phi}/{\rm d}z$, radial velocity $v_{\rm R}$, vertical velocity $v_{\rm z}$ and velocity dispersion $\sigma$).

Our results can be summarised as follows:
\begin{enumerate}
\item The EPG is ubiquitous in late-type galaxies. 
We have failed to detect it only in 2 galaxies (NGC\,4274 and NGC\,4448) out of 15, arguably because of the poor spatial resolution that complicates the disc/EPG separation. We have also excluded NGC\,1003 from our analysis due to the nearly edge-on orientation of its outermost \hi\ component.
\item The EPG typically accounts for $\sim10-15\%$ of the total \hi\ mass, reaching up to $\sim25\%$ in a small number of individual cases.
\item The scale-height of the EPG is in the range $1-3\kpc$. 
These values are not compatible with expectations based on hydrostatic equilibrium.
\item The kinematics of the EPG is vastly dominated by rotation. All systems show a vertical \emph{lag} in rotational speed, ${\rm d}v_{\phi}/{\rm d}z$, with an average value of about $-10\kmskpc$. There are no galaxies with ${\rm d}v_{\phi}/{\rm d}z\ge0$.
\item In most galaxies (9 out of 12) the EPG appears to be globally \emph{inflowing} towards the disc/centre, with typical speeds of $\sim20-30\kms$ in both the radial and vertical direction. 
\end{enumerate}

A galactic fountain appears to be the most probable origin for the EPG in late-type galaxies, although gas accretion is likely to contribute \citep{Sancisi+08, Fraternali17}.
The EPG masses of the HALOGAS galaxies are in good agreement with theoretical expectations based on a simple galactic fountain model.
The kinematics of the anomalous \hi\ can be explained qualitatively by two processes: photo-ionisation of the gas outflowing from the disc, and interaction between the fountain material and a CGM with a lower specific angular momentum.
Both these processes were key to the success of the dynamical models of \citet{Marasco+12} at reproducing the properties of the Galactic EPG.
Our analysis shows that the EPG of most late-type galaxies has properties analogous to those of the Milky Way.
The application of dynamical models of the galactic fountain to the HALOGAS data will allow us to understand better the physics of the disc/CGM interface and constitutes a natural follow-up to the current study.

\begin{acknowledgements}
The authors thank Renzo Sancisi, Richard Rand and Paolo Serra for providing constructing comments to the manuscript.
FF and AM thank Martina Valleri who, in her Master's thesis in 2013, had already found the presence of extraplanar gas in several nearby galaxies from the THINGS survey.
The Westerbork Synthesis Radio Telescope is operated by the ASTRON (Netherlands Foundation for Research in Astronomy) with support from the Netherlands Foundation for Scientific Research NWO.
\end{acknowledgements}


\bibliographystyle{aa} 
\bibliography{extrahalogas} 

\appendix
\section{The optimal sampling}\label{app:npart}
A quantity that is relevant in our modelling procedure is the number of particles $N$ used in the stochastic realisation of the model.
It is convenient to express $N$ as $\eta\times n_{\rm res}$, with $\eta$ being the mean number of particles per resolution element and $n_{\rm res}$ being the number of independent resolution elements in the cube.
The latter depends on the angular and kinematical `size' of the system and on the resolution, and can be approximated as
\begin{multline}\label{eq:nres}
n_{\rm res} \simeq 1.18\times10^5 \left(\frac{R_{\rm out}}{\kpc}\right)^2 \left(\frac{v_{\rm max}\sin(i)+50}{\kms}\right)\times \\ 
\times\left[\left(\frac{\Delta_{\rm v}}{\kms}\right) \left(\frac{D}{\Mpc}\right) \left(\frac{B_{\rm maj} B_{\rm min}}{{\rm arcsec}^2}\right)\right]^{-1}
\end{multline}
where $R_{\rm out}$ is the outermost radius that we model (in our case the outermost point of the rotation curve), v$_{\rm max}$ is the maximum rotation velocity, $i$ is the inclination, and the $50\kms$ term represents a typical FWHM of the \hi\ line broadening for the EPG component. 
Thus $N$ can be readily computed for any choice of $\eta$ via eq.\,\ref{eq:nres}.

\begin{figure}
\begin{center}
\includegraphics[width=0.45\textwidth]{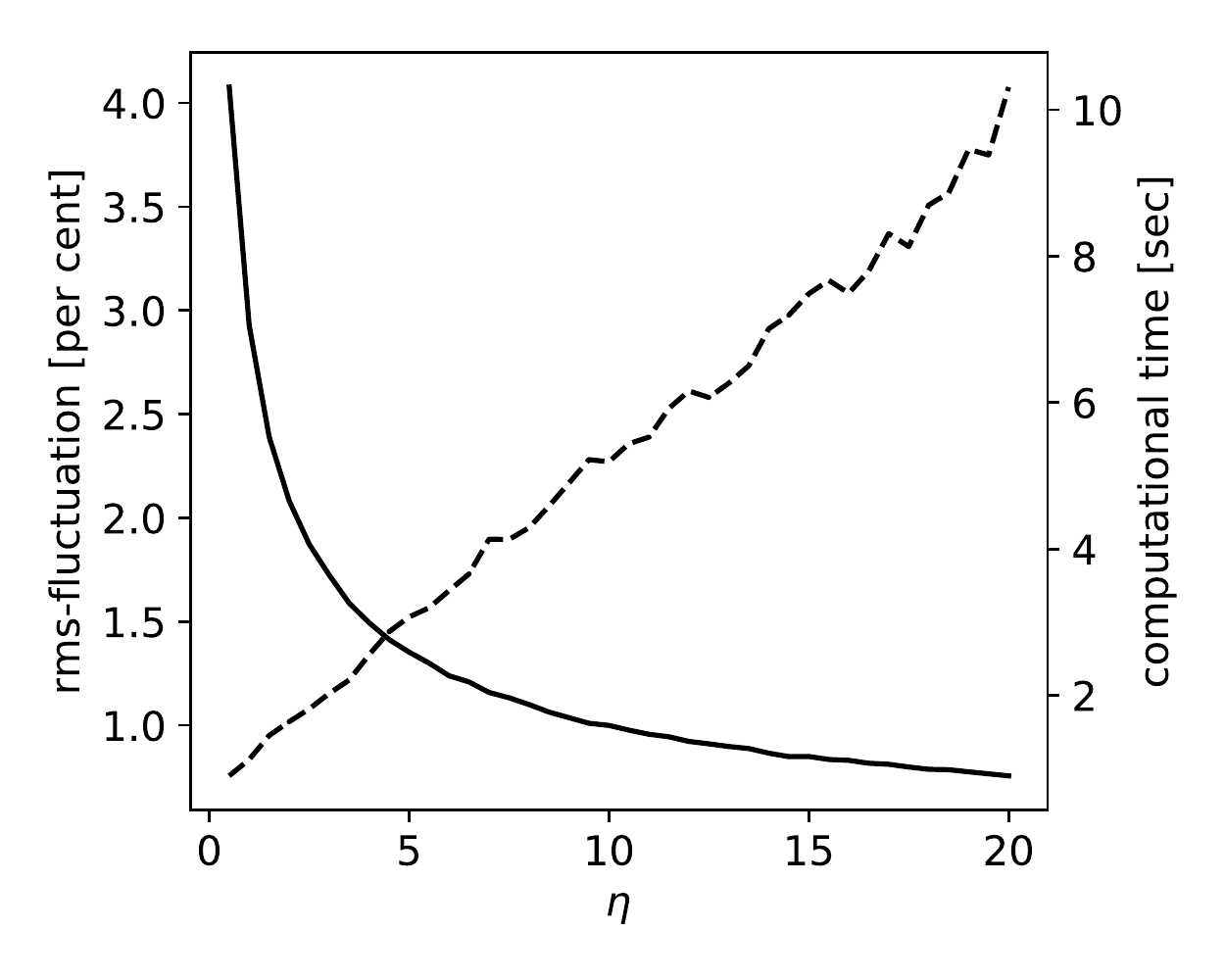}
\caption{Rms-fluctuation in the cubes (solid line, left-hand axis) and median computational time per model (dashed line, right-hand axis) as a function of $\eta$, the mean number of particle per resolution element used in the modelling. The improvements in the modelling accuracy are negligible for $\eta\!\gtrsim\!4$.}
\label{fig:eta}
\end{center}
\end{figure}

Clearly, the larger $\eta$ is, the smaller the stochastic fluctuations between different realisations of the same model are, at the expenses of the computational time.
In order to test how the magnitude of these fluctuations varies as a function of $\eta$ we proceed as follows.
As a fiducial case we consider the best-fit parameters found for the EPG of NGC\,3198 and reported in Table \ref{tab:propertiesEPG}.
Using these parameters we first construct a very-high resolution cube using $\eta\!=\!50$, which is used as a reference.
Then we explore the range $0.5\!<\!\eta\!<\!20$ and, for any given $\eta$, we build multiple realisations of the same model, for each storing the absolute residual between the current and the reference cube.
Fig.\,\ref{fig:eta} shows how the rms-fluctuation of these residuals decreases (solid line) and how the median computational time grows (dashed line) for increasing values of $\eta$.
While the computational time increases linearly with $\eta$, the rms-fluctuations show an inflection point around $\eta\sim4$ above which the accuracy in the model improves very slowly with increasing the number of particles.
For this reason we set $\eta\!=\!4$ for all models used in this work.

\section{Testing the method on mock data}\label{app:mcmctest}
We now test whether our method can recover the input parameters in mock \hi\ observations.
We first focus on a well resolved galaxy with a prominent EPG component, and then discuss a less resolved, more difficult system.
For both cases we build models made of two \hi\ components: a regularly rotating thin disc and a layer of EPG.

\begin{figure*}
\begin{center}
\includegraphics[width=0.85\textwidth]{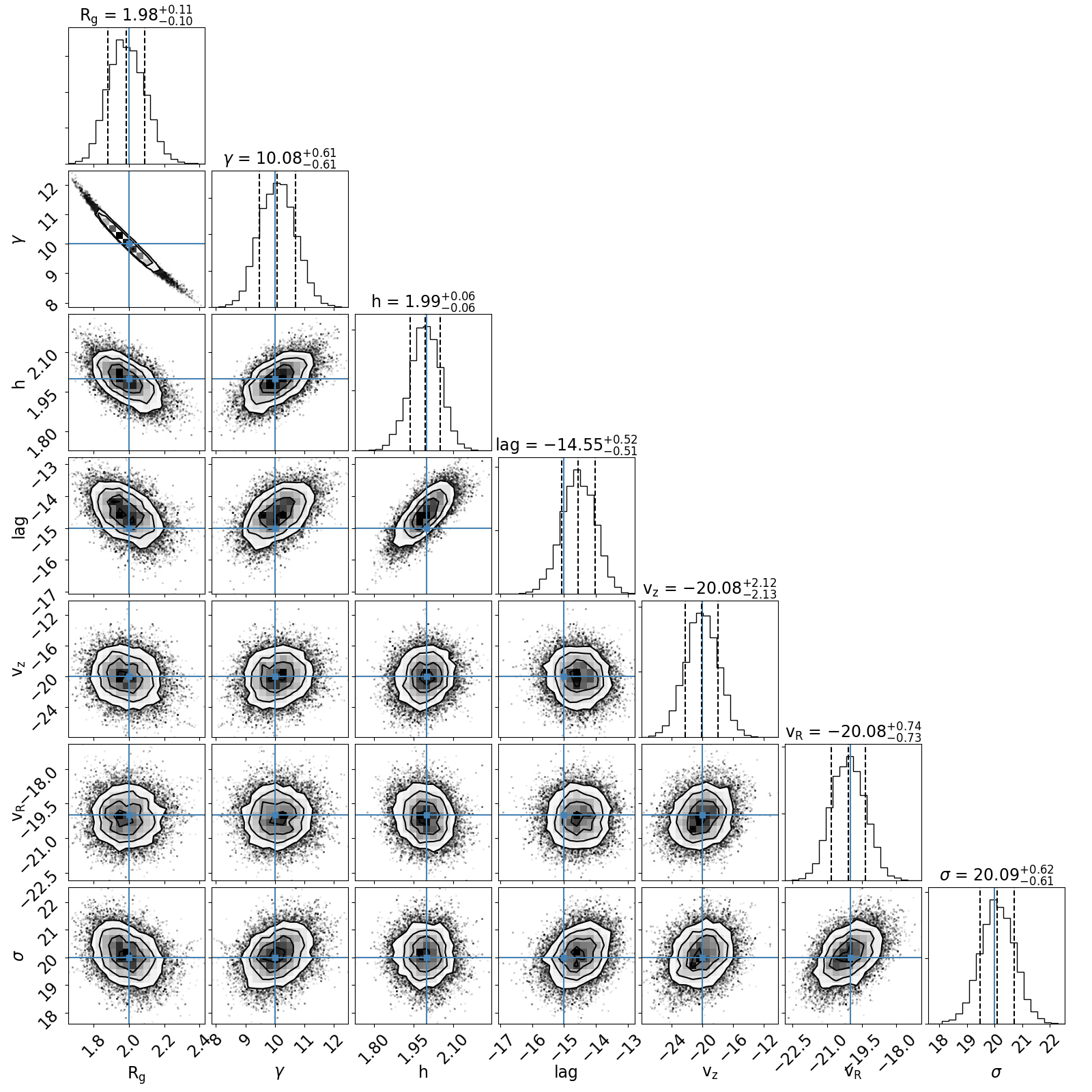}
\includegraphics[width=1.0\textwidth]{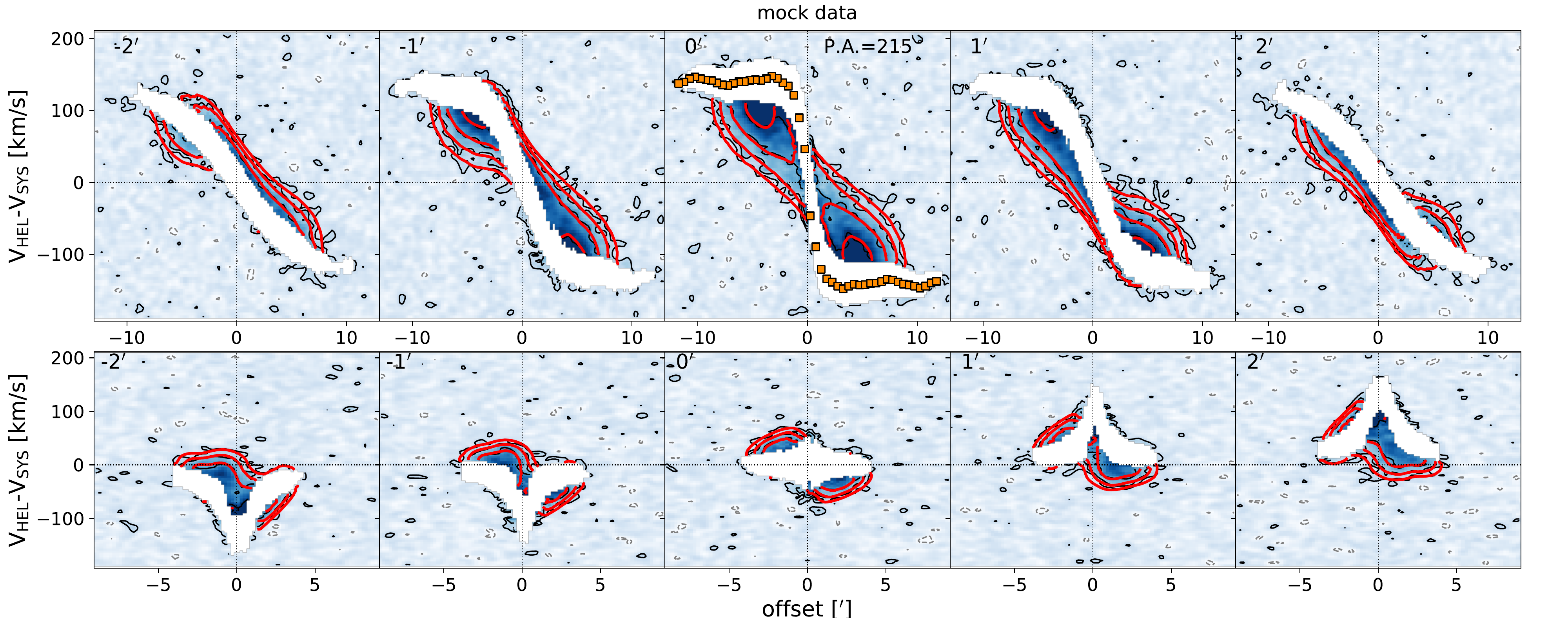}
\caption{As in for Fig.\,\ref{fig:corner_n3198}, but using the mock data as described in Appendix \ref{app:mcmctest}.} 
\label{fig:corner_test}
\end{center}
\end{figure*}

For the first case we take as a reference NGC\,3198 and build our model using the same distance and rotation curve of this galaxy.
The synthetic cube is also based on the same resolution and grid size of the HALOGAS data of NGC\,3198.
For the thin disc, we assume Gaussian distributions for the radial surface density and the vertical density profiles, with standard deviations of $15\kpc$ and $0.1\kpc$ respectively.
For the EPG we adopt $h\!=\!2\kpc$, ${\rm d}v_{\phi}/{\rm d}z\!=\!-15\kmskpc$, $v_{\rm R}=v_{\rm z}\!=\!-20\kms$, $\sigma\!=\!20\kms$, and use the same best-fit values for $R_{\rm g}$ and $\gamma$ found for NGC\,3198.
The \hi\ mass of the system is $10^{10}\msun$ with the EPG accounting for $15\%$ of the total, similar to the median $f_{\rm EPG}$ in our sample (but a factor $1.7$ larger than the $f_{\rm EPG}$ found for NGC\,3198).
This choice for the parameters produces quite prominent `beards' in the pv slices, as shown by the bottom panel of Fig.\,\ref{fig:corner_test}).

We further inject artificial noise in our synthetic cube with the following procedure. We first create a separate cube made of Gaussian noise, which we then convolve with the observation beam and perform Hanning smoothing. 
We then re-scale the intensities so that their standard deviation matches the rms-noise of the HALOGAS data, and finally sum it to our synthetic.

We analyse the mock data with the procedure described in Section \ref{sec:method}.
Fig.\,\ref{fig:corner_test} shows the inferred posterior probability of the EPG parameters.
The input parameters, marked with blue squares, always fall within the 16th and 84th percentiles (corresponding to a $1\sigma$ uncertainty) of the posterior distribution, indicating that our method can robustly mask the emission coming from the disc and efficiently recover the properties of the extra-planar gas.

How much can we degrade our resolution and signal-to-noise and still be able to recover the correct input parameters?
To answer this question, we have repeated the exercise above but used as a reference galaxy NGC\,4062, the most distant and less resolved galaxy in our sample for which we are still able to characterise the EPG component.
We have used EPG parameters analogous to those inferred for this galaxy\footnote{we assumed $R_{\rm g}=3\kpc$ as this parameter is unconstrained.}, a total \hi\ mass of $1.4\times10^9\msun$ with an $f_{\rm EPG}$ of $15\%$, a thin disc scale length of $4.7\kpc$, and the same resolution and grid size of the HALOGAS data for this system.
In this case the input and output parameters were compatible with each other only within the $2\sigma$ uncertainty, indicating that our procedure is still quite reliable at this resolution and signal-to-noise but should not be applied to data of lower quality.

\section{Column density maps}\label{app:maps}
\begin{figure*}
\begin{center}
\includegraphics[width=0.75\textwidth]{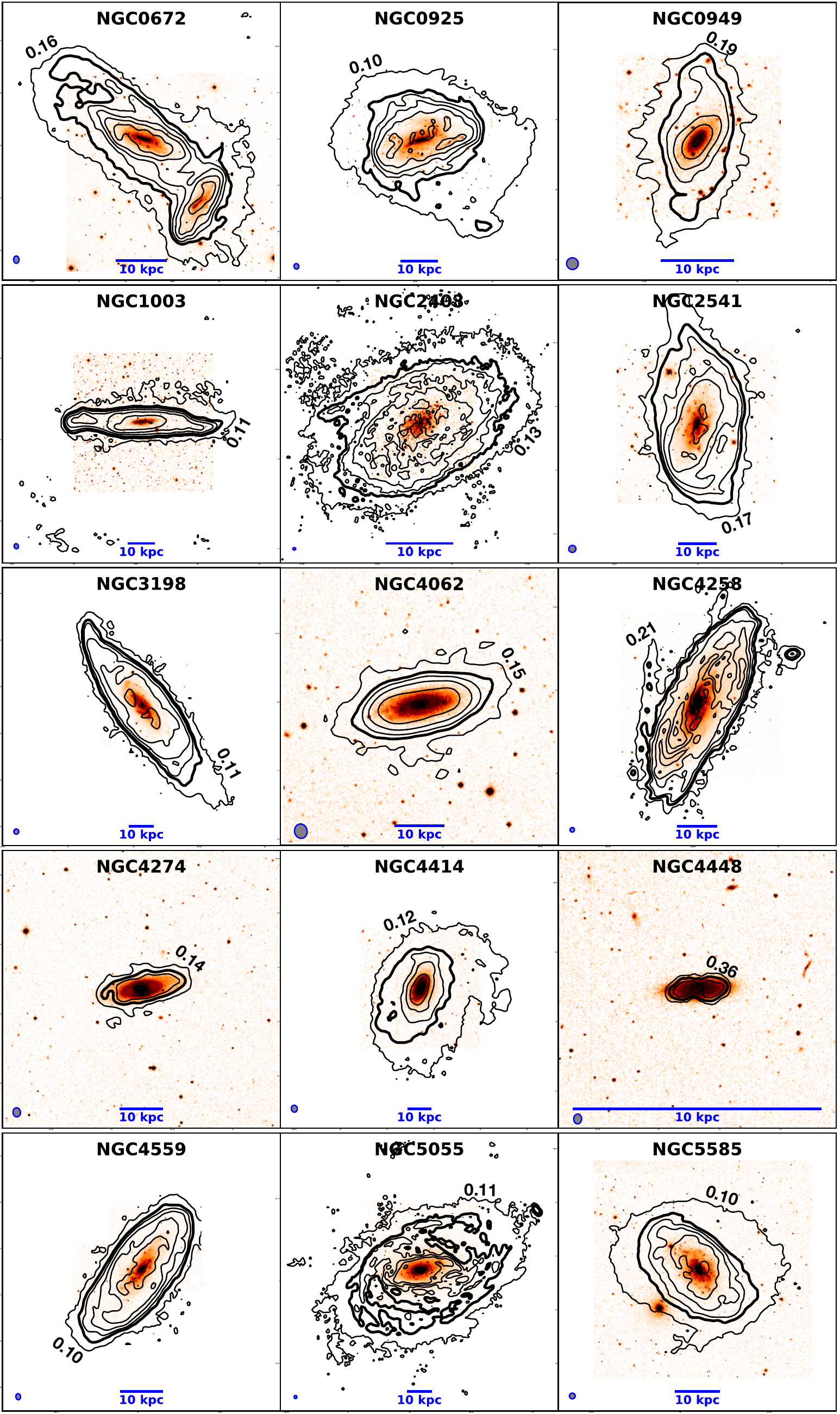}
\caption{\hicap\ column density maps for the 15 HALOGAS galaxies in our sample (black contours) overlaid on top of their optical images. 
The outermost contour is set to a S/N of 3, its value varies and is indicated on each panel in units of $\msunsqp$. The remaining contours are spaced by powers of 2, the first (thicker line) being at a column density of $1\msunsqp$. The beam size (FWHM) of the \hicap\ data is indicated in the bottom-left corner of each panel.}
\label{fig:maps}
\end{center}
\end{figure*}

Fig.\,\ref{fig:maps} shows the \hi\ column density maps for the 15 HALOGAS systems studied here (black contours) overlaid on top of their optical images.

The \hi\ column density at a given sight-line $(i,j)$, $\Sigma_{\rm HI}(i,j)$, is derived by integrating the \hi\ profile in velocity after the application of the external mask (see Section \ref{ssec:separation}), and is converted to physical units following \citet{Roberts75} as
\begin{equation}\label{eq:sigmaHI}
\frac{\Sigma_{\rm HI}(i,j)}{\msunsqp} = 8794 \sum_{k=0}^{\rm n.chan.} \left(\frac{I(i,j,k)}{\rm Jy/beam}\right) \left(\frac{\Delta {\rm v}}{\kms}\right) \left(\frac{B_{\rm maj} B_{\rm min}}{{\rm arcsec}^2}\right)^{-1} \,,
\end{equation}
where $I(i,j,k)$ is the voxel intensity, $\Delta {\rm v}$ is the channel separation and the sum is extended to all velocity channels.

The outermost column density contours in Fig.\,\ref{fig:maps} are determined under the assumption of Gaussian noise by substituting the sum that appears in eq.\,(\ref{eq:sigmaHI}) with the term $1.6\times{\rm (S/N)}\,\sigma_{\rm rms}\,\mathcal{N}^\frac{1}{2}$, where (S/N) is the desired signal-to-noise level ($3$ in our case), $\sigma_{\rm rms}$ is the rms-noise in the datacube, $\mathcal{N}$ is the median number of channels used in the computation of the column density map (which we derive from the external mask) and the factor $1.6$ accounts for Hanning smoothing.
This gives column density sensitivities in the range $0.1-0.2\msunsqp$, depending on the system.

Optical images come from various ground-based telescopes and are taken from \citet[][NGC\,0672, NGC\,4258, NGC\,5055, NGC\,5585]{Cook+14}, \citet[][NGC\,0925, NGC\,3198, NGC\,4559]{Kennicutt+03}, the DSS (NGC\,0949, NGC\,1003, NGC\,4062, NGC\,4274, NGC\,4414, NGC\,4448), \citet[][NGC\,2403]{Brown+14} and \citet[][NGC\,2541]{Baillard+11}.

Additional visualisation of these and all other HALOGAS galaxies, including presentation of ancillary multiwavelength data products, will be provided in the HALOGAS Atlas paper (Heald et al, in prep).

\end{document}